\documentclass[a4paper,11pt,reqno]{article}
\usepackage{a4wide}

\usepackage{amsmath,amsfonts,amssymb}
\usepackage[T1]{fontenc}
\usepackage[p,osf]{cochineal}
\usepackage[varqu,varl,var0]{inconsolata}
\usepackage[scale=.95,type1]{cabin}
\usepackage[vvarbb]{newtxmath}
\usepackage[cal=boondoxo]{mathalfa}
\usepackage[mathscr]{euscript}
\usepackage{bbold} 
\usepackage{cite}
\usepackage[toc,page]{appendix}
\usepackage{graphicx}
\usepackage{musicography}
 \usepackage{soul}

\usepackage{lettrine} 
\usepackage{pict2e}

\makeatletter
\DeclareRobustCommand{\loplus}{\mathbin{\mathpalette\dog@lsemi{+}}}
\DeclareRobustCommand{\lotimes}{\mathbin{\mathpalette\dog@lsemi{\times}}}
\DeclareRobustCommand{\roplus}{\mathbin{\mathpalette\dog@rsemi{+}}}
\DeclareRobustCommand{\rotimes}{\mathbin{\mathpalette\dog@rsemi{\times}}}

\newcommand{\dog@rsemi}[2]{\dog@semi{#1}{#2}{-90,90}}
\newcommand{\dog@lsemi}[2]{\dog@semi{#1}{#2}{270,90}}
\newcommand{\dog@semi}[3]{%
  \begingroup
  \sbox\z@{$\m@th#1#2$}%
  \setlength{\unitlength}{\dimexpr\ht\z@+\dp\z@\relax}%
  \makebox[\wd\z@]{\raisebox{-\dp\z@}{%
    \begin{picture}(1,1)
    \linethickness{\variable@rule{#1}}
    \roundcap
    \put(0.5,0.5){\makebox(0,0){\raisebox{\dp\z@}{$\m@th#1#2$}}}
    \put(0.5,0.5){\arc[#3]{0.5}}
    \end{picture}%
  }}%
  \endgroup
}
\newcommand{\variable@rule}[1]{%
  \fontdimen8  
  \ifx#1\displaystyle\textfont3\else
    \ifx#1\textstyle\textfont3\else
      \ifx#1\scriptstyle\scriptfont3\else
        \scriptscriptfont3\relax
  \fi\fi\fi
}
\makeatother

\newcommand{\p}{\partial}
\newcommand{\scri}{{\cal I}}
\newcommand{\nn}{\nonumber}

\renewcommand{\textbf}[1]{\begingroup\bfseries\mathversion{bold}#1\endgroup}

 \usepackage{braket}

\usepackage{pict2e}


\newcommand{\be}{\begin{equation}}
\newcommand{\ee}{\end{equation}}
\newcommand{\barray}{\begin{array}}
\newcommand{\earray}{\end{array}}
\newcommand{\bea}{\begin{eqnarray}}
\newcommand{\eea}{\end{eqnarray}}
\newcommand{\bs}{\begin{subequations}}
\newcommand{\es}{\end{subequations}}

\newcommand{\bit}{\begin{itemize}}
\newcommand{\eit}{\end{itemize}}
\newcommand{\bd}{\begin{description}}
\newcommand{\ed}{\end{description}}

\def\w{\wedge}

\DeclareMathOperator{\Tr}{Tr}
\newcommand{\f}{\frac}

\renewcommand{\a}{\alpha}    
  \newcommand{\eps}{\epsilon}

\let\m=\mu        \let\om=\omega

 \let\Om=\Omega

\usepackage{nicefrac}
\usepackage{setspace}
\usepackage{datetime}

\usepackage{comment}

\usepackage[euler]{textgreek}
\usepackage[breaklinks=true]{hyperref}

\numberwithin{equation}{section}

\hypersetup{
			colorlinks=true,
			urlcolor=blue,
			citecolor=magenta,
			linkcolor=blue,
     }

\let\oldsqrt\sqrt
\def\sqrt{\mathpalette\DHLhksqrt}
\def\DHLhksqrt#1#2{%
\setbox0=\hbox{$#1\oldsqrt{#2\,}$}\dimen0=\ht0
\advance\dimen0-0.2\ht0
\setbox2=\hbox{\vrule height\ht0 depth -\dimen0}%
{\box0\lower0.4pt\box2}}

\newcommand{\RNum}[1]{\uppercase\expandafter{\romannumeral #1\relax}}

\author{
  \begin{minipage}{.97\linewidth}
    \vspace{1cm}
       \begin{center}
      \begin{small}
             \textbf{Antoine Rignon-Bret}$^{1,}$\footnote{\href{arignonbret@gmail.com}{arignonbret@gmail.com}} 
     \, and 
      \textbf{Matthieu Vilatte}$^{2,}$\footnote{\href{matthieu.vilatte@umons.ac.be}{matthieu.vilatte@umons.ac.be}}
              \end{small}
    \end{center}
    \vspace{0.5cm}
    \hspace{2.4cm}\begin{minipage}{.7\linewidth}
\begin{center}     {\it \begin{footnotesize}
\hbox{\kern-1.8cm\vbox{\vskip0cm
 \begin{itemize}
               \item[$^1$]Universit\'e de Lorraine, CNRS, \\ Laboratoire de Physique et de Chimie Th\'eoriques,\\
               F-54000, Nancy, France\\
                           \vskip0.25cm
      \end{itemize}}
\kern-3.2cm\vbox{
\begin{itemize}
 \item[$^2$]Service de Physique de l'Univers, Champs et Gravitation, \\
        Université de Mons -- UMONS, \\ 
        Place du Parc 20, 7000 Mons, Belgium         
      \end{itemize}
      \vskip0.cm
}}
     \end{footnotesize}}
\end{center}
    \end{minipage}
  \end{minipage}
}

\title{\vspace{1.5cm}
 \boldmath \begin{LARGE}
    \textbf{\textsc{Black hole thermodynamics at null infinity \\
    Part 1: Dual Generalized Second Law}}
  \end{LARGE} \unboldmath
}

\date{}

\begin{document}

\begin{titlepage}
\maketitle
\thispagestyle{empty}

\begin{center}
\textsc{Abstract}\\  
\vspace{1. cm}	
\begin{minipage}{1.0\linewidth}

The generalized second law (GSL) of black hole thermodynamics asserts the monotonic increase of the generalized entropy combining the black hole area and the entropy of quantum fields outside the horizon. Modern proofs of the GSL rely on information-theoretic methods and are typically formulated using algebras of observables defined on the event horizon together with a vacuum state invariant under horizon symmetries, inducing a geometric modular flow. In this work, we formulate a dual version of the generalized second law from the perspective of asymptotic observers at future null infinity, who do not have access to the black hole area. Our approach exploits the dependence of the second law on the choice of algebra of observables and of a reference state invariant under suitable symmetries, in close analogy with open quantum thermodynamics. Using algebraic quantum field theory and modular theory, we analyze several physically motivated vacuum states, including the Hartle–Hawking state and two classes of regularized vacua. We show that, at null infinity, the monotonic quantity governing an irreversible evolution is no longer the generalized entropy, but rather a thermodynamic potential constructed from asymptotic observables. Depending on the chosen vacuum, this potential takes the form of the free energy or of a generalized grand potential built from the Bondi mass and additional (angular) mode-dependent chemical potentials. The resulting inequalities define a dual generalized second law at future null infinity, which can be consistently combined with the standard GSL involving variations of the black hole area.

\end{minipage}
\end{center}

\end{titlepage}

\onehalfspace

\begingroup
\hypersetup{linkcolor=black}
\tableofcontents
\endgroup
\noindent\rule{\textwidth}{0.6pt}



\section{Introduction}
\label{sec: intro}

For more than two centuries, thermodynamics, and in particular its second law, has been one of the cornerstones of physics. It has repeatedly guided the development of fundamental theories, from the birth of quantum mechanics -- which resolved the ultraviolet catastrophe arising from the equipartition theorem for the electromagnetic field -- to the modern search for a quantum theory of gravity \cite{planck1978gesetz, einstein1905erzeugung}. A remarkable indication of this connection is that black holes behave as thermodynamic systems and possess an entropy given by the celebrated Bekenstein-Hawking formula \cite{bekenstein1973black, Hawking:1974rv, hawking1975particle}
\be \label{bakeinsteinhawentropy}
    S_{\text{BH}} = \frac{A}{4 G \hbar},
\ee
where $A$ denotes the area of a spacelike section of the event horizon.

Unlike in classical gravity, however, quantum effects such as Hawking evaporation allow the horizon area to decrease, apparently violating the classical area theorem \cite{Hawking:1971tu, bardeen1973four}. This observation led Bekenstein to propose the generalized entropy
\be \label{eq: gslintro}
    S_{\text{gen}} = \frac{A}{4 G \hbar} + S_{\text{out}}
\ee
whose monotonic increase constitutes the generalized second law (GSL) \cite{bekenstein1974generalized, hawking1975particle, bekenstein2020black}. Understanding the microscopic origin of this law remains one of the central questions in black-hole thermodynamics \cite{leutheusser2023causal, Witten:2021unn, Chandrasekaran:2022cip, Chandrasekaran:2022eqq, Jensen:2023yxy, Kudler-Flam:2023qfl, ali2024crossed, Ali:2024jkx, DeVuyst:2024khu, DeVuyst:2024grw, Faulkner:2024gst, Kudler-Flam:2024psh, DeVuyst:2024fxc, Chen:2024rpx, Speranza:2025joj}.

Modern proofs of the GSL \cite{sorkin1998statistical, Wall:2009wm, Wall:2010cj, Wall:2011hj, Wall:2015raa, Faulkner:2024gst, Kirklin:2024gyl} rely on the monotonicity of the relative entropy in quantum field theory \cite{lindblad1975completely, uhlmann1977relative, araki1975relative, petz2003monotonicity, ohya2004quantum}. In its most systematic form, due to Wall \cite{Wall:2011hj}, one compares an arbitrary state with a preferred vacuum and exploits the monotonicity of the relative entropy under restriction to nested algebras of observables. Through the geometric interpretation of the modular Hamiltonian and the semiclassical Einstein equations, this information-theoretic statement becomes precisely the generalized second law.

Besides providing elegant proofs of the GSL, the relative entropy \cite{araki1975relative, umegaki1962conditional} has emerged as one of the fundamental quantities of quantum field theory. Unlike the von Neumann entropy, it remains finite for local algebras and has played a central role in subjects ranging from the Bekenstein bound to quantum energy inequalities and semiclassical gravity \cite{casini2008relative, Casini:2019qst, Kudler-Flam:2023hkl, Bousso:2015wca, Wall:2017blw, ceyhan2020recovering, Hollands:2025glm, Hollands:2025vpc, Dorau:2025hmq}. These developments suggest that the second law should not be regarded as a peculiarity of a particular spacetime construction, but rather as a general consequence of the algebraic structure of quantum field theory.

A useful perspective on the GSL comes from open quantum thermodynamics  \cite{spohn1978entropy, schumacher2000relative, breuer2002theory, vedral2002role, goold2016role, alicki2019introduction, landi2021irreversible}. There, changing the environment with which a system exchanges conserved quantities modifies the stationary state of the dynamics and changes the appropriate thermodynamic potential. An isolated system evolves towards states of increasing entropy, whereas a system coupled to a thermal reservoir \cite{kubo1957statistical, martin1959theory} is governed instead by the monotonic decrease of the Helmholtz free energy \cite{landau2013statistical}. More general reservoirs exchanging particles or other conserved quantities lead naturally to grand potentials and generalized Gibbs ensembles \cite{Jaynes1957Information, Jaynes1957InformationII, Rigol2007Relaxation, Ilievski2015Complete, Essler2016Quench, YungerHalpern2019BeyondHeatBaths}. From this viewpoint, the freedom in choosing the vacuum state in quantum field theory strongly resembles the freedom of choosing the thermodynamic environment. Different vacua should therefore be expected to define different thermodynamic potentials and, consequently, different formulations of the second law.

An interesting setting in which to explore these ideas is future null infinity. Although future null infinity and the future event horizon are both non-expanding null hypersurfaces and admit closely related algebraic descriptions \cite{Ashtekar:2024mme, Ashtekar:2024bpi, Ashtekar:2024stm, ARB24, Agrawal:2025fsv, Ruzziconi:2025fct, Ruzziconi:2025fuy, Tan:2025mre}, they correspond to different physical observers and therefore naturally suggest different thermodynamic descriptions. While the generalized entropy governs the evolution from the viewpoint of horizon observers, asymptotic observers only have access to the Bondi mass and the outgoing Hawking radiation. This raises the question of whether a second law can be formulated entirely in terms of asymptotic observables.

A first attempt was proposed by one of us in \cite{ARB24}, where it was argued that asymptotic observers satisfy a \emph{dual generalized second law}. In the idealized limit of a perfectly thermal Hawking flux, the relevant thermodynamic potential is the Helmholtz free energy built from the Bondi mass. More generally, graybody factors modify the stationary state seen at null infinity, producing additional non-geometric contributions that naturally take the form of effective chemical potentials. From the perspective advocated here, these corrections simply reflect that the asymptotic vacuum defines a thermodynamic environment which differs from an ideal thermal reservoir.

A realistic description of Hawking radiation therefore requires going beyond the Hartle-Hawking vacuum \cite{israel1976thermo, hartle1976path, Unruh:1976db}. Indeed, this state, which describes a black hole in thermal equilibrium, is incompatible with asymptotic flatness, as it corresponds to a thermal bath extending all the way to spatial infinity and carrying an infinite energy flux. Moreover, although the quantum field is thermal near the horizon, the effective potential surrounding the black hole backscatters part of the outgoing radiation, so that the spectrum observed at future null infinity is no longer exactly thermal. In particular, high-angular-momentum modes are strongly suppressed. 
Motivated by these observations, we introduce two regularizations of the Hartle-Hawking vacuum. A hard regularization suppresses all modes above a prescribed angular momentum cutoff, whereas a soft regularization assigns an effective temperature to each angular momentum mode, thereby mimicking the angular momentum dependent transmission probabilities associated with graybody factors.

Within these three frameworks, i.e. the Hartle-Hawking vacuum and its hard and soft regularizations, we derive the corresponding dual generalized second laws. For the hard regularization, the relevant thermodynamic potential is the Helmholtz free energy,
\be \label{eq: freee1}
    \Delta M - T_H \Delta S \leq 0,
\ee
where $M$ denotes the Bondi mass, $S$ the entropy of the quantum fields between two cuts of future null infinity, and $T_H$ the Hawking temperature. Combining this inequality with the generalized second law on the horizon yields
\be \label{eq: freee2}
    \Delta M - T_H \Delta \left(\frac{A}{4G} + S \right) \leq 0
\ee
which reduces to the ordinary generalized second law when the 
asymptotic slices terminate on the same cut. For the soft regularization, the effective chemical potentials generated by graybody factors replace the Helmholtz free energy by a generalized grand potential,
\be \label{eq: freee1s}
    \Delta M - \sum_{lm} \int_0^{+ \infty}  \m_{\om l} \Delta n_{\om l m} \text{d} \om - T_H \Delta S \leq 0,
\ee
whose precise form depends on the chosen regularization. The detailed assumptions underlying these inequalities are discussed in the main text.

In particular, compared with \cite{ARB24}, the present work develops the dual generalized second law from a fully algebraic perspective. We construct explicitly the relevant algebras of observables, identify the corresponding vacuum states, derive their modular Hamiltonians, and establish the relation between the resulting modular energies and the geometric quantities entering the semiclassical Einstein equations. In addition, we introduce physically motivated regularizations of the Hartle-Hawking vacuum that account for graybody effects while preserving the algebraic structure of the proof.

To keep the paper reasonably self-contained, we complement the main text with six appendices reviewing material scattered throughout the literature. Appendix \ref{app: unruh} proves the Unruh effect for arbitrary non-expanding null hypersurfaces, making it applicable both to horizons and future null infinity. Appendix \ref{app: QFT curved} reviews the construction of vacuum states and one-particle Hilbert spaces from positive-frequency solutions. Appendix \ref{app: AQFT} introduces the basics of algebraic quantum field theory. Appendix \ref{app: modular theory} reviews von Neumann algebras and modular theory, while Appendix \ref{app: modular boost energy} relates the modular Hamiltonian to the integral of the normal-ordered stress-energy tensor. Finally, Appendix \ref{App: MinkowskivacGSL} complements the analysis of the paper by proving the dual GSL in the Minkowski vacuum, thereby connecting our framework with the results of \cite{Kapec:2016aqd, Bousso:2016vlt, Bousso:2016wwu}.

The paper is organized as follows. Section~\ref{sec: max scri and einstein} reviews the Schwarzschild geometry, introduces the matter content (namely a scalar field) on the horizon and at future null infinity, and derives the corresponding semiclassical Einstein equations. Section~\ref{sec: quantization} constructs the algebra of observables at null infinity, quantizes the field, and studies the Hartle-Hawking vacuum together with its hard and soft regularizations. Section~\ref{sec: GSL proof} establishes the dual generalized second law from the monotonicity of the relative entropy and derives the associated thermodynamic potentials in each of these settings. Throughout this work we set $c = k_B = \hbar = G = 1$.


\section{Setting the stage: kinematics and semiclassical dynamics}
\label{sec: max scri and einstein}

\begin{figure}[ht]
        \center
        
    \includegraphics[width=0.65\textwidth]{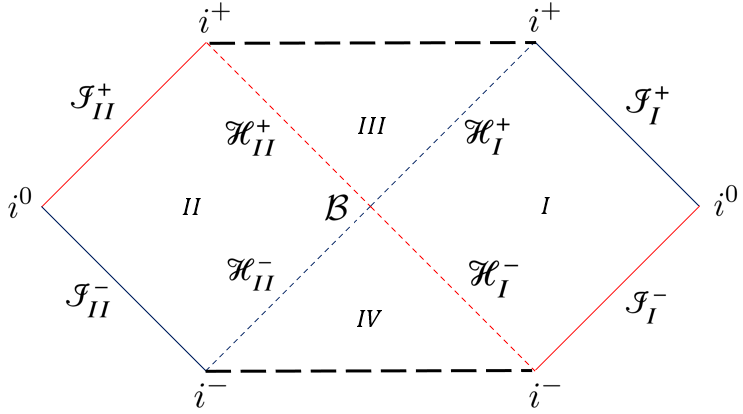}
    \caption{Penrose diagram of the maximal extension of the Schwarschild solution. The solid blue (reps. red) regions depict $\scri_R$ (resp. $\scri_L$), whereas the dashed blue (resp. red) regions depict $\mathcal{H}_L$ (resp. $\mathcal{H}_R$). Hence, the whole blue (resp. red) region represent $\Sigma_f$ (resp. $\Sigma_i$).}
    \label{fig: penrose eternal schwars}
    \end{figure}

We consider in this work the maximal extension of an eternal Schwarzschild black hole in $D$-dimensions, with $D \geq 2$. Most of the time in this work we will be in $D = 4$ dimensions. Upon conformal compactification, one gets the Penrose diagram depicted in Figure \ref{fig: penrose eternal schwars}. While regions $I$ and $II$ represents the space outside the event horizons, region $III$ and region $IV$ are respectively the black and white hole. Region $I$ is 
bounded by the white hole horizon $\mathcal{H}^{-}_{I}$, the black hole horizon $\mathcal{H}^{+}_{I}$, past $\scri^{-}_{I}$ and future $\scri_{I}^{+}$ null infinity (and similarly for region $II$). The reunion of several of these regions will be widely used in the remaining of the paper so we define them now once and for all.
\begin{itemize}
    \item The reunion $\mathcal{H}^{-}_{I} \cup \mathcal{H}^{+}_{II}$ will be called \emph{the right horizon} and denoted $\mathcal{H}_R$.

    \item The reunion $\mathcal{H}^{-}_{II} \cup \mathcal{H}^{+}_{I}$ will be called \emph{the left horizon} and denoted $\mathcal{H}_L$.

    \item The reunion $\scri^{+}_{I} \cup \scri^{-}_{II}$ will be called \emph{right maximally extended null infinity} and denoted $\scri_R$.

    \item The reunion $\scri^{+}_{II} \cup \scri^{-}_{I}$ will be called \emph{left maximally extended null infinity} and denoted $\scri_L$.
\end{itemize}
The reason why these regions are introduced will become clearer in the next paragraph. For the time being we remark that Kruskal coordinates, which we denote $(\tilde{U}, \tilde{V})$, are a suitable pair of coordinates from which we are able to describe the entire spacetime (we also have to complete these coordinates with a set of angular coordinates $(x^A), A = 2,\dots,D$ in $D > 2$). These coordinates, which are inertial and affine on $\mathcal{H}_R$ and $\mathcal{H}_L$ respectively, go to infinity at spacelike infinity $\iota^0$ and vanish at the bifurcation surface $\mathcal{B}$. Also, on $\scri_R$, we have that $\tilde{V} = \pm \infty$ and on $\scri_L$, we have $\tilde{U} = \pm \infty$. The affine coordinates on $\scri^+_I$ and $\scri^-_{II}$ are denoted $u_+$ and $u_{-}$ while $v_+$ and $v_{-}$ are respectively affine on $\scri^-_{I}$ and $\scri^{+}_{II}$. The relations between the affine and Kruskal coordinates are
\be
\label{relationKruskalaffine}
\tilde{U} = \left\{ \begin{array}{ll}
        -e^{-\kappa u_+} & \mbox{on} \quad \scri^+_I \\
        e^{\kappa u_-} & \mbox{on} \quad \scri^-_{II}
    \end{array}
    \right.
    \quad \text{and} \quad \tilde{V} = \left\{\begin{array}{ll}
        e^{\kappa v_+} & \mbox{on} \quad \scri^-_I \\
        -e^{-\kappa v_-} & \mbox{on} \quad \scri^+_{II} 
    \end{array}
    \right. \, ,
\ee
where $\kappa = \frac{1}{4M}$ is the surface gravity.

This spacetime exhibits four isometries, one of which is generated by a Killing vector that is timelike in the regions $I$ and $II$ and can be written in terms of the Kruskal coordinates as 
\begin{equation}
\label{timelikekilling}
    \xi = \kappa (\tilde{V} \partial_{\tilde{V}} - \tilde{U} \partial_{\tilde{U}}) \, .
\end{equation}
This Killing field generates the natural time translations in the aforementioned regions. Moreover the left horizon $\mathcal{H}_L$ of the black hole is a Killing horizon for that vector (evaluated at $\tilde{U} = 0$) while $\mathcal{H}_R$ is Killing for the same vector but evaluated at $\tilde{V} = 0$.

\subsection{A scalar field on the Schwarzschild background}
\label{subsectionschback}

We study the propagation and properties of a massless quantum field on such a background. For a massless scalar field $\Phi = \frac{\phi}{r}$ the Klein-Gordon equation reads in the external regions
\begin{equation} 
\label{reggewheelereq}
    \left(\partial_{u_\pm} \partial_{v_\pm} + V(r, x^A)\right) \phi(u_\pm, v_\pm, x^A) = 0
\end{equation}
where $V(r, x^A)$ is a linear operator that vanishes in a Schwarzschild background at $r = 2M$ and $r = + \infty$, i.e on the horizon and at null infinity. Also $u_{\pm} = t_{\pm} - r^\ast_{\pm}$ and $v_\pm = t_\pm + r^\ast_\pm$ are\footnote{Recall that $r^* = r + 2M\ln \left( \frac{|r-2M|}{2M}\right)$.} the retarded and advanced time coordinates which are finite in the region $I$ and $II$. Note that a similar equation holds for massless fields with higher spins, especially for $s = 1,2$, even though we focus in this work on scalar fields. Solving the equations of motion \eqref{reggewheelereq} requires to specify an initial Cauchy slice as well as a final one. Since in Schwarzschild spacetime, there are no stable bounded solutions for massless fields (they either fall into the black hole or escape at null infinity)\footnote{However, there are unstable ones, for instance the photon sphere at $r = 3M$.}\cite{Price:1971fb, Price:1972pw, chandrasekhar1998mathematical, dafermos2005proof} a good example of complete initial data surface on which we can specify the massless field's initial data is
\begin{equation}
\label{cauchyinitial}
    \Sigma_i = \mathcal{H}_R \cup \scri_L \, ,
\end{equation}
while a final null data surface can be given equivalently by 
\begin{equation}
\label{cauchyfinal}
    \Sigma_f = \mathcal{H}_L \cup \scri_R \, ,
\end{equation}
see Figure \ref{fig: penrose eternal schwars}.

Since the linear operator $V$ vanishes on the horizon and null infinity, it vanishes on $\Sigma_i$ and $\Sigma_f$, so that \eqref{reggewheelereq} reads
\begin{align} \label{reggewheelercauchy}
    \partial_{u_\pm} \partial_{v_\pm} \left(\phi(u_\pm, v_\pm, x^A)\right) = 0 \quad \text{on $\Sigma_i$ and $\Sigma_f$} .
\end{align}
The solutions of \eqref{reggewheelercauchy} on $\Sigma_{i,f}$ can be written as a sum separating left and right moving modes 
\begin{equation} \label{soleqofmot}
    \phi(u_\pm, v_\pm, x^A) = f (u_\pm, x^A) + g (v_\pm, x^A) 
\end{equation}
where $f (u_\pm, x^A)$ and $g (v_\pm, x^A)$ are arbitrary smooth functions. In the two dimensional case, this separation occurs at any point, but in the four dimensional case, we only have it on the horizon or at null infinity, because the potential vanishes there. These regions can be covered both by the affine or Kruskal coordinates \eqref{relationKruskalaffine}. Therefore, we can express the arbitrary functions $f(u_\pm, x^A)$ (respectively $g(v_\pm, x^A)$) in terms of $\tilde{U}$ (respectively $\tilde{V}$) instead of $u_\pm$ (respectively $v_\pm$). In fact, we can perform any smooth arbitrary coordinate transformations $U = U(u_\pm)$ and $V = V(v_\pm)$ and still have a valid solution of the equations of motion on $\Sigma_{i,f}$. 
This is reminiscent of a two dimensional conformal field theory, that is classically invariant under any holomorphic (respectively anti-holomorphic), transformation $Z = Z(z)$ and $\bar{Z} = \bar{Z}(\bar{z})$. In particular, we can expand the solutions on $\Sigma_{i,f}$ in terms of the (yet unormalized) modes
\begin{equation}
\label{eq: modes omega lm}
    \phi_{\Omega l m} = e^{- i \Omega U(u_\pm)} Y_m^l(x^A), \qquad \phi_{\Omega l m} = e^{- i \Omega V(v_\pm)} Y_m^l(x^A) \, 
\end{equation}
where $U(u_\pm)$ and $V(v_\pm)$ denote arbitrary choices of time coordinates. Although these different mode decompositions are classically equivalent, they generally lead to inequivalent notions of positive frequency in quantum theory, making the choice of time coordinate physically significant.

\subsection{Maximally extended null infinity}

The region that will be of most interest in the remaining of the paper is the conformal infinity appearing in the final null data surface \eqref{cauchyfinal}, namely right maximally extended null infinity $\scri_R$. It can be covered by the coordinates $(U, x^A)$ where $U$ is defined to be
\begin{align}
\label{eq: def of U}
    U &:= -\frac{1}{\tilde{U}} = \left\{ 
    \begin{array}{ll}
        e^{\kappa u_+} & \mbox{on} \quad \scri^+_R := \scri^+_I \\
        - e^{-\kappa u_-} & \mbox{on} \quad \scri^-_R : = \scri^-_{II}
    \end{array}
\right.
\end{align}
that is the inverse of the Kruskal coordinate. The asymptotic Killing field \eqref{timelikekilling} reduces to $\xi = \kappa U \p_U = \p_{u_+}$ (resp. $\xi = \p_{u_-}$) on $\scri^+_R$ (resp. $\scri^-_{R}$). The advantage of these new coordinates is that we cover $\scri_R$ using a single coordinate $U$ instead of the pair $(u_+, u_-)$, so that it is more natural to use it for a maximally extended black hole, and contrary to the Kruskal coordinate $\tilde{U}$, the generator of the "time" translation with respect to the coordinate $U$ is future oriented on $\scri^+_R$, while $\partial_{\tilde U}$ would have been past oriented. Similarly, we chart left maximally extended null infinity with
\begin{equation}
\label{eq: def of V}
    V := -\frac{1}{\tilde{V}} = \left\{ 
    \begin{array}{ll}
        -e^{-\kappa v_+} & \mbox{on} \quad \scri^-_L := \scri^-_I \\
        e^{\kappa v_-} & \mbox{on} \quad \scri^+_L := \scri^+_{II}
    \end{array}
\right. \, ,
\end{equation}
and therefore \eqref{timelikekilling} can also be expressed in the bulk as
\begin{equation}
    \xi = \kappa(U \partial_U - V \partial_V ) \, .
\end{equation}


Since $\scri_R$ is the union of two null hypersurfaces (in the conformally compactified spacetime), it is also a null hypersurface, and it consists of two copies of $\mathbb{R} \times S^2$, i.e two three-dimensional cylinders. One of these cylinders is covered by the patch $U > 0$ while the other one is covered by the patch $U < 0$. Therefore, even if the locus $U = 0$ has not been well defined yet, we add it to the description of $\scri_R$. This means that we glue $\scri_I^+$ to $\scri_{II}^{-}$ together through the sphere at $U = 0$, that we must morally identify to spacelike  infinity $\iota^0$. The final configurations of the field should now be specified on this null hypersurface (that we continue to call $\scri_R$, since it is different from what we previously called $\scri_R$ through a null measure set). Once we have done it, the topology of $\scri_R$ remains the topology of the three-dimensional  cylinder $\mathbb{R} \times S^2$. Analogous considerations apply also to $\scri_L$ for the initial configurations of the bulk field.

\begin{figure}[ht]
    \centering
    \includegraphics[width=0.7\linewidth]{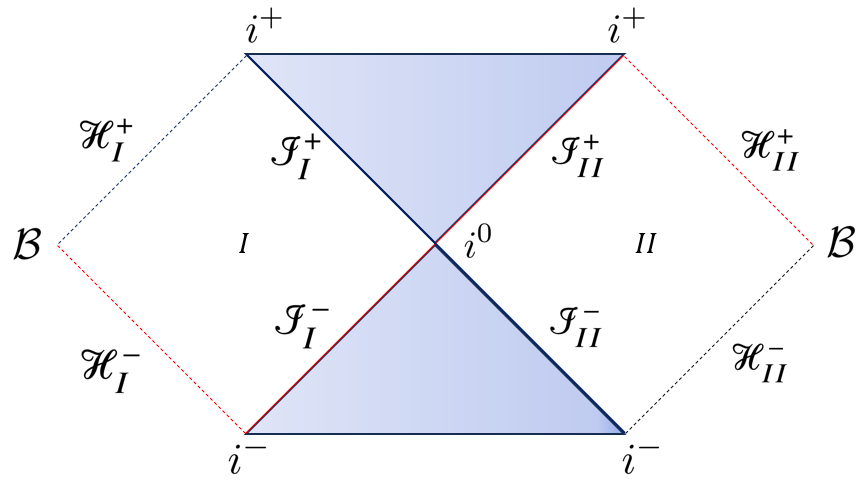}
    \caption{Conformal extension of the black hole spacetime centered on spacelike infinity. Regions $III$ and $IV$ (respectively the black and white hole regions) are not depicted here, since they lie beyond the horizons. The region shaded in blue is called $\mathcal{R}_\infty$ (see \cite{Faulkner:2024gst}) and does not exist in the physical spacetime, but can be constructed in the conformal spacetime, although it may contain singularities. It is the dual of the black hole regions in the conformal spacetime.}
    \label{fig: gluing at iota zero}
\end{figure}

The set of coordinates $(U, V, x^A)$ describes regions $I$ and $II$, so that spacelike infinity (on both regions) is located at $U = V = 0$ and the bifurcation surface is now at $U = V = \infty$, see the Penrose diagram Figure \ref{fig: gluing at iota zero}. Therefore, it is natural to identify these two regions and see $\scri_R$ and $\scri_L$ as complete null hypersurfaces in the conformal compactification, but now centered on $\iota^0$ (see Figure \ref{fig: gluing at iota zero}).
This construction is very similar to the conformal extension of Schwarzschild spacetime by Faulkner and Speranza \cite{Faulkner:2024gst}. The regions $\mathcal{R}_\infty$ (located beyond null infinity in the conformal spacetime) that they consider in their work would be covered by the coordinates $U > 0, V < 0$ and $U < 0, V > 0$.\footnote{The authors of \cite{Faulkner:2024gst} needed these regions so that they could smear the field around infinity and get a well defined algebra of observables at infinity. Here, alternatively, we will bypass this step by quantizing the field directly on the three dimensional null hypersurface $\scri_R$ to obtain a suitable algebra.} However, notice that the coordinates $U$ and $V$ cannot cover the black hole and white hole interiors. Similarly, the Kruskal coordinates $(\tilde{U}, \tilde{V})$ did not cover their conformally extended region $\mathcal{R}_\infty$. It was also outlined in \cite{Faulkner:2024gst} that the algebra of observables at $\scri^+_{I}$ was entangled with the algebra of observables at $\scri^{-}_{II}$, and that the restriction of the Hartle-Hawking state was a pure state when restricted to this algebra. \footnote{Therefore, the union of the algebra of observables at $\scri^+_I$ and at $\scri^-_{II}$ is type I von Neumann algebra, see Appendix \ref{app: modular theory} for some basics of von Neumann algebras.} These considerations align with ours, as we will see in the next sections of this paper. The take home message of our construction is that maximally extended null infinity $\scri_R$ and $\scri_L$ are suitable null initial data surfaces in a black hole background and that it makes sense to identify them as overcomplete null hypersurfaces with the topology of the cylinder in the conformal spacetime. It renders the situation symmetric with respect to the horizon. However, it is possible to think about the later constructions from a purely algebraic point of view and define the algebra of observable on $\scri_R$ (resp. $\scri_L$) as the algebraic union of the algebra on $\scri^+_I$ and the algebra on $\scri^-_{II}$ (resp. the union of the algebra on $\scri^+_{II}$ and the one on $\scri^-_I$) without referring to the gluing.

\subsection{Semiclassical Einstein equations}
\label{subsec: einstein eq}

Having defined the kinematics of our problem we now focus on dynamics, both on the future horizon $\mathcal{H}^+_I$ charted by the patch $ \tilde V > 0$ and at $\scri^+_I$ charted by $U > 0$ or the affine $u_+$ (plus angles). These are the two regions of interest from the point of view of an external observer in region $I$ who studies the radiation falling across the horizon or escaping to infinity. The aim of this paragraph is to relate the variation of the canonical charges of interest, namely the area of the horizon $A$ or the Bondi mass $M$, to an integral of the matter stress-energy tensor.

\subsubsection{On the horizon}
\label{athorizoneesc}

The dynamics of a black hole horizon is contained in the completely longitudinal and the mixed projections of Einstein's equations, dubbed respectively the Raychaudhuri \cite{Raychaudhuri} and Damour \cite{Damour:1979wya} equations. When the horizon is perturbed at order $\mathcal{O}(\varepsilon)$ by matter (with $\varepsilon \ll 1)$, we neglect any higher order contributions like e.g. the shear square $\sigma^2$ and the (affine) expansion square $\theta^2_l$. Under this hypotheses the Raychaudhuri equation becomes 
\be \label{firstrayperphor}
    \frac{d \theta_l}{d \tilde V} = - 8 \pi T_{\tilde V \tilde V} + O(\varepsilon^2)
\ee
where $\boldsymbol{l} = \p_{\tilde V}$ is an affine normal to the horizon. If we go to the semiclassical regime, the perturbation in stress energy is of order $O(\hbar)$, and we have to replace the classical stress energy component $T_{\tilde V \tilde V}$ by the average in the quantum state $\ket{\Psi}$ of its (properly renormalized) stress energy tensor $\langle T_{\tilde V \tilde V} \rangle_\Psi$, so that we can rewrite \eqref{firstrayperphor} as 
\be \label{secondrayperphor}
    \frac{d^2 \eps_S}{d\tilde V^2} = - 8 \pi \langle T_{\tilde V \tilde V} \rangle_\Psi \eps_S + O(\hbar^2)
\ee
where $\eps_S$ is the volume form of a co-dimension one cross-section of the horizon.\footnote{The latter is obtained via the pullback of the ambient space volume form $\eps_\mathcal{M}$
\begin{equation}
    \label{pullbackvolform}
    \eps_S = \iota_{\boldsymbol{l}}\iota_{\boldsymbol{k}}\eps_{\mathcal{M}}
\end{equation}
where $\boldsymbol{k} = k^{\tilde U} \partial_{\tilde U} + k^A \partial_A$ is an auxiliary vector field such that its associated one-form via the ambient space musical isomorphism reads $\boldsymbol{k}^{\musFlat} = -\text{d}\tilde V$.}
Then, for an arbitrary cross-section at $\tilde V = \tilde V_0$, we can multiply both sides of \eqref{secondrayperphor} by the factor $-(\tilde V - \tilde V_0)$ and after a partial integration on a cross section $S$ of constant affine parameter $\tilde V$, we get
\be \label{thirdrayperphor}
    \frac{d}{d\tilde V} \left[ \f14 \left(A - (\tilde V - \tilde V_0) \frac{dA}{d\tilde V}\right) \right] = 2 \pi \int_{S} (\tilde V - \tilde V_0) \langle T_{\tilde V \tilde V} \rangle_\Psi \eps_S
\ee
where $A$ is the area of the cross section $S$. The quantity 
\be
    S_D := \f14 \left(A - (\tilde V - \tilde V_0) \frac{d A}{d\tilde V}\right)
\ee
has been referred as the Dirichlet entropy in \cite{Rignon-Bret:2023fjq} and as the \emph{dynamical entropy} of the Killing horizon in the case where $\tilde V_0 = 0$ \cite{Hollands:2024vbe, Visser:2024pwz, Odak:2023pga, Ciambelli:2023mir}. It is the quantity that satisfies a local version of the physical process first law \cite{Rignon-Bret:2023lyn, Rignon-Bret:2023fjq, Hollands:2024vbe, Visser:2024pwz}. Consider now the three-dimensional domain $\mathcal{D}_0^{\mathcal{H}} = (V_0, + \infty) \times S^2$. The variation of $S_D$ between the cut $\tilde V = \tilde V_0$ and $\tilde V \to +\infty$ is given by the null time integration of \eqref{thirdrayperphor}
\be
    \Delta S_D = 2 \pi \int_{\mathcal{D}_0^\mathcal{H}} (\tilde V - \tilde V_0) \langle T_{\tilde V \tilde V} \rangle_{\Psi} \text{d}\tilde V \w \eps_S
\ee
that is positive if the null energy conditions are satisfied.\footnote{They are usually satisfied classically but not in quantum field theory, where the energy density can be in general arbitrary low.} Assuming\footnote{Wall qualified in \cite{Wall:2011hj} this late time boundary condition as ``theological".} $\underset{\tilde V \rightarrow + \infty}{\lim} \tilde V \theta_l  = 0$, we get 
\be \label{differenceareasde}
   \f14 (A\rvert_{+ \infty} - A \rvert_{\tilde V = \tilde V_0}) = 2 \pi \int_{\mathcal{D}_0^\mathcal{H}} (\tilde V - \tilde V_0) \langle T_{\tilde V \tilde V} \rangle_\Psi \text{d}V \wedge \eps_S := 2 \pi \langle K_{\tilde V_0} \rangle_\Psi
\ee
and we see that the quantity that appears inside the integral is the local boost energy above the cut $\tilde V = \tilde V_0$. Indeed, the boost field preserving the region delimited by the cut $\tilde V = \tilde V_0$ of the horizon is 
\be \label{localbosstfieldv}
    \xi = (\tilde V - \tilde V_0) \p_{\tilde V}
\ee
and is different from the action of the Killing field on the horizon (i.e. $\tilde V\p_{\tilde V}$), that corresponds to $\tilde V_0 = 0$ (since the bifurcation surface is located at $\tilde V = 0$). Note that the right hand side of \eqref{differenceareasde} is the quantity that appears in \cite{Wall:2011hj} and is needed to prove the generalized second law on the horizon. 

The quantity denoted $K_{\tilde V_0}$, the boost energy of the region $\tilde V > \tilde V_0$ of the horizon, is closely related to the notion of modular Hamiltonian coming from Tomita-Takesaki's theory, reviewed in Appendices \ref{app: modular theory} and \ref{app: modular boost energy}. Note however that we are being slightly imprecise here by omitting the reference to any notion of algebra of observables and Hilbert space representation built upon a reference vacuum state $\ket{\Omega}$, in which our state $\ket{\Psi}$ belongs to. Everything will become clearer in the next Section but for the time being it is important to observe that \eqref{differenceareasde} is written in terms of the covariant stress-energy tensor $T_{\tilde{V} \tilde{V}}$, while the modular hamilonian of Tomita-Takesaki involves the normal-ordered version (w.r.t. $\ket{\Omega}$). This will play a prominent role in our proof of the dual GSL in Section \ref{sec: GSL proof}.

Considering now two cuts $\tilde V = \tilde V_1$ and $\tilde V = \tilde V_2 > \tilde V_1$ we can write the variation of area as
\be
    \f14(A_{\tilde V_2} - A_{\tilde V_1}) = -2 \pi \left(\langle K_{\tilde V_2} \rangle_\Psi - \langle K_{\tilde V_1} \rangle_\Psi \right) \, .
\ee
At this point, since we look at a \textit{difference} of boost energies $\langle K_{\tilde V_i} \rangle_\Psi$, the condition $\tilde V \theta_l \underset{\tilde V \rightarrow + \infty}{\rightarrow} 0$ is not strictly necessary, all we need is that it converges to a constant.\footnote{However, if we consider that the semi classical equations apply up to $\tilde V = + \infty$, then we expect that caustics form at a finite affine parameter $\tilde V$ if we do not assume that $\tilde V \theta_l \underset{\tilde V \rightarrow + \infty}{\to} 0$.}

\subsubsection{At null infinity} \label{atnullinfeesc}

A analogous story runs at $\scri^+_R$ where Raychaudhuri's and Damour's equations are replaced by the BMS equations for the mass and angular momentum \cite{Bondi:1960jsa, sachs1962asymptotic, penrose1982quasi}. The mass equation  can be written as 
\be \label{bondimasslossform}
    \frac{d M}{d u} = - \int_S T_{uu} \, \eps_S 
\ee
where $M$ is the integrated Bondi mass,\footnote{This quantity is related to the integral on a compact cross section $S$ of the Weyl curvature coefficient $\psi_2$ and the local shear.} $u = u_+$ is the affine time on $\scri^+_R$, and $S$ is a cross section of the latter. Here $T_{uu} = T_{uu}^{\text{matter}} + \frac{1}{32 \pi} N_{AB} N^{AB}$ is the total stress energy tensor at null infinity, including also the contribution of the gravitons through the new tensor $N_{AB}$. Now, we are looking for an equivalent of \eqref{thirdrayperphor} at null infinity when we we consider perturbations with respect to the black hole background. First, the semi-classical version of \eqref{bondimasslossform} is obtained by replacing the stress energy $T_{uu}$ by the average of the (properly renormalized) stress energy tensor operator evaluated in the quantum state $\ket{\Psi}$, so that \eqref{bondimasslossform} is now
\be \label{bondimasslossform2}
    \frac{d M}{d u} = - \int_S \langle T_{uu} \rangle_\Psi \, \eps_S.
\ee
Second, we write the above equation in terms of the coordinate $U = e^{\kappa u}$ that covers $\mathcal{\scri}_R$ (exactly as the coordinate $\tilde V$ was covering $\mathcal{H}_L$ in the previous paragraph), which will turns out to be very practical in the following. Therefore, we change $\frac{d M}{du}$ into $\kappa U \frac{d M}{dU}$ and take its derivative with respect to $\p_u$. We get
\be \label{Bml33hscri}
    \kappa U^2 \frac{d^2 M}{dU^2} = \int_S \left(-\kappa^{-1} \p_u \langle T_{uu} \rangle_\Psi +  \langle T_{uu} \rangle_\Psi \right) \, \eps_S
\ee
and now we assume to have 
\be \label{assumpthor}
\kappa^{-1} \p_u \langle T_{uu} \rangle_\Psi \ll \langle T_{uu} \rangle_\Psi,
\ee
that is the equivalent on a black hole background of the condition $\theta^2_l \ll \langle T_{\tilde V \tilde V} \rangle_\Psi$. Indeed, if we assume that the small perturbations do not perturb the horizon a lot, in a black hole background, we have a constant flux of particle at future null infinity that carries stress energy. The condition \eqref{assumpthor} is verified in most of the known vacuum states one deals with in black hole thermodynamics. All of them will be carefully redefined later in the paper but we say now that for example in the Hartle-Hawking state, the particle flux is actually infinite on $\scri^+_R$ \footnote{However, it is actually finite in the physical spacetime, since the Hartle-Hawking vacuum is a Hadamard state on the whole maximally extended Schwarzschild spacetime. The divergences at null infinity come from the fact that the fall-off of the physical stress energy tensor (which is constant near infinity) are incompatible with the usual assumptions of asymptotic flatness, see Section \ref{sec: quantization}.} because we have a thermal spectrum in all the directions. Similarly, in the Unruh vacuum, we have a constant flux of particles at $\scri^+_R$,\footnote{In fact, in both the Hartle-Hawking and the Unruh vacuum, the ADM mass is formally infinite because of this constant flux of particle on $\scri^+_R$. However, in the Unruh vacuum the flux on $\scri^+_R$ is not infinite.} and therefore in both cases the condition \eqref{assumpthor} is obviously satisfied, since the left hand side is equal to zero. Excitations on top of these two vacua that leave the assumption \eqref{assumpthor} valid, correspond to "small perturbations", in the same sense as the small perturbations we considered on the horizon. Then, taking into account this approximation, and using the fact that the covariant stress energy tensor transforms as $\langle T_{uu} \rangle_\Psi = \kappa^2 U^2 \langle T_{UU} \rangle_\Psi$ under the change $u \to U$, \eqref{Bml33hscri} becomes
\be \label{bbml4}
    \frac{d^2 M}{d U^2} = \int_S \kappa \langle T_{UU} \rangle_\Psi \eps_S \, .
\ee
Consider then a cross-section of constant parameter $U = U_0$ (with affine parameter $u_0 = \kappa^{-1} \ln{U_0}$) on $\scri^+_R$, and then multiply both sides of \eqref{bbml4} by $-(U - U_0)$, similarly to what we did on the horizon, so that \eqref{bbml4} becomes
\be \label{bbml5}
    \frac{d}{dU} \left(M - (U - U_0) \frac{d M}{d U} \right) = -\int_S \kappa (U - U_0) \langle T_{UU} \rangle_\Psi \eps_S
\ee
that is similar to \eqref{thirdrayperphor}. The quantity
\be
    M_D := M - (U - U_0) \frac{d M}{d U}
\ee
is analogous to the dynamical entropy $S_D$, and we will refer to it as the \emph{dynamical Bondi mass}. Now, if we assume that
\be
\label{finitemasslatetimes}
M_D(\infty) = M \lvert_{+\infty} - \kappa^{-1} \p_u M \lvert_{+ \infty} =  M \lvert_{+\infty}
\ee
is finite, the last equality being true because in this case we need $\p_u M \lvert_{\infty} = 0$, we can define the local boost energy 
\be \label{kuzerodef}
    M_D \lvert_{+ \infty} - M_D (U_0) = M \lvert_{+ \infty} - M (U_0) = - \int_{\mathcal{D}^{\scri}_0} \kappa (U - U_0) \langle T_{UU} \rangle_\Psi \text{d}U \wedge \eps_S := - \kappa \langle K_{U_0} \rangle_\Psi
\ee
where $\mathcal{D}^{\scri}_0 = (U_0, +\infty) \times S_2$. The expression \eqref{kuzerodef} is similar to what we had on a perturbed Killing horizon. Again we remark that $K_{U_0}$ is written in terms of the covariant stress-energy tensor at $\scri_R^+$. In addition, if we consider $U = U_1$ and $U = U_2 > U_1$ to be two arbitrary cross sections of $\scri^+_R$, we have that 
\be \label{diffbondimass}
    M (U_2) - M(U_1) = \kappa (\langle K_{U_2} \rangle_\Psi - \langle K_{U_1} \rangle_\Psi) 
\ee
and it is also interesting to notice at this point that the variation of Bondi mass only requires the difference $\langle K_{U_2} \rangle_\Psi - \langle K_{U_1} \rangle_\Psi$ to be finite, which can be true even if $\langle K_{U_1} \rangle_\Psi$ and $\langle K_{U_2} \rangle_\Psi$ are not finite by themselves. In a vacuum state aiming to reproduce boundary conditions at late time compatible with a black hole collapse, we do not expect to make sense of the Bondi mass or the dynamical Bondi mass at late time, since we have a constant flux of particle at null infinity. It is the case for the example of the Unruh vacuum. However, we will still be able to make sense of the difference in \eqref{diffbondimass} between any arbitrary cross sections by considering states lying in Hilbert spaces representations built upon a vacuum state compatible with the late time boundary conditions. 

\begin{figure}[ht]
    \centering
    \includegraphics[width=0.5\linewidth]{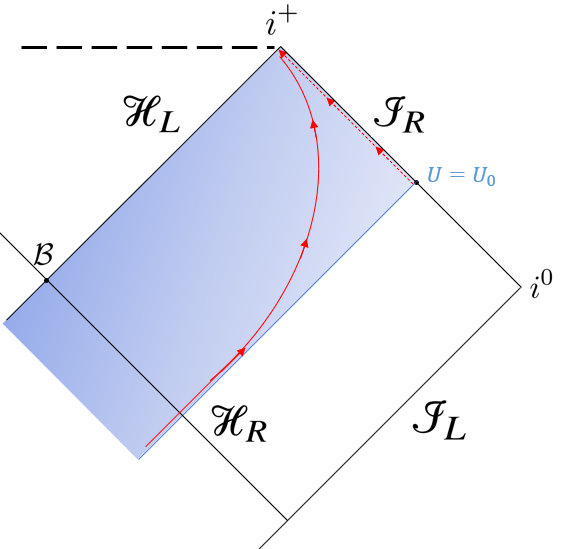}
    \caption{The trajectory in red is generated by a vector $\bar{\chi}$ in the bulk whose vertical component is given by the asymptotic vector $\chi$ (generating the dashed red trajectory in the region $U \geq U_0$ of $\scri_R$). It resembles a boost near the cut $U = U_0$ but it looks like a usual time translation at late times. The blue-shaded region represents the portion of spacetime where an observer traveling along the red curve can determine the physics using only the local measurements they can perform. It is delimited by the hypersurfaces $U = U_0$ and $U = + \infty$ (the event horizon).}
    \label{fig:placeholder}
\end{figure}

To pursue the analogy with physics on a perturbed Killing horizon further, we will introduce the local boost field 
\be \label{localbfscri}
    \chi = \kappa (U - U_0) \p_U
\ee
which vanishes at $U = U_0$ while when $U_0 = 0$, we recover the time translation Killing field 
\be \label{Killinffieldsch}
    \chi = \kappa U \p_U = \p_u \, ,
\ee
that is timelike in regions $I$ and $II$ of the black hole background. Around $U \rightarrow + \infty$ (i.e $u \rightarrow + \infty$), the vector field \eqref{localbfscri} behaves as
\be
    \chi \underset{U \rightarrow + \infty}{\sim} \kappa U \p_U = \p_u 
\ee
which is a usual time translation \eqref{Killinffieldsch} while around $U \rightarrow U_0$ (i.e $u \rightarrow u_0$), we have that 
\be
    \chi \underset{U \rightarrow U_0}{\sim} \kappa (u - u_0) \p_u
\ee
that is proportional to the vertical component of a usual boost field, vanishing on the cross section $u = u_0$. Therefore, an observer following a trajectory in (conformal) spacetime spanned by the local boost field $\chi$ will be an inertial observer at late time, since in this limit $\chi \sim \p_u$, while they will start accelerating moving backward in time, so that they never cross their past horizon at $U = U_0$ and remain causally disconnected from the region $U < U_0$ of $\scri_R$. They do not have access to the interior of the black hole region and to the patch $U < U_0$ (see Figure \ref{fig:placeholder}).

It is also worthy to notice that we can write \eqref{localbfscri} as 
\be \label{otherformforxi}
    \chi = (1 - e^{- \kappa (u - u_0)}) \p_u
\ee
so that $\chi \sim \p_u$ when $(u - u_0) \gg \kappa^{-1}$ is more explicit. Using \eqref{otherformforxi}, we can write \eqref{kuzerodef} as
\be
    \kappa \langle K_{U_0} \rangle_\Psi = \int_{u_0}^{+ \infty} \int_S (1 - e^{- \kappa(u - u_0)}) \langle T_{uu} \rangle_\Psi \text{d}u \w \eps_S \, ,
\ee
showing that $\langle K_{U_0} \rangle_\Psi $ is directly related to the energy flux.\footnote{Recall that the later is encoded in the covariant stress-tensor written in affine null time.} It is also useful to consider the time 
\be \label{modaffinetimeu0}
    \bar{u} := \kappa^{-1} \ln{(U - U_0)} = u + \kappa^{-1} \ln{(1 - e^{- \kappa(u - u_0)})}
\ee
that runs on $\mathbb{R} = (- \infty, + \infty)$ when the inertial time $u$ runs from $u_0$ to $+ \infty$. It is a nice choice of time since it covers entirely the region $U > U_0$ (i.e. $u > u_0$), and so it is a natural time used for the quantization in this region that is of most interest in the following of the paper. This concludes the study of the dynamics at $\scri_R^+$. It is worth mentioning again that we obtained similar equations for the dynamical mass than for the dynamical entropy thanks to the hypothesis \eqref{assumpthor}. The latter possesses a nice interpretation in terms of Markovian processes in open quantum thermodynamics, on which we will come back in \cite{ARBMVsoon}. 

\subsubsection{From the horizon to null infinity}

We saw in the previous two paragraphs that similar relations were displayed at the horizon and null infinity of a black hole background. An explicit comparison is summarized in Table \ref{tablethermo}. There, $T_H$ is the Hawking temperature, vN means ``von Neumann" and NEC stands for ``Null Energy Conditions", meaning that $T_{ab} n^a n^b \geq 0$ where $n^a$ is a null normal. Finally ``QF" means Quantum Fields.
\begin{table}\begin{center}\begin{spacing}{1.3} 
\begin{tabular}{|l|l|l|}
\hline \emph{Quantity} & \emph{On the future horizon} & \emph{On future null infinity} \\\hline
Canonical charge & Black hole area $\frac{\kappa}{8 \pi} A$  & Bondi mass $M$  \\
Improved charge & Dyn. Black hole area $\frac{\kappa}{8 \pi} A_D$
&  Dyn. Bondi mass $M_D$  \\
Monotonic flux (by imposing the NEC) & Hawking area theorem & Bondi mass loss formula \\
Thermodynamic interpretation & Black hole entropy (geometric) &  Black hole energy (geometric) \\
Thermodynamic potential (quantum) & 
The generalized entropy & The generalized free energy \\
Quantum corrections & $+$ vN entropy of QF & $-T_H$ $\times$ vN entropy of QF
\\\hline
\end{tabular}\end{spacing}\end{center}\vspace{-1cm}
\caption{\label{tablethermo}\emph{\small{Comparing the dynamical quantities of interest and their thermodynamic counterparts on the horizon and at future null infinity. Note that we have anticipated some of the results of the next sections.} }}
\end{table}


\section{Quantizing free fields at null infinity}
\label{sec: quantization}

In this section, we quantize the scalar field on $\scri_R$ in $D=4$. We first consider general notions of time and the corresponding decompositions of the solution space into positive and negative frequency modes, from which we construct the associated one-particle Hilbert space and vacuum state. We then investigate the symmetries of these vacuum states, focusing in particular on the transformations that leave the field's two-point function on $\scri_R$ invariant. Finally, we introduce four physically relevant (classes of) vacuum states that will play a central role throughout this work.

Readers unfamiliar with algebraic quantization or the notion of a Weyl algebra are referred to Appendices \ref{app: QFT curved} and \ref{app: AQFT} for a brief review.


\subsection{Positive frequency solutions}
\label{subsec: 4D quantization}

\subsubsection*{Generalities}

We go back to the Klein-Gordon equation for the bulk scalar field $\Phi$ in Schwarzschild coordinates, which we decompose on a basis of complex spherical harmonics\footnote{They satisfy among other properties $\bar Y^{l}_{m} = (-1)^m Y^{l}_{-m}$.}
\begin{equation}
    \label{eq: scalr decomposed 2D and harmo}
    \Phi(t,r,x^A) = \frac{1}{r}f(t,r)Y_{lm}(x^A) \, .
\end{equation}
The Klein-Gordon equation for the radial part reads
\begin{equation}
    \label{eq: KG for our field}
    \left(\partial_{t}^{2} - \partial_{r^*}^{2} + V_l(r) \right)f(t,r) = 0, \qquad V_l(r) =  \left(1 - \frac{2M}{r}\right)\left(\frac{l(l+1)}{r^2} + \frac{2M}{r^{3}} \right)
\end{equation}
with $r^*$ the tortoise coordinate, so that \eqref{eq: KG for our field} has the same form as \eqref{reggewheelereq}. When $r \to +\infty$ (or equivalently $r^* \rightarrow + \infty$), the solutions are 
given by \eqref{soleqofmot} and if we focus on $\scri_R$ we are only interested in the right moving modes of the space of solutions $\mathscr{S}$. In general, we need that the solutions decay fast enough so that we restrict the space of solutions $\mathscr{S}$ to the space of Schwartz functions on $\scri_R$, $f(U, x^A) \in \mathcal{T}(\mathbb{R} \times S^2)$. Thus, the most general decomposition of the asymptotic field $\phi = \underset{r \rightarrow + \infty}{\lim} r \Phi$ is 
\begin{equation}
    \label{eq: full quantum field}
   \phi ( U, x^A) = \sum_{l = 0}^{+ \infty} \sum_{m= - l}^{m=+l} \int_{0}^{+ \infty} \frac{d \Omega}{\sqrt{4\pi \Omega}} \left(Y_m^l(x^A) a_{\Omega l m} e^{-i \Omega U} + \bar{Y}_m^l(x^A) \bar{a}_{\Omega l m} e^{i \Omega U}\right) \, ,
\end{equation}
where we used the modes $f(U,x^A) = \frac{Y_m^l(x^A)}{\sqrt{4 \pi \Om}} e^{-i \Om U}$, $\Om > 0$, for the decomposition. 
The latter are normalized with respect to the asymptotic Klein-Gordon product 
\begin{equation}
        \label{symplecticformsu}
    \forall (f, g) \in \mathscr{S}^{\mathbb{C}}, \qquad \Om^{\mathbb{C}} (f, g) = i\int_{\scri_R} (\bar{f} \p_U g - g \p_U \bar{f}) \text{d}U \wedge \eps_S
\end{equation}
which is parametrization independent. 
Here, $\mathscr{S}^{\mathbb{C}}$ denotes the complexification of the solution space $\mathscr{S}$ of the equations of motion. Then, we promote the Fourier coefficients to quantum ladder operators $\left(\hat a_{\Omega l m}, \hat a_{\Omega l m}^{\dag}\right)$ satisfying the commutation relations 
\be 
\label{eq: lad4D}
    \left[\hat a_{\Om l m}, \hat a_{\Om' l' m'}^\dag\right] = \Om^{\mathbb{C}}\left(\frac{Y_m^l(x^A)}{\sqrt{4 \pi \Om}} e^{-i \Om U}, \frac{Y_{m'}^{l'}(x^A)}{\sqrt{4 \pi \Om}} e^{-i \Om' U}\right) = \delta(\Om - \Om') \delta_{ll'} \delta_{mm'}
\ee
so, given the canonical momenta  $\pi = \p_U \phi$, we deduce that \footnote{The Dirac distribution $\delta^{(2)}(x^A_1 - x^A_2)$ is normalized so that 
\be
    \int_S \delta^{(2)}(x^A_1 - x^A_2) \eps_S(x^A_2) = 1
\ee
where $S$ is the two dimensional unit sphere.
}
\be
    \left[\hat \phi(U_1, x^A_1) , \hat \pi(U_2, x^A_2)\right] = \frac{i}{2} \delta(U - U') \delta^{(2)}(x^A_1 - x^A_2)
\ee
On top of this, the Weyl algebra $\mathcal{A}^{\scri}$ is 
obtained by smearing the momentum field $\pi$ with a Schwartz function $f(U, x^A) \in \mathcal{T}(\mathbb{R} \times S^2)$
\be
\label{Weyl algebra}
    \mathcal{A}^{\scri} = \left\{ e^{i \pi(f)} \; \middle| \;  \forall f \in \mathcal{T}(\mathbb{R} \times S^2) \right\}.
\ee
Note that the Weyl algebra on $\scri_R$ is generated by the smeared momentum field $\pi$, rather than by the smeared field $\phi$; see Appendices \ref{app: QFT curved} and \ref{app: AQFT} for further details.
The algebra $\mathcal{A}_0$ (resp. $\mathcal{A}_0'$) is the subalgebra of $\mathcal{A}^{\scri}$ spanned by functions $f$ with support on $\scri^+_R$ (resp. $\scri^-_R$) only. An other interesting Weyl algebra is
\be
\label{4DLsmearing}
    \mathcal{A}^{\scri}_L = \left\{ e^{i \pi(f)} \; \middle| \;  \forall f \in \mathcal{T}(\mathbb{R} \times S^2), \quad \forall g \in \mathscr{S}_{2D}, \quad \int_{\scri_R} f(U, x^A) g(U) Y_m^l \text{d}U \wedge \eps_S = 0, \quad \text{if} \quad l \geq L \right\}
\ee
where $\mathscr{S}_{2D}$ is the two-dimensional space of solutions that we studied in the previous subsection. The definition \eqref{4DLsmearing} amounts to considering smearings of the field $\pi$ such that in the general decomposition \eqref{eq: full quantum fieldlm} we do not see the angular modes higher than the cutoff $L$.

The one-particle Hilbert space is built from a notion of positive frequency solutions. Simple examples are analogous to the two dimensional case as we can choose a decomposition with respect to the null time $U$ which covers entirely $\scri_R$. However, in four dimensions, we have some additional freedoms. We may for instance consider the following decomposition of the field 
\begin{equation}
    \label{eq: full quantum fieldlm}
   \hat \phi ( U, x^A) = \sum_{l = 0}^{+ \infty} \sum_{m= - l}^{m=+l} \int_{0}^{+ \infty} \frac{d \Omega}{\sqrt{4\pi \Omega}} \left(Y_m^l(x^A) \hat a_{\Omega l m} e^{-i \Omega U^{(l,m)}(U)} + \bar{Y}_m^l(x^A) \hat a_{\Omega l m}^{\dagger} e^{i \Omega U^{(l,m)}(U)}\right) \, ,
\end{equation}
where $U^{(l,m)} = U^{(l,m)}(U)$ is a choice of time attached to a particular sector defined by the angular mode $(l,m)$, so that the time $U^{(l,m)}$ can be different for two different sectors $(l, m) \neq (l', m')$. Indeed, we can check that still
\be \label{eq: kgulmmodedecom}
    \Om^{\mathbb{C}}\left(Y_m^l \frac{e^{- i \Om U^{(l,m)}}}{\sqrt{4 \pi \Om}}, Y_{m'}^{l'} \frac{e^{- i \Om' U^{(l',m')}}}{\sqrt{4 \pi \Om'}}\right) = \delta(\Om - \Om') \delta_{ll'} \delta_{mm'}, \qquad \Om, \Om' > 0
\ee
since the Klein Gordon product of two modes with different angular momentum numbers $(l,m)$ vanishes because of the orthogonality of the spherical harmonics. Calling $K_{\{ lm \}}$ the projector into the subspaces of positive frequency solutions $\mathscr{S}^{>0}_{K_{\{lm\}}}$  spanned by the set of orthonormal modes $\{ Y_m^l \frac{e^{- i \Om U^{(l,m)}}}{\sqrt{4 \pi \Om}} \}$, we have
\be
    \langle f, g \rangle_{K_{\{ lm \}}} = \Om^{\mathbb{C}}(K_{\{ lm \}}f, K_{\{ lm \}}g)
\ee
that is our positive definite inner product on the one-particle Hilbert space. The Fock space is then obtained from the one-particle Hilbert space by the usual construction, see \eqref{eq: sym fock space}. In particular, the vacuum state $\ket{\Om^{\{ U^{( l,m )} \}}}$, where $\{U^{(l,m)}\}$ represents a sequence of times $U^{(l,m)}$ indexed by $l$ and $m$, is the unique state such that
\be
\label{lmvac}
    \forall \Om > 0, \quad \forall (l,m), \qquad \hat a_{\Om l m} \ket{\Om^{\{ U^{( l,m )} \}}} = 0 \, .
\ee
Anticipating a bit we state here that $\ket{\Om^{\{ U^{( l,m )} \}}}$ can be seen as a tensor product of the vacua associated to each sector with angular momentum numbers $(l,m)$.

\subsubsection*{Symmetries of the vacua}

We look at the transformations preserving $\ket{\Om^{\{ U^{(l,m)} \}}}$. First, we decompose the field operator $\hat \phi$ into effective two dimensional quantum scalar fields $\hat \phi_{lm}$ so that
\be \label{philmdecomp}
    \hat \phi = \sum_{lm} \hat \phi_{lm}, \qquad \hat \phi_{lm} = \int_{0}^{+ \infty} \frac{d \Om}{\sqrt{4 \pi \Om}} \left(Y_m^l \hat a_{\Om l m} e^{- i \Om U^{(l,m)}} + \bar{Y}_m^l \hat a_{\Om l m}^\dag e^{i \Om U^{(l,m)}}\right)
\ee
so that it is clear from \eqref{philmdecomp} that an affine transformation
\be \label{affine4D}
    U^{(l,m)} \rightarrow \tilde{U}^{(l,m)} = a_{lm} U^{(l,m)} + b_{lm}
\ee
does not mix the upper and lower operators, so that they leave the vacuum state $\ket{\Om^{\{ U^{(l,m)} \}}}$ invariant. However, it is not clear if \eqref{affine4D} are the only symmetries allowed. To go further and get a complementary point of view, we can look at the transformations of the two-point function \footnote{Since the states $\ket{\Om^{\{ U^{(l,m)} \}}}$  are gaussian states, they are entirely determined by their two-point function.} under $U^{(l,m)} \to \tilde{U}^{(l,m)}$ with $\tilde{U}^{(l,m)} = \tilde{U}^{(l,m)}(U^{(l,m)})$. 
\footnote{More rigorously, we should include the $- i 0^+$ prescription:
\begin{align}
    \bra{\Om^{\{ U^{(l,m)} \}}} &\p_u \phi(u_1, x_1^A) \p_u \phi (u_2, x_2^A)\ket{\Om^{\{ U^{(l,m)} \}}} = -\frac{1}{4 \pi} \sum_{lm} Y_m^l (x_1^A) \bar{Y}_m^l (x_2^A) \frac{d U^{(l,m)}}{du}(u_1) \frac{d U^{(l,m)}}{du}(u_2) \frac{1}{(U^{(l,m)}_1 - U^{(l,m)}_2 - i0^+)^2}
\end{align}
}
\begin{align} \label{gentwopointlm}
    \bra{\Om^{\{ U^{(l,m)} \}}} &\p_u \phi(u_1, x_1^A) \p_u \phi (u_2, x_2^A)\ket{\Om^{\{ U^{(l,m)} \}}} = \sum_{lm} \bra{\Om^{\{ U^{(l,m)} \}}} \p_u \phi_{lm} \p_u \phi_{lm} \ket{\Om^{\{ U^{(l,m)} \}}} \nn \\
    &= -\frac{1}{4 \pi} \sum_{lm} Y_m^l (x_1^A) \bar{Y}_m^l (x_2^A) \frac{d U^{(l,m)}}{du}(u_1) \frac{d U^{(l,m)}}{du}(u_2) \frac{1}{(U^{(l,m)}_1 - U^{(l,m)}_2)^2}
\end{align}
and then it is clear that we have invariance when changing the vacuum state $\ket{\Om^{\{ U^{(l,m)} \}}}$ into the vacuum state $\ket{\Om^{\{\tilde{U}^{(l,m)}\}}}$
if and only if
\be
    \forall (l,m), \qquad \frac{d \tilde{U}^{(l,m)}}{dU^{(l,m)}} (U^{(l,m)}_1) \frac{d \tilde{U}^{(l,m)}}{dU^{(l,m)}} (U^{(l,m)}_2) \frac{1}{(\tilde{U}^{(l,m)}_1 - \tilde{U}^{(l,m)}_2)^2} = \frac{1}{(U^{(l,m)}_1 - U^{(l,m)}_2)^2}
\ee
which is satisfied if and only if
\be \label{mobius4D}
    \forall (l,m):
    U^{(l,m)} \rightarrow \tilde{U}^{(l,m)} = \frac{a_{lm} U^{(l,m)} + b_{lm}}{c_{lm} U^{(l,m)} + d_{lm}}, \qquad a_{lm} d_{lm} - b_{lm} c_{lm} \neq 0,
\ee
similarly to the two dimensional case \cite{navarro2005modeling}, but now for each sector $(l,m)$. Then, from the discussion on the two dimensional case, we conclude that the vacua $\ket{\Om^{ \{U^{  (l,m) } \}}}$ are invariant under local M\"obius transformations for any sector $(l,m)$.

Note that the situation is very similar to what happens a non-expanding horizon. In his proof of the generalized second law \cite{Wall:2011hj}, Wall showed that the vacuum associated to the affine coordinate $V$ on the left horizon, namely the projection of the Hartle-Hawking vacuum, was invariant under local M\"obius transformations of the affine parameter. He therefore restricted his attention to one of the (infinitely many) generators of the horizon. He also argued that the quantization of the scalar field on any complete null geodesic was equivalent to the left moving sector of a two dimensional CFT. Indeed, because of ultralocality, we can define a vacuum state attached to each null geodesic, and the total vacuum as the tensor product over all generators of all these vacua. We have something similar in this work, except that we do not attach a vacuum state to any geodesic, but instead to any sector $(l,m)$. It is possible because we have the commutation relations 
\be
    \left[\phi_{lm}(U), \pi_{l'm'}(U')\right] = \frac{i}{2} \delta(U - U') \delta_{ll'} \delta_{mm'} \bar{Y}_{m'}^{l'} Y_{m}^l
\ee
where $\pi_{lm} = \p_U \phi_{lm}$, so that we can consider for each sector $(l,m)$ an effective two-dimensional chiral conformal field theory described by the field $\phi_{lm}$ and a vacuum state $\ket{\Om^{U^{(l,m)}}}$ associated to the positive frequency modes $\frac{Y_m^l}{\sqrt{4 \pi \Om}} e^{- i \Om U^{(l,m)}}, \Om > 0 $ for a given time $U^{(l,m)}$ at fixed $l$ and $m$. Then, our total vacuum \eqref{lmvac} is obtained as a tensor product of the vacua associated to each sector $(l,m)$
\be \label{vallmten}
    \ket{\Om^{\{U^{(l,m)} \}}} = \bigotimes_{lm} \ket{\Om^{U^{(l,m)}}} \, .
\ee
Vacua of the form \eqref{vallmten} are relevant for us since we are interested in the physics at null infinity in a black hole background. Due to the black hole potential, the modes emitted at the horizon undergo a back-scattering and we expect a different behavior depending on the angular momentum quantum numbers $(l,m)$. Therefore, vacua of defined angular momentum, like \eqref{vallmten}, are more relevant to our purpose than those defined on the null generators of $\scri_R$. However, it is insightful to appreciate the analogy between \eqref{mobius4D} and the local M\"obius transformations on the horizon 
\be
    V(x^A) \rightarrow V'(x^A) = \frac{a(x^A) V(x^A) + b(x^A)}{c(x^A)V(x^A) + d(x^A)}, \qquad a(x^A)d(x^A) - b(x^A)c(x^A) \neq 0
\ee
leaving the restriction of the Hartle-Hawking state to the horizon invariant, a property needed to prove the generalized second law. \footnote{In fact, in \cite{Wall:2011hj}, considering only the subset of symmetries $c(x) = 0, d(x) = 1$ was enough.} 


\subsection{Examples of vacua}
\label{subsec: examples vacua}

In this subsection we look at different examples of vacua defined on the algebra of observables $\mathcal{A}^{\scri}$. All the vacua considered below are examples of the form \eqref{vallmten}. When needed, the reader is advised to look at Appendix \ref{app: modular theory} where the notion of von Neumann algebra, cyclic and separating vectors are defined.

\subsubsection*{The Minkowski vacuum}
\label{subsect: minkowskivac}

For our first example, we take 
\be
    \forall(l,m), \quad U^{(l,m)} = \left\{ \begin{array}{ll}
        u_+ & \mbox{on} \quad \scri^+_R  \\
        u_- & \mbox{on} \quad \scri^-_R
    \end{array}
    \right. \, ,
\ee
where we recall that $u_+$ (resp. $u_-$) is an affine coordinate on $\scri^+_R$ (resp. $\scri^-_R$). We identify the modes $\left(\frac{Y_m^l}{\sqrt{4 \pi \om_+}} e^{- i \om_+ u_+},  \frac{Y_m^l}{\sqrt{4 \pi \om_-}} e^{- i \om_- u_-}\right)$ with $\om_+, \om_- > 0 $ to be the positive frequency solutions. The vacuum state $\ket{\Om^u} : = \ket{0}_M$ associated to these positive frequency solutions is the restriction of the Minkowski vacuum to $\scri^+_R$ (and $\scri^-_R$). It is therefore the unique state annihilated by all the operators $\hat{a}_{\omega_+ lm}$ and $\hat{a}_{\omega_- lm}$. Note that it is a pure state with respect to the algebra $\mathcal{A}_0^\scri$. When the latter is seen as a subalgebra of $\mathcal{B}(\mathscr{H}_{\Omega^u})$ i.e. the bounded linear operators of the Gelfand-Naimark-Segal (GNS) \cite{gelfand1943imbedding, segal1947irreducible} Hilbert space $\mathscr{H}_{\Omega^u}$ built upon $\ket{0}_M$, one finds that its double commutant is a Type \text{I} von Neumann algebra, see Appendix \ref{app: modular theory} for related discussions and details. It will not be the case of the other representations of the algebra that we consider below, whose double commutant will be of Type \text{III}.

Then, we can compute the two-point function in the state $\ket{0}_M$ for any pair of points of $\scri_R^+$, $(u_1, x_1^A)$ and $ (u_2, x_2^A)$ (for the rest of the paragraph $u_+ = u$ to avoid clutter) that is, from \eqref{gentwopointlm}, 
\be \label{tptfunu}
    \bra{0}_M \p_u \phi(u_1, x_1^A) \p_u \phi(u_2, x_2^A) \ket{0}_M = - \frac{1}{4 \pi} \frac{\delta^{(2)}(x_1^A - x_2^A)}{(u_1 - u_2)^2}
\ee
where we used the completeness relation of the spherical harmonics
\be
    \sum_{lm} \bar{Y}_m^l (x_1^A) Y_m^l (x_2^A) = \delta^{(2)}(x_1^A - x_2^A)
\ee
to simplify the sum over the different sectors $(l,m)$. Notice that the two point function \eqref{tptfunu} coincides with standard expressions appearing in \cite{kay1991theorems, moretti2006uniqueness, dappiaggi2017hadamard}. As previously explained, we can apply the M\"obius transformations on each $(l,m)$ sector
\be
    u \rightarrow u' = \frac{a_{lm} u + b_{lm}}{c_{lm} u + d_{lm}}, \qquad a_{lm} d_{lm} - b_{lm} c_{lm} \neq 0
\ee
but since the time $u$ is independent of the sector $(l,m)$, we can see from the structure of the two point function \eqref{tptfunu} that the transformations 
\be \label{spactimesymmxmink}
    u \rightarrow u' = \frac{a(x) u + b(x)}{c(x) u + d(x)}, \qquad ad - bc \neq 0
\ee
are also symmetries of the Minkowski vacuum restricted to the algebra $\mathcal{A}_{0}^\scri$. Then, the normal-ordered (wrt $\ket{0}_M$) stress energy tensor is obtained by subtracting the vacuum expectation value \eqref{tptfunu} to the momentum square
\be \label{minkstressenergy}
    :T_{uu}:_{0_M} = \underset{(u_2, x_2) \rightarrow (u_1, x_1)}{\lim} \left( \p_u \phi (u_1,x_1^A) \p_u \phi (u_2,x_2^A) - \bra{0}_M \p_u \phi(u_1, x_1^A) \p_u \phi(u_2, x_2^A) \ket{0}_M \right)
\ee
that is a sesquilinear form for a dense set of states in $\mathscr{H}_{\Om^u}$. In general, the normal-ordered stress energy tensor cannot be identified as a good component of the stress energy tensor, because such a good component had a vanishing expectation value for the Minkowski vacuum.\footnote{This requirement is actually one of Wald's four axioms a stress-tensor should satisfy, see \cite{wald1994quantum}.} However, the expectation value of \eqref{minkstressenergy} in the Minkowski vacuum vanishes by construction, and \eqref{minkstressenergy} has finite expectation value for Hadamard states at null infinity. Therefore, \eqref{minkstressenergy} is indeed a component of the renormalized stress energy tensor that is coupled to the geometry
\be
\label{eq: equiv principle for stress tensor}
    T_{uu} = :T_{uu}:_{0_M} \, ,
\ee
meaning that the energy flux at $\scri_R^+$ in the Minkowski vacuum i.e. $\langle T_{uu}\rangle_{0_M}$ is vanishing. Therefore $\ket{0}_M$ physically represents the absence of radiation (as it should). 

\subsubsection*{The Hartle-Hawking vacuum}

Consider now the vacuum state defined using a time coordinate that is complete on $\scri_R$ i.e. so that it runs from $-\infty$ to $+ \infty$. From \eqref{eq: def of U} we take
\be \label{timehhbigU}
    U^{(l,m)} = U =  \left\{ 
    \begin{array}{ll}
        e^{\kappa u_+}  & \mbox{on} \quad \scri^+_R \\
        -e^{-\kappa u_-} & \mbox{on} \quad \scri^-_R
    \end{array}
    \right .
\ee
where $\kappa$ is the surface gravity. Then, we can construct a Hilbert space $\mathscr{H}_{\Om_H}$ using the positive frequency modes associated to the time \eqref{timehhbigU} and we denote the vacuum state $\ket{\Om_H}$. This state is cycling and separating with respect to the representation of the algebra $\mathcal{A}_0^\scri$ as it will turn out to be thermal. The state $\ket{\Om_H}$ can be interpreted as the restriction of the Hartle-Hawking vacuum at null infinity. More precisely, the 
Hartle-Hawking state is defined using the positive frequency mode decomposition associated to the Kruskal coordinates $(\tilde U, \tilde V)$, (related to the coordinates $(U,V)$ by the relations \eqref{eq: def of U} and \eqref{eq: def of V}) on the horizons $\mathcal{H}_R \cup \mathcal{H}_L$ and its restriction to an algebra of observables at null infinity can be deduced using a similar decomposition.
\footnote{In fact it is easy to show that the state $\ket{\Om_H}$ is a KMS state with respect to the modular Hamiltonian generating the Killing flow in the subalgerbas $\mathcal{A}_0^\scri$ and $\mathcal{A}_{0}^{\scri'}$ using the results of Appendices \ref{app: unruh} and \ref{app: modular theory}. Therefore, the restriction of the state $\ket{\Om_H}$ and the Hartle-Hawking state to the algebras $\mathcal{A}_0^\scri$ and $\mathcal{A}_{0}^{\scri'}$ are identical. The restriction of the Hartle-Hawking state to the full algebra $\mathcal{A}^{\scri}$ might be different, since the correlations might not be the same. However, since we will restrict ourselves to $\mathcal{A}_0^\scri$ or to one of its subalgebras in the subsequent section, we are not interested in the correlations and we will refer for simplicity to the state $\ket{\Om_H}$ as the Hartle-Hawking vacuum in the following.}

The two point function associated to the Hartle-Hawking vacuum on $\scri_R$ is therefore given by 
\be \label{tptfchhvach}
    \bra{\Om_H} \p_U \phi(U_1, x_1^A) \p_U \phi(U_2, x_2^A) \ket{\Om_H} = - \frac{1}{4 \pi} \frac{\delta^{(2)}(x_1^A - x_2^A)}{(U_1 - U_2)^2}
\ee
and is similar to the result on the horizon (Eq.(1.1) of \cite{kay1991theorems}). In addition, similarly to the Minkowski vacuum, the transformations
\be
    U \rightarrow U' = \frac{a_{lm} U + b_{lm}}{c_{lm} U + d_{lm}}, \qquad a_{lm} d_{lm} - b_{lm} c_{lm} \neq 0
\ee
are not the only symmetries. From the structure of the two point function \eqref{tptfchhvach}, we can see that the local transformations 
\be
\label{eq: sym HH state}
    U \rightarrow U' = \frac{a(x) U + b(x)}{c(x) U + d(x)}, \qquad a(x) d(x) - b(x) c(x) \neq 0
\ee
are also symmetries of $\ket{\Omega_H}$. However, there is an important difference with what happens on the horizons: while on the horizon the Kruskal coordinate is an affine coordinate, at null infinity the affine coordinate is given by $u$, not $U$. The ``good" notion of energy flux is therefore encoded into the component $T_{uu}$ of the renormalized stress-energy tensor. In terms of the $U$-coordinate, the doubly null component of the normal-ordered stress energy tensor is  
\be \label{stressenergyhh}
    :T_{UU}:_{\Om_H} = \underset{(U_2, x_2) \rightarrow (U_1, x_1)}{\lim} \left( \p_U \phi (U_1,x_1^A) \p_U \phi (U_2,x_2^A) - \bra{\Om_H} \p_U \phi(U_1, x_1^A) \p_U \phi(U_2, x_2^A) \ket{\Om_H} \right)
\ee
which is well defined as a sesquilinear form on the dense subset of $\mathscr{H}_{\Om_H}$ spanned by $\ket{\Omega_H}$.\footnote{This dense subspace comes from the GNS construction, see Appendix \ref{app: AQFT}.} However, the energy flux is given by
\begin{align}
    \label{eq: normal ordered HH}
    \bra{\Om_H} T_{uu} \ket{\Om_H} &= \bra{\Om_H} :T_{uu}:_{0_M} \ket{\Om_H} \\
    \nonumber
    &= -\frac{1}{24 \pi}  \underset{(U_2, x_2) \rightarrow (U_1, x_1)}{\lim} S(u, U) \delta^2(x_1^A - x_2^A) \\
    \nonumber
    &= \underset{(U_2, x_2) \rightarrow (U_1, x_1)}{\lim} \frac{\kappa^2}{48 \pi} \delta^2(x_1^A - x_2^A) = +\infty
\end{align}
where we used \eqref{eq: equiv principle for stress tensor} in the first line and in the second line the general transformation law of a normal-ordered stress tensor (which includes a Schwarzian) together with the fact that $\langle :T_{UU}:_{\Omega_H}\rangle_{\Omega_H} = 0$. From \eqref{eq: normal ordered HH} we see that the restriction of the Hartle-Hawking state is singular at null infinity. The Hartle-Hawking state being a Hadamard state, the expectation value of the renormalized\footnote{Renormalization has to be understand here from the point of view of the matter theory.} stress energy has finite expectation values on the physical spacetime. However, the energy density $\langle \hat{T}_{uu} \rangle_{\Om_H}$ tends, following Stefan's law, to a constant very far from the black hole \footnote{Note that in the Hartle-Hawking state the outgoing radiation current is exactly compensated by the ingoing radiation current, so that $ \langle \hat{T}_t^r \rangle_{\Om_H} = 0$ in the bulk.} \cite{page1982thermal, howard1984quantum}
\be \label{falloffsetens}
    \langle \hat{T}_{uu} \rangle_{\Om_H} \underset{r \rightarrow + \infty}{\sim} T_H^4 = \left( \frac{\kappa}{2\pi} \right)^4 \, .
\ee
Therefore, we see that even though the energy density is finite, its asymptotic behavior is incompatible with the fall-offs required to have asymptotic flatness, which are \cite{ashtekar2014geometry}
\be \label{asymptotflat}
    r^2 \langle \hat{T}_{uu} \rangle \underset{r \rightarrow + \infty}{=} O(1) \, .
\ee
This occurs because the radiation comes uniformly from all directions at infinity, rendering the black hole indistinguishable from the surrounding radiation field. Indeed, the Hartle–Hawking state describes a black hole in thermal equilibrium with incoming radiation at the Hawking temperature, so that the black hole remains in equilibrium  \footnote{\label{foot: unstable equilib}However, because of the negative heat capacity of the black hole, this equilibrium is in fact unstable.} since the outgoing Hawking radiation is exactly balanced by incoming radiation at the same temperature. Moreover, because the Schwarzschild potential vanishes both at the horizon and at infinity, thermal radiation propagates just as easily from past null infinity to the future horizon as outgoing radiation propagates from the past horizon to future null infinity. As a consequence, all outgoing modes are thermally populated at the Hawking temperature, for every value $(l,m)$ of the angular momentum. In particular, although high angular-momentum modes emitted from the past horizon $\mathcal{H}^-_R$ are partially back-scattered by the potential barrier, their contribution is exactly compensated by high–angular-momentum modes coming from $\scri^-_L$, which are reflected in a symmetric manner and therefore do not reach the future horizon. By contrast, a physical black hole is not in thermal equilibrium with incoming radiation and emits predominantly low angular-momentum modes, since the Hawking radiation originates only from the near-horizon region. We take this crucial physical insight into account in the next two paragraphs.

\subsubsection*{Hard regularization of the Hartle-Hawking state}

Since restricting the Hartle–Hawking state $\ket{\Omega_H}$ to the algebra $\mathcal{A}_0^\scri$ leads to a constant thermal flux therefore to a divergent energy density, we are naturally led to consider alternative vacuum states, constructed in order to obtain finite energy densities at $\scri^+_R$. It is important to emphasize that this is not merely a mathematical pathology: the singular behavior reflects the fact that the physical description itself is inappropriate. A real black hole is never immersed in a thermal bath at the Hawking temperature; rather, the radiation originates only from the black hole itself. While this radiation can be regarded as thermal near the horizon at early times,\footnote{Or, equivalently, on the entire horizon if one follows Unruh’s construction \cite{Unruh:1976db}.} it is certainly \emph{not} thermal at $\scri^+_R$. Indeed, as noted in the previous paragraph, the gravitational potential \eqref{eq: KG for our field} becomes very large for modes with high angular momentum. Intuitively, this reflects the fact that the latter are not directed sufficiently “towards” the black hole and therefore cannot be efficiently absorbed—or, by time-reversal symmetry, emitted.

One possible approach is to choose a vacuum state $\ket{\Omega_H^L}$ that appears thermal (with respect to the Killing flow \eqref{localbfscri}) for low angular-momentum modes $l < L$, with $L$ fixed, while coinciding with the Minkowski vacuum for high angular-momentum modes $l \geq L$. Hence, we choose on $\scri_R$
\be \label{timehhbigULvac}
    U^{(l,m)} = \left\{ 
    \begin{array}{ll}
        U  & \mbox{if} \quad l < L \\
        u_{\pm} = \kappa^{-1}\ln{\pm U} & \mbox{if} \quad  l \geq L
    \end{array}
    \right .
\ee
and we get a bunch of new vacuum states. The space of positive frequency solutions can be written as
\begin{equation}
    \label{eq: L-vac positive solutions}
    \mathscr{S}^{>0}_{L} :=\left\{f \in \mathscr{S}^{\mathbb{C}} \Bigg| f(U, x^A) = \sum_{l=0}^{L-1} \sum_{m = -l}^{l} \int_{0}^{+\infty} \text{d}\Omega a_{\Omega lm} Y^{l}_{m}e^{-i\Omega U} + \sum_{l=L}^{+\infty} \sum_{m = -l}^{l} \int_{0}^{+\infty} \text{d}\omega a_{\omega lm} Y^{l}_{m}e^{-i\omega u_+} \right\} \, ,
\end{equation}
on $\scri_R^+$ for example. For $L = 0$ we recover the Minkowski vacuum and for $L = +\infty$ the Hartle-Hawking vacuum. The case $L = 1$ corresponds to the $s$-wave approximation, for which we consider only spherically symmetric radiation waves.  The rationale for the choice \eqref{timehhbigULvac} is that, for any finite choice of $L$, only a finite subset of modes is excited relative to the Minkowski vacuum, ensuring that the stress–energy density on the horizon remains finite. Moreover, the vacuum state $\ket{\Omega_H^L}$ is physically more relevant at future null infinity than the Hartle–Hawking state $\ket{\Omega_H}$, as it provides a more realistic description of an isolated black hole. We call this state the $L$\emph{-vacuum} state. 

This state is a natural interpolation between the Hartle–Hawking vacuum and the Unruh vacuum: modes with low angular momentum $l < L$ are in a thermal state, as in the Hartle–Hawking vacuum, while modes with large angular momentum remain unexcited, leading to a finite energy density on $\scri^+_R$, as in the Unruh vacuum. Physically, the state $\ket{\Omega_H^L}$ may be interpreted (approximately) as arising from an initial condition in which an incoming thermal flux of modes with angular momentum $l < L$ at the Hawking temperature exactly compensates the outgoing Hawking radiation in those same modes, so that the $l < L$ sector remains in thermal equilibrium. If $L$ is chosen sufficiently large, the contribution of outgoing modes with $l \geq L$ to the flux at null infinity may be neglected, since such modes are strongly suppressed by the gravitational potential barrier and therefore do not efficiently escape.

Using \eqref{eq: L-vac positive solutions}, we can construct the Hilbert space $\mathscr{H}_{\Om_H^L}$ for the $L$-vacuum state. The two point function between two points of $\scri^+_R$ in $\ket{\Omega_H^L}$ is given by ($u = u_+$ again)
\begin{align}
    \bra{\Om^L_H} \p_u \phi (u_1, x^A_1) \p_u \phi (u_2, x^A_2) \ket{\Om^L_H} =& - \frac{1}{4 \pi} \sum_{l < L, m} \bar{Y}_m^l(x_1^A) Y_m^l(x_2^A) \frac{dU}{du}(u_1) \frac{dU}{du}(u_2) \frac{1}{(U_1 - U_2)^2} \nn \\
    &- \frac{1}{4 \pi} \sum_{l \geq L, m} \bar{Y}_m^l(x_1^A) Y_m^l(x_2^A) \frac{1}{(u_1 - u_2)^2}  \nn \\
    =& - \frac{1}{4 \pi} \sum_{l < L, m} \kappa^2 \bar{Y}_m^l(x_1^A) Y_m^l(x_2^A) \frac{e^{\kappa u_1} e^{\kappa u_2}}{(e^{\kappa u_1} - e^{\kappa u_2})^2} \nn \\
    &- \frac{1}{4 \pi} \sum_{l \geq L, m} \bar{Y}_m^l(x_1^A) Y_m^l(x_2^A) \frac{1}{(u_1 - u_2)^2}
\end{align}
and all the symmetries given by \eqref{mobius4D}, where $U^{(l,m)} = U$ for $l < L$ and $U^{(l,m)} = u$ for $l \geq L$, leave the $L$-vacuum invariant. The doubly null component of the normal-ordered affine stress-energy tensor is given by 
\be \label{stressenergyhhl}
    :T_{uu}:_{\Om_H^L} = \underset{(u_2, x_2) \rightarrow (u_1, x_1)}{\lim} \left( \p_u \phi (u_1, x_1^A) \p_u \phi (u_2, x_2^A) - \bra{\Om_H^L} \p_u \phi(u_1, x_1^A) \p_u \phi(u_2, x_2^A) \ket{\Om_H^L} \right)
\ee
so that the doubly null component of the covariant affine stress energy tensor is given by
\be \label{stressenooml}
     T_{uu} = :T_{uu}:_{\Om_H^L} + \frac{\kappa^2}{48 \pi} \sum_{l < L, m} \bar{Y}_m^l(x^A) Y_m^l(x^A) < +\infty
\ee
which is finite. Therefore there is a dense subspace of Hadamard states in $\mathscr{H}_{\Om_H^L}$ giving finite energy densities, namely the ones of the form $A\ket{\Omega_H^L}$ for $A \in \mathcal{A}^{\scri}_0$. It is because the sum appearing in \eqref{stressenooml} only contained a finite number of angular momentum modes $(l,m)$. Note that the thermal flux at infinity in the $L$-vacuum is therefore effectively two dimensional, since it is a sum of thermal fluxes associated to the Hartle-Hawking vacuum of a finite number of two dimensional scalar CFT. 

\subsubsection*{Soft regularization of the Hartle-Hawking state}

The $L$-vacuum studied in the previous section was used to model a finite thermal flux at future null infinity. The state was exactly thermal for the low angular-momentum modes. However, as we already said, the state to which an actual black hole relaxes is \textit{not} thermal at all. Indeed, we have to take into account the corrections due to the gray-body factors, bringing a much richer physics than a mere thermal driving. It is indeed well-known that the correct vacuum state to describe the physics of a collapsing black hole from the point of view of late time asymptotic observers is the Unruh vacuum. In this paragraph, we are not considering the Unruh vacuum directly, but instead another family of vacua which will help us to appreciate the better features of the Unruh vacuum compared to the Hartle-Hawking vacuum.

In order to do this, we rely on the observation that different angular momentum modes are scattered in a way that depends on their angular momentum $l$, as we explained before. However, while for the $L$-vacuum we introduced a "hard" cutoff on the angular momentum modes, here we soften the regularization by introducing a series of vacuum states so that all the different angular momentum modes have their own effective temperature\footnote{The precise expression for $\kappa_l$ is phenomenological, see a proposal in footnote \ref{foot: proposal kappa l}.} 
\be \label{effectivetl} 
    T_l := \frac{\kappa_l}{2 \pi}.
\ee
Physically, since the higher the angular momentum, the more back-scattered is the mode, we will take $\kappa \geq \kappa_l \geq 0$, so that if there is no back-scattering the effective temperature is the Hawking temperature, and $\kappa_l \underset{l \rightarrow + \infty}{\longrightarrow} 0$, the latter being necessary in order to get a finite flux at $\scri_R^+$. To construct the Hilbert space, we choose a decomposition into positive frequency modes associated to the times
\be \label{ulmsoftdef}
U^{(l,m)} = \left\{ 
    \begin{array}{ll}
        U^l := e^{\kappa_l u_+} & \mbox{on} \quad \scri^+_R \\
        U^l := -e^{-\kappa_l u_-} & \mbox{on} \quad \scri^-_R
    \end{array}
\right.
\ee
so that $U^l = U$ if $\kappa_l = \kappa$ and $ \lvert U^l \lvert = \lvert U \lvert^{\frac{\kappa_l}{\kappa}} $. The space of positive modes is therefore given by
\begin{equation}
    \label{eq: positive space kappa vacuum}
    \mathscr{S}^{>0}_{\kappa_l} := \left\{f \in \mathscr{S}^{\mathbb{C}} \Bigg| f(U,x^A) = \sum_{l=0}^{+\infty} \sum_{m=-l}^{l} \int_{0}^{+\infty} \text{d}\Omega \, a_{\Omega l m}Y^{l}_{m}e^{-i\Omega U^l}\right\} \, .
\end{equation}
The associated vacuum state $\ket{\Om^{\{ \kappa_l \}}_H}$ is called the $\kappa_l$-\emph{vacuum state} and its Hilbert space is denoted $\mathscr{H}_{\Om_H^{\kappa_l}}$. Each sequence of positive terms $\{ \kappa_l \}$ generates a different Hilbert space, and if any of the $\kappa_l$ is changed, we get a representation of the algebra of observables which is unitarily inequivalent. 

The two point function on this vacuum state between two-point of $\scri^+_R$ is given by 
\begin{align} \label{gentwopointlmkl}
    \bra{\Om^{\{ \kappa_l \}}_H} \p_u \phi(u_1, x_1^A) \p_u \phi (u_2, x_2^A)\ket{\Om^{\{\kappa_l\}}_H} &= -\frac{1}{4 \pi} \sum_{lm} \bar{Y}_m^l (x_1^A) Y_m^l (x_2^A) \frac{d U^l}{du}(u_1) \frac{d U^l}{du}(u_2)\frac{1}{(U^l_1 - U^l_2)^2} \nn \\
    &= -\frac{1}{4 \pi} \sum_{lm} \kappa_l^2 \bar{Y}_m^l (x_1^A) Y_m^l (x_2^A) \frac{e^{\kappa_l u_1} e^{\kappa_l u_2}}{(e^{\kappa_l u_1} - e^{\kappa_l u_2})^2}
\end{align}
and the doubly null component of the normal-ordered stress-energy tensor is given by 
\be \label{stressenergykl}
    :T_{uu}:_{\Om_H^{\{ \kappa_l \}}} = \underset{(u_2, x_2) \rightarrow (u_1, x_1)}{\lim} \left( \p_u \phi (u_1, x_1^A) \p_u \phi (u_2, x_2^A) - \bra{\Om_H^{\{ \kappa_l \}}} \p_u \phi(u_1, x_1^A) \p_u \phi(u_2, x_2^A) \ket{\Om_H^{\{ \kappa_l \}}} \right)
\ee
so that the doubly null component of the affine stress energy tensor is given by
\be \label{stressenoomlkl}
     T_{uu} = :T_{uu}:_{\Om_H^{\{ \kappa_l \}}} + \sum_{lm} \frac{\kappa^2_l}{48 \pi} \bar{Y}_m^l(x^A) Y_m^l(x^A)
\ee
which converges if the sequence $\{ \kappa_l \}$ decays sufficiently quickly.\footnote{The shall see in Section \ref{sec: GSL proof} that it is enough to ask $\sum_{l=0}^{+\infty}(2l+1) \kappa_l < +\infty$.} In the latter case, there exists a dense set of Hadanard states in $\mathscr{H}_{\Om_H^{\{ \kappa_l \}}}$ for which the energy density on $\scri^+_R$ given by \eqref{stressenoomlkl} in finite. Of course, all the symmetries \eqref{mobius4D}, where here $U^{(l,m)} = U^l$, leave the $\kappa_l$-vacuum invariant. Finally note that $\ket{\Om_H^{\{ \kappa_l \}}}$ admits the tensorial structure \eqref{vallmten} with $U^{(l,m)} = U^l$.

\subsubsection*{Why so many vacuum states?}

One may wonder why we have introduced that many different vacuum states. From a mathematical point of view, the GNS representation theorem asserts that given an algebraic state (the vacuum state) one can always construct a Hilbert-space representation of a unital $\mathbb{C}^\ast$-algebra like the Weyl algebra. The representation is tight to the choice of state, as two different states generically give rise to two Hilbert spaces non related by a unitary transformation. Mathematics however do not tell how to choose the vacuum state. This depends on the physical situation at hand, what kind of process one wants to model. The study of black hole radiation and evaporation led to the construction of various vacuum states among which the Hartle-Hawking and the Unruh state. Each of them proposes a different way to model a black hole and its radiation, with more or less relevance depending on the degree of accuracy with which we want to study that process. If the Hawking radiation at late time is modeled by the Unruh vacuum, the latter cannot be written at future null infinity using a positive frequency decomposition, and its characterization under an algebra of observables at null infinity is left as an open problem. However, it is still possible to get an expression of the modular Hamiltonian starting from the thermal character of the Unruh vacuum on the past horizon and using transmission/reflexion coefficients (see the second part of this work \cite{ARBMVsoon}). However, as we have repeatedly emphasized, the Hartle–Hawking state is far from providing a realistic description of an evaporating black hole. This led us to propose the hard and soft regularizations of the Hartle-Hawking state at $\scri_R$. The next Section will show how the thermodynamical potential, whose monotonicity at $\scri_R^+$ is the dual GSL, gets modified when we consider the latter vacua.


\subsection{Beyond the scalar field}

It is of main interest to extend the formalism of this Section to massless field of higher spins or even to interacting field theories. We give some brief words about it in this subsection. We only focus on $\scri_R^+$ here, for simplicity.

\subsubsection*{Spin-$1$ field}

We take $A_a$ to be a massless spin-1 free field. We can consider it to be the electromagnetic field. Since our construction for the scalar field started from the symplectic form, we have to write it in the case at hand. We can make a choice of gauge so that $A_u = 0$ on $\scri_R^+$ \cite{strominger2018lectures, Prabhu:2022zcr, Kudler-Flam:2025pol}. In this case, we have that (with $a= u, A$)
\be \label{EMsympstru}
    \Om \left(A^1_a, A^2_b \right) = - \frac{1}{e^2} \int_{\scri_R^+} q^{AB} \left(E^1_A A_B^2 - E^2_A A_B^1 \right) \text{d}u \wedge \eps_S
\ee
where $E_A^i = \p_u A_A^i$ is the electric field and $q_{AB}$ is the conformal metric (that is finite at $\scri_R^+$), $e$ the coupling constant, and $A_a^i$, $i= 1,2$ are solutions to the classical Maxwell equations. Therefore, the symplectic form is exactly similar to the one for the massless scalar field \eqref{symplecticformsu} at $\scri_R^+$, and the electric field is the conjugate momentum variable. Therefore, we can define an algebra of operators using smearings of the electric field against test functions, and choose a class of vacua states similar to the one that we introduced in the previous paragraphs of this section. Therefore, we do not expect to face serious difficulties when using a free spin-$1$ field instead of a free spin-$0$ field for any of the arguments in the remaining of the paper.

\subsubsection*{Spin-$2$ field}

We can now treat the case of the spin-$2$ field. In this case, we take the example of the gravitational field and fix a gauge so that $g_{au} = 0$ on $\scri_R^+$, $a = u,A$. Then, it is well known that in this case the symplectic form reduces to \cite{ashtekar1981symplectic}
\be \label{ASsympform}
    \Om \left(C^1_{AB}, C^2_{A'B'} \right) = -\frac{1}{16 \pi} \int_{\scri_R^+} q^{AA'} q^{BB'} \left(N^1_{AB} C_{A'B'}^2 - N^2_{A' B'} C_{AB}^1\right) \text{d}u \wedge \eps_S
\ee
where $N_{AB} = - \p_u C_{AB}$ is the asymptotic news, and $C_{AB}$ is the asymptotic shear. The news is the conjugate momentum of the shear which is the asymptotic configuration variable. An algebra of observables at $\scri_R^+$ can be obtained from smearings of the news, similarly to what we did for the massless scalar field and the electromagnetic field. The symplectic structure \eqref{ASsympform} is of the same form as the one of the electromagnetic field \eqref{EMsympstru} and of the scalar field \eqref{symplecticformsu} so the construction is similar to the previous cases that we have encountered. Therefore, we do not expect to face serious difficulties when dealing with the spin-$2$ field.

\subsubsection*{Interacting field theories}

For interacting field theories, it is much more subtle in general. However, near $\scri_R^+$ interacting field theories tend to behave as free field theories since the interactions spread out \cite{strominger2018lectures, Prabhu:2022zcr}. For instance, gravity is by any means an interacting field theory. However, its symplectic structure at infinity is very similar to the one of a free theory, like the massless scalar field or the electromagnetic field. Therefore, we expect to be able to treat interacting field theories on $\scri_R^+$ as we treat free field theories, and do not plan to see any additional difficulty as long as we restrict ourselves to the algebra of observables there. However, as long as we take into account the algebra on the horizon, we should be much more careful in treating interacting field theories since an algebra on the horizon does not necessarily exist in this case \cite{bousso2015entropy}.   


\section{Dual Generalized Second Law}
\label{sec: GSL proof}

This Section constitutes the core of our work, namely to prove the dual GSL at future null infinity. As stated in the Introduction, the GSL is proven on the horizon using the monotonicity of the relative entropy between the restriction of the Hartle-Hawking vacuum state and an arbitrary state belonging to the GNS Hilbert space representing, from that vacuum state, the algebra of observables of the horizon. The two spacelike hypersurfaces between which the GSL was proven in \cite{Wall:2011hj} were both starting at spatial infinity $\iota^0$ and ending at two different cuts of the future horizon. In that case the relevant thermodynamic potential is the generalized entropy. Black hole thermodynamics has therefore been unveil from a near-horizon perspective. We treat here a dual picture, focusing on asymptotic observers.

The dual GSL corresponds to the monotonicity of a thermodynamical potential $\mathcal{G}$ (free energy or grand potential, depending on the vacuum state) between two spacelike hypersurfaces starting from the same cut on the horizon but ending at different cuts of $\scri_R^+$. A proof involving an effective description of the process was published by one of us in \cite{ARB24}. This work improves the latter, as we give rigorous algebraic foundations of all the steps. In particular we defined precisely the algebra of observables at $\scri_R$ and introduced three vacua, the Hartle-Hawking state and its hard and soft regularizations, from which we constructed the associated GNS Hilbert spaces in Section \ref{sec: quantization}. Once this is done the proof basically amounts to apply again the monotonicity of the relative entropy. In the course of the proof, appears the notion of modular Hamiltonian, which we construct in Appendix \ref{app: modular theory} while in Appendix \ref{app: modular boost energy} we relate it to the integral of some normal-ordered stress-energy tensor. Given this relations, Einstein's equations are used to make contact with the physical quantities (e.g. the Bondi mass $M$). Therefore all the steps of the subsequent proof are rigorously proven from the point of view of Algebraic QFT and Modular Theory. The reader is advised to look at these Appendices for more details.

Once the dual GSL is proven at future null infinity for the Hartle-Hawking state (subsection \ref{subsec: GSL HH state}) and its regularizations (subsections \ref{subsec: GSL hard} and \ref{subsec: GSL soft}), we consider spacelike hypersurfaces ending at different cuts on the horizon in subsection \ref{subsec: open horizon}. The black hole area enters the game and imposes to generalize the definition of the thermodynamic potential $\mathcal{G}$.

\subsubsection*{Setup and notations}

\begin{figure}[ht]
    \centering
    \includegraphics[width=0.5\linewidth]{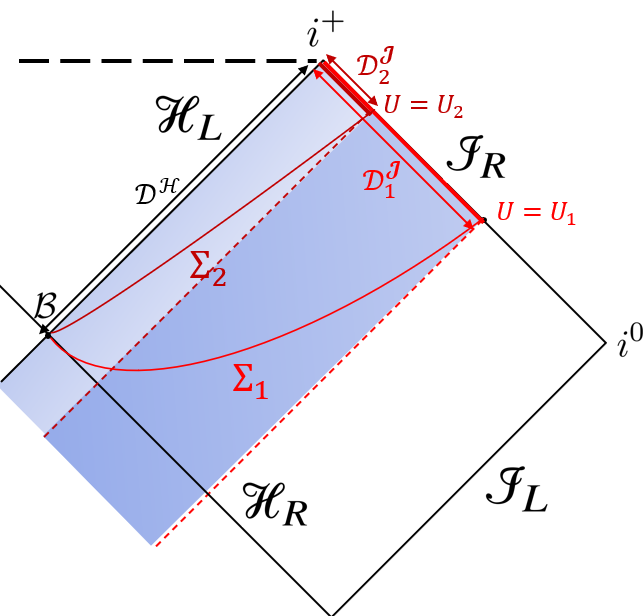}
    \caption{Setup in which the hypersurfaces $\Sigma_1$ and $\Sigma_2$ both starts at the horizon bifurcation surface $\mathcal{B}$ and end at different cuts $U = U_1$ and $U = U_2$ respectively at $\scri_R^+$. Are also depicted the regions $\mathcal{D}^{\mathcal{H}}$ and $\mathcal{D}_i^{\scri}$.}
    \label{fig: semi setup GSL}
\end{figure}

The first part of the proof focuses only on $\scri_R^+$. Consider two spacelike hypersurfaces $\Sigma_1$ and $\Sigma_2$ starting at the bifurcation surface $\mathcal{B}$ i.e. at the cut $V = 0$ on the horizon $\mathcal{H}_L^+$, and ending at two different cuts $\mathcal{C}_1$ and $\mathcal{C}_2$ at $\scri_R^+$, say, at $U = U_1$ and $U = U_2$ with $U_2 > U_1 > 0$ (see Figure \ref{fig: semi setup GSL}).\footnote{\label{foot: part to add spacelike}Strictly speaking, in order to be spacelike, we should add to these regions the part between $\iota^0$ and the cut $\mathcal{C}_i$. As the proof is based on differences these regions do not matter.}
Following \cite{Wall:2011hj, ARB24}, instead of considering directly $\Sigma_i$ we push these hypersurfaces to infinity so that they cover a certain region $\mathcal{D}_i^{\mathcal{B}}$ of $\mathcal{H}_L^+ \cup \scri_R^+$ namely
\begin{equation}
    \label{eq: def domain total}
    \mathcal{D}_i^{\mathcal{B}} = \underbrace{\left( (0, +\infty) \times S^2_{\mathcal{H}}\right)}_{:= \mathcal{D}^{\mathcal{H}}} \cup \underbrace{\left( (U_i, +\infty) \times S^2_{\scri}\right)}_{:= \mathcal{D}_{i}^{\scri}} \, ,
\end{equation}
as both the horizon and null infinity have topology $\mathbb{R}\times S^2$. If we show the dual GSL between $\mathcal{D}_1^{\mathcal{B}}$ and $\mathcal{D}_2^{\mathcal{B}}$, then unitarity ensures that the proof extends between $\Sigma_1$ and $\Sigma_2$. At the algebraic level our notations are the following. We denote $\mathcal{A}_i^{\scri}$ the algebra of observables of $\mathcal{D}_i^{\scri}$ and similarly $\mathcal{A}^{\mathcal{H}}$ is the one of $\mathcal{D}^{\mathcal{H}}$ so that the total algebra of $\mathcal{D}_i^{\mathcal{B}}$ is $\mathcal{A}_i^{\mathcal{B}} = \mathcal{A}^{\mathcal{H}} \otimes \mathcal{A}_i^{\scri}$.\footnote{This factorization occurs for massless fields.}
In this Section we are using all the vacuum states defined and analyzed in Section \ref{sec: quantization}, at the exception of the Minkowski vacuum since we are interested in black hole backgrounds. However the dual GSL from the Minkowski vacuum is treated in Appendix \ref{App: MinkowskivacGSL} for completeness.

\subsubsection*{Strategy of the proof}

Let $\ket \Omega$ be the Hartle-Hawking state, or one of its hard or soft regularization, 
and $\mathscr{H}_{\Omega}$ the Hilbert space built upon it via the GNS construction (see Theorem 1. of Appendix \ref{app: AQFT}). One can also built this Hilbert space from a set of positive frequency modes as we have done in Section \ref{sec: quantization} (see also Appendix \ref{app: QFT curved}). This $\mathscr{H}_{\Omega}$ represents the algebra of observables $\mathcal{A}^{\scri}$ which we can consider to be a von Neumann algebra (see Appendix \ref{app: modular theory}). Upon restriction to the subalgebra $\mathcal{A}_0^{\scri}$ (when $U_i = 0$), the state $\ket \Omega$ becomes cyclic and separating (it can be seen as a straightforward extension of the Reeh-Schlieder theorem \cite{reeh1961bemerkungen}). Using the symmetries we derived in the last Section, it will be actually possible to show that $\ket \Omega$ is also cyclic and separating for any subalgebra $\mathcal{A}_i^{\scri}$. Considering then $\ket \Psi \in \mathscr{H}_{\Omega}$ another cyclic and separating state for $\mathcal{A}_i^{\scri}$ one can compute the relative entropy \eqref{eq: relative entropy} between them.

Proving the GSL amounts to use the monotonicity of the relative entropy \eqref{eq: monotonicity} between the algebras $\mathcal{A}_1^{\scri}$ and $\mathcal{A}_2^{\scri} \subset \mathcal{A}_1^{\scri}$ (as $\mathcal{D}_2^{\scri} \subset \mathcal{D}_1^{\scri}$ as spacetime regions) which states that
\begin{equation}
    \label{eq: consequence relat entropy}
    S_{\mathcal{A}_1^{\scri}}(\Psi || \Omega) \geq S_{\mathcal{A}_2^{\scri}}(\Psi||\Omega) \, .
\end{equation}
Using the results of Appendix \ref{app: modular theory} especially \eqref{eq: relative in terms of von neumann} we find that \eqref{eq: consequence relat entropy} is equivalent to
\begin{equation}
    \label{eq: monotonicity in term of von Neumann}
    -S^{\text{v.N}, \mathcal{A}_1^{\scri}}_{\Psi|\Omega} + \langle K^{\mathcal{A}_1^{\scri}}_{\Omega} \rangle_{\Psi} \geq -S^{\text{v.N}, \mathcal{A}_2^{\scri}}_{\Psi|\Omega} + \langle K^{\mathcal{A}_2^{\scri}}_{\Omega} \rangle_{\Psi} \, ,
\end{equation}
with $S^{\text{v.N}, \mathcal{A}_i^{\scri}}_{\Psi|\Omega}$ the renormalized von Neumann entropy of $\ket \Psi$ with respect to $\ket \Omega$ (see \eqref{eq: def von neumann entropy}) and with $K^{\mathcal{A}_i^{\scri}}_{\Omega}$ the one-sided modular Hamiltonian of $\ket \Omega$ (see \eqref{eq: def one sided mod ham}). Consequently we get
\begin{equation}
    \label{eq: fundamental inequality}
    \Delta S^{\text{v.N.}, \mathcal{A}_i^{\scri}}_{\Psi|\Omega} - \Delta \langle K^{\mathcal{A}_i^{\scri}}_{\Omega}\rangle_{\Psi} \geq 0 \, ,
\end{equation}
with
\begin{equation}
    \label{eq: def des delta}
    \Delta S^{\text{v.N.}, \mathcal{A}_i^{\scri}}_{\Psi|\Omega} := S^{\text{v.N.}, \mathcal{A}_2^{\scri}}_{\Psi|\Omega} - S^{\text{v.N.}, \mathcal{A}_1^{\scri}}_{\Psi|\Omega} \quad \text{and} \quad \Delta \langle K^{\mathcal{A}_i^{\scri}}_{\Omega}\rangle_{\Psi} :=  \langle K^{\mathcal{A}_2^{\scri}}_{\Omega}\rangle_{\Psi} - \langle K^{\mathcal{A}_1^{\scri}}_{\Omega}\rangle_{\Psi} \, .
\end{equation}
An inequality analogous to \eqref{eq: fundamental inequality} can be also written for the total algebra $\mathcal{A}_i^{\mathcal{B}} = \mathcal{A}^{\mathcal{H}} \otimes \mathcal{A}_i^{\scri}$. One just has to take $\ket {\Omega^{\mathcal{H}}}$ an arbitrary faithful state on $\mathcal{H}_L^+$ (e.g. the restriction of the Hartle-Hawking state on the horizon) and consider the total vacuum state as
\begin{equation}
    \label{eq: total vac}
    \ket{\bar \Omega} := \ket{\Omega^\mathcal{H}} \otimes \ket{\Omega} \, ,
\end{equation}
which is cyclic and separating for $\mathcal{A}_i^{\mathcal{B}}$. From \eqref{eq: total vac} we construct the GNS Hilbert space $\mathscr{H}_{\bar \Omega}$ and consider a second cyclic and separating vector $\ket{\bar \Psi} \in \mathscr{H}_{\bar \Omega}$. Given the tensorial decomposition of the total algebra one can repeat the steps from \eqref{eq: consequence relat entropy} to \eqref{eq: fundamental inequality} and arrive at\footnote{Note that due to the product state structure of \eqref{eq: total vac}, the constant terms drop out from all the inequalities, and we have that $\Delta \langle K_{\bar{\Om}}^{\mathcal{A}_i^{\mathcal{B}}} \rangle_{\bar{\Psi}} = \Delta \langle K_{\Om}^{\mathcal{A}_i^{\scri}} \rangle_{\bar{\Psi}}$.} 
\begin{equation}
    \label{eq: total fundamental inequality}
    \Delta S^{\text{v.N.}, \mathcal{A}^\mathcal{B}_i}_{\bar \Psi|\bar \Omega} - \Delta \langle K^{\mathcal{A}_i^{\scri}}_{ \Omega}\rangle_{\bar \Psi} \geq 0 \, ,
\end{equation}
which is an inequality between the total algebras $\mathcal{A}^{\mathcal{B}}_1$ and $\mathcal{A}^{\mathcal{B}}_2$ i.e. between the spacetime domain $\mathcal{D}^{\mathcal{B}}_1$ and $\mathcal{D}^{\mathcal{B}}_2$.

Eq. \eqref{eq: total fundamental inequality} will be called the \emph{fundamental inequality} and admits the same form whatever the cyclic and separating states $\ket{\bar{\Omega}}$ and $\ket{\bar{\Psi}} \in \mathscr{H}_{\bar \Omega}$. What changes from a vacuum state to another is the relation between the vev of the one-sided modular Hamiltonian and the physical quantities of interest like the Bondi mass $M$ (at $\scri_R^+$) or the area of the horizon $A$ (on $\mathcal{H}_L^+$). The link is done in two steps. First, using Appendix \ref{app: modular boost energy}, one can relate the modular Hamiltonian to the integral of the normal-ordered boost energy (see \eqref{eq: normal ordered and stress tensor}). Second, one relates this normal-ordered quantity to the covariant one and uses the results of paragraph \ref{subsec: einstein eq} to make $M$ appear. Finally one can write the thermodynamic potential $\mathcal{G}_{\Omega}$ whose monotonicity at $\scri_R^+$ is the dual GSL.

\subsection{Hartle-Hawking state}
\label{subsec: GSL HH state}

First, we choose $\ket{\Om}$ to be the restriction of the Hartle-Hawking vacuum to $\scri^+_R$, and we denote it by $\ket{\Om_H}$. It has been constructed in Section \ref{sec: quantization} and it is a thermal state when restricted to the algebra $\mathcal{A}_0^{\scri}$ i.e. any pair of observables $(A, B) \in \left(\mathcal{A}_0^\scri\right)^2$ satisfies the Kubo-Martin-Schwinger (KMS) conditions \eqref{eq: KMS condition}. That state is natural to consider at null infinity for the dual GSL as Wall \cite{Wall:2011hj} used the restriction of the Hartle-Hawking state to the black hole horizon $\mathcal{H}_L$.\footnote{In particular, the state $\ket{\Om_H}$ can be defined at null infinity even if the Hartle-Hawking state does not exist, as it is the case for the Kerr black hole, similarly to what had been noted on the horizon \cite{Wall:2011hj}.} It is shown in Appendix \ref{app: modular boost energy} that the one-sided modular Hamiltonian wrt to the restricted algebra $\mathcal{A}_0^{\scri}$ is given by
\be \label{totmodhamnormord}
    K_{\Om_H}^{\mathcal{A}_0^{\scri}} = 2 \pi \int_{\scri_R^+} U :T_{UU}:_{\Om_H} \text{d}U \wedge \eps_S
\ee
so that formally, the restriction of $\ket{\Om_H}$ to $\mathcal{A}_0^{\scri}$ can be written as
\be \label{modhamhh0}
    \rho_{\Om_H}^{\mathcal{A}_0^\scri} = \frac{e^{- K_{\Om_H}^{\mathcal{A}_0^\scri}}}{\Tr \left(e^{- K_{\Om_H}^{\mathcal{A}_0^\scri}}\right)}
\ee
where $:T_{UU}:_{\Om_H}$ is the normal-ordered stress energy tensor introduced in \eqref{stressenergyhh}. 

Using \eqref{eq: sym HH state}, namely the symmetries of the state $\ket{\Om_H}$, we can restrict it to the future domain $\mathcal{D}_i^\scri$ of any cut $U = U_i$ of $\scri^+_R$ (via the change $U \to U - U_i$) and still get a thermal state wrt  the restricted algebra of observables $\mathcal{A}_i^\scri$. Therefore, the one-sided modular Hamiltonian associated to $\mathcal{A}_i^\scri$ is given by
\be \label{modhamnormordi}
    K_{\Om_H}^{\mathcal{A}_i^\scri} = 2 \pi \int_{\mathcal{D}_i^\scri} (U - U_i) :T_{UU}:_{\Om_H} \text{d}U \wedge \eps_S \, ,
\ee
and the formal density matrix is
\be \label{modhamhhi}
    \rho_{\Om_H}^{\mathcal{A}_i^\scri} = \frac{e^{- K_{\Om_H}^{\mathcal{A}_i^\scri}}}{\Tr{\left(e^{- K_{\Om_H}^{\mathcal{A}_i^\scri}}\right)}} \, .
\ee
Hence we can apply the fundamental inequality \eqref{eq: total fundamental inequality} with \eqref{modhamhhi}. It remains to relate the latter to the Bondi mass $M$, using Einstein's equations \eqref{kuzerodef}. However these equations uses the covariant stress-tensor $T_{UU}$ instead of the normal-ordered version $:T_{UU}:_{\Omega_H}$. The relation between the two can be obtained from the equivalence principle which states that
\begin{equation}
    \label{eq: equiv principle minko}
    T_{u_+u_+} = :T_{u_+u_+}:_{0_M} \, ,
\end{equation}
with $u_+ = \kappa^{-1} \ln U$ the affine coordinate on $\scri_R^+$ and $\ket{0}_M$ the Minkowski vacuum. Under the change $u_+ \to U$ the lhs of \eqref{eq: equiv principle minko} transforms covariantly while the rhs transforms with the Schwarzian derivative \eqref{eq: normal ordered HH}. The latter physically corresponds to the zero point energy, i.e the energy density of the Hartle-Hawking vacuum $\ket{\Om_H}$ compared to the one of the Minkowski vacuum $\ket{0}_M$ on $\scri^+_R$. This quantity, as we said precisely in \eqref{eq: normal ordered HH}, is infinite, and that was why regularizations of the Hartle-Hawking state were necessary to introduce. However, an important proof of concepts can be performed if we allow ourself for a moment to manipulate this infinite Schwarzian. Formally then, the relation between the covariant and normal-ordered stress-energy tensor admits the following form
\begin{equation}
    \label{eq: relation cov and normal}
    T_{UU} = \, :T_{UU}:_{\Omega_H} + \, \text{Schwarzian term} \, ,
\end{equation}
so that from Einstein's equations \eqref{kuzerodef} we get (the state $\ket {\bar \Psi} \in \mathscr{H}_{\bar \Omega_H}$, with $\ket{\bar \Omega_H} = \ket{\Omega^\mathcal H} \otimes \ket{\Omega_H}$, has to be chosen such that \eqref{assumpthor} holds)
\begin{equation}
    \label{eq: modular HH and bondi mass}
     \Delta \langle K_{\Om_H}^{\mathcal{A}_i^{\scri}} \rangle_{\bar \Psi} = \Delta M - \Delta M_{\Om_H}
\end{equation}
where $\Delta M_{\Omega_H}$ is the variation of Bondi mass in the Hartle-Hawking state coming from the Schwarzian. It is infinite.\footnote{Notice that even if there is no net energy flux in the bulk in the Hartle-Hawking vacuum, there is an outgoing energy flux at $\scri^+_R$ since at future null infinity the outgoing radiation is not compensated by the incoming one.} From the fundamental inequality \eqref{eq: total fundamental inequality} we get
\begin{equation}
    \label{eq: incomplete inequality HH}
    \Delta M - \Delta M_{\Om_H} - T_H \Delta S_{\bar{\Psi} \lvert \bar{\Om}_H}^{\text{v.N.},\mathcal{A}_i^{\mathcal{B}}} \leq 0 \, .
\end{equation}
The term $\Delta M_{\Omega_H}$ can actually be related to a notion of entropy using the following reasoning. The outgoing radiation in the state $\ket{\Om_H}$ not only carries energy but also entropy, and it is clear that $S_{\bar{\Om}_H \lvert \bar{\Om}_H}^{\text{v.N.},\mathcal{A}_i^{\mathcal{B}}} = 0 $, so that we also need to add the entropy variation of the reference state $\Delta S_{\Om_H}$. However, the Hartle-Hawking state is a thermal state. On the slices $\Sigma_i$ of constant volume, \footnote{The volume of the two spacelike slices $\Sigma_1$ and $\Sigma_2$ is infinite but fixed.} the entropy variation can only come from the energy variation between $\Sigma_1$ and $\Sigma_2$, given by  $\Delta M_{\Om_H}$. Using the equilibrium relation
\be
    \frac{\p S}{\p E} \Big|_V = \frac{1}{T_H}
\ee
we conclude \footnote{Equivalently, we can use the Clausius relation $\Delta S = \frac{Q}{T_H}$, valid at first order in perturbation around an equilibrium state. Since the radiation in the Hartle-Hawking state at $\scri^+_R$ is exactly thermal, there is no work that can be extracted by an asymptotic observer, so that $W = 0$ and we have $\Delta S = \frac{\Delta E}{T_H}$ using the first law of thermodynamics.} that
\be \label{zeropointenergyentropyrel}
   \Delta S_{\Om_H} = \frac{\Delta M_{\Om_H}}{T_H} \, .
\ee
Therefore by adding \eqref{zeropointenergyentropyrel} to \eqref{eq: incomplete inequality HH} we get 
\be \label{free energy decreases hh}
    \Delta M - T_H \Delta S \leq 0
\ee
where
\be
    \Delta S = \Delta S_{\bar{\Psi} \lvert \bar{\Om}_H}^{\text{v.N.},\mathcal{A}_i^\mathcal{B}} +  \Delta S_{\Om_H}
\ee
would be the total entropy variation. The thermodynamic potential of interest in the Hartle-Hawking state is therefore the free-energy
\begin{equation}
    \label{eq: potential HH}
    \mathcal{G}_{\Omega_H} = M - T_{H} S := \mathcal{F} \, .
\end{equation}
It seems that we have proven in \eqref{free energy decreases hh} the decreasing of the free energy \eqref{eq: potential HH} in the Hartle-Hawking state. A similar relation has been argued to hold in Section 3 of \cite{ARB24}. Notice that the variation $\Delta \mathcal{F}$ is finite and well-defined, and vanishes if $\ket{\bar \Psi} = \ket{\bar \Om_H}$.
However, both quantities appearing in \eqref{zeropointenergyentropyrel} are \emph{stricto sensus} infinite in that state. 
It is because the radiation does not only come from the black hole but also from every direction in space, so that the energy density of the physical stress energy tensor is constant and does not decay at infinity. In particular, the black hole is in thermal equilibrium with the surrounding radiation and cannot be distinguished from the latter. Still, a real black hole is not in equilibrium with some incoming radiation and only emits low angular momenta modes, because the Hawking radiation comes \emph{only} from the neighborhood of the black hole horizon. Therefore, in order to interpret \
\eqref{free energy decreases hh}, we need to regulate the high angular momentum modes. This is what we do now by considering the hard and soft regularizations we introduced in Section \ref{sec: quantization}.

\subsection{Hard regularization: the $L$-vacuum}
\label{subsec: GSL hard}

We consider in this paragraph the $L$-vacuum $\ket{\Omega_H^L}$ constructed from the notion of time \eqref{timehhbigULvac} and the space of positive solutions \eqref{eq: L-vac positive solutions}. Before proceeding with the expression of the modular Hamiltonian, we prove a general result.

\subsubsection*{A useful lemma}
\label{secnonvanishingzeropointdiv}

Assume that on any section $S$ of $\scri^+_R$ there exist a constant $C_\Om$ (which depends on the vacuum state $\ket{\Om}$) such that the relation between the integrated stress tensor (on $S$) and its integrated normal-ordered version is of the form
\be \label{splittingrealnosch}
    \int_{S} T_{UU} \text{d}U \wedge \eps_S = \int_{S} :T_{UU}:_{\Om} \text{d}U \wedge \eps_S +  \frac{C_\Om}{U^2} \text{d}U \, .
\ee
Then it follows that
\be \label{IRregularizationlambda}
    \int_{\mathcal{D}_i^\Lambda} (U - U_i) T_{UU} \text{d}U \wedge \eps_S  =  \int_{\mathcal{D}_i^\Lambda} (U - U_i) :T_{UU}:_{\Om} \text{d}U \wedge \eps_S + \int_{u_i}^\lambda C_\Om  (1 - e^{- \kappa(u - u_i)}) \kappa  \text{d}u
\ee
where $\mathcal{D}_i^\Lambda = (U_i, \Lambda) \times S^2$ with $\lambda = \kappa^{-1} \ln{\Lambda}$, and $U = \Lambda$ (i.e. $u = \lambda$) is a cross-section of $\scri^+_R$ at (almost) arbitrary late time, as we assume that $\lambda - u_i \gg \kappa^{-1}$. Then, in the limit $\lambda \rightarrow + \infty$ (equivalently $\Lambda \rightarrow + \infty $), the second term in the right hand side of \eqref{IRregularizationlambda} diverges while
\be
    \int_{\mathcal{D}_i^\Lambda} (U - U_i) :T_{UU}:_{\Om} \text{d}U \wedge \eps_S \underset{\Lambda \rightarrow + \infty}{\longrightarrow} \frac{1}{2 \pi} K_{\Om}^{\mathcal{A}_i^\scri}
\ee
converges to the one-sided modular Hamiltonian. Therefore, we conclude that the left hand side of \eqref{IRregularizationlambda} necessarily diverges. However, computations related to the dual GSL only involve differences like e.g.
\be
    \Delta K_{U_i}^{\Lambda} := \int_{\mathcal{D}_2^\Lambda} (U - U_i) T_{UU} \text{d}U \wedge \eps_S - \int_{\mathcal{D}_1^\Lambda} (U - U_i) T_{UU} \text{d}U \wedge \eps_S
\ee
which is well defined for $\Lambda > U_2$ and converges when $\Lambda \rightarrow + \infty$. Indeed, 
\be
    \Delta K_{U_i}^{+\infty} = \frac{1}{2 \pi}\Delta K_{\Om}^{\mathcal{A}_i^\scri} - (u_2 - u_1) \kappa C_\Om \, ,
\ee
so that for any $\ket{\Psi} \in \mathscr{H}_\Om$, we have that 
\be
\label{lemma 1}
    \Delta \langle K_{U_i}^{+\infty} \rangle_{\Psi} = \frac{1}{2\pi}\Delta \langle K_{\Om}^{\mathcal{A}_i^{\scri}} \rangle_\Psi - (u_2 - u_1) \kappa C_\Om
\ee
that is finite. Recalling \eqref{diffbondimass} i.e.  
\be
\label{lemma 2}
     \Delta \langle K_{U_i}^{+\infty} \rangle_{\Psi} := \Delta \langle K_{U_i} \rangle_{\Psi} = \frac{\Delta M}{\kappa}
\ee
we find that now the Bondi mass variation is finite. Finally when \eqref{splittingrealnosch} holds the relation between the variation of the one-sided modular Hamiltonian and the variation of Bondi mass reads
\begin{equation}
    \label{eq: link mod Bondi lemma}
    \Delta M = T_H \Delta \langle K_{\Om}^{\mathcal{A}_i^{\scri}} \rangle_\Psi - (u_2 - u_1) \kappa^2 C_\Om \, ,
\end{equation}
and we define, by analogy with \eqref{eq: modular HH and bondi mass},
\begin{equation}
    \label{eq: bond mass vacuum variation}
    \Delta M_{\Omega} = -(u_2 -u_1) \kappa^2 C_\Om \,,
\end{equation}
the variation of Bondi mass in the vacuum state $\ket{\Omega}$. The main message of this paragraph is that having a non vanishing Schwarzian (i.e. an energy density in the vacuum state $\ket \Om$) is fine as long as the integrated stress energy tensor on a cross section $S$ can be written is a form \eqref{splittingrealnosch}. Under such an hypothesis the variation of Bondi mass will be finite between two arbitrary cross sections. Note that \eqref{lemma 1}, \eqref{lemma 2} and \eqref{eq: link mod Bondi lemma} remain valid upon replacing $\ket{\Psi}$ by $\ket{\bar \Psi} \in \mathscr{H}_{\bar \Omega}$ with $\ket{\bar \Omega}$ defined in \eqref{eq: total vac}.

\subsubsection*{Modular Hamiltonian of the $L$-vacuum}

To make contact with the last lemma and to avoid rendering the notations even more heavy, we shall consider in this paragraph the integrated version of the stress-energy tensor over the spatial cross-sections of $\scri_R^+$. When this is the case we replace the canonical capital $T$ by a lowercase $t$ e.g.
\begin{equation}
    \label{eq: integrated stress 1}
    :t_{U U}:_{\Omega} = \int_S :T_{U U}:_{\Omega} \eps_S \, .
\end{equation}
From the general results of Appendix \ref{app: modular boost energy} and using the symmetries \eqref{mobius4D}, it can be shown that the one-sided modular Hamiltonian of $\ket {\Om_H^L}$ restricted to the algebra $\mathcal{A}_i^\scri$ is 
\be \label{modhamlvacuum}
    K_{\Om_H^L}^{\mathcal{A}_i^\scri} := \sum_{l=0}^{+ \infty} \int_{\mathcal{D}_i^\scri} k_{l} = 2 \pi \sum_{l = 0}^{L-1} \int_{U_i}^{+ \infty} (U - U_i) :t_{UU}^l:_{\Om_H^L} \text{d}U + 2 \pi  \sum_{l = L}^{+ \infty} \int_{u_i}^{+ \infty} (u - u_i) :t_{uu}^l:_{\Om_H^L} \text{d}u
\ee
where (we omit the hat as it is clear that we are dealing with quantum fields)
\be
    \phi = \sum_{l = 0}^{+ \infty} \phi_l, \qquad \phi_l = \sum_{m=-l}^{m = +l} \int_0^{+ \infty} \frac{d \Om}{\sqrt{4 \pi \Om}} \left(a_{\Om l m} Y_{m}^l e^{- i \Om U} + \bar{Y}_m^l a_{\Om l m}^\dag e^{i \Om U}\right)
\ee
so that
\begin{align}
    :t_{uu}^l:_{\Om_H^L} &= \int_{S^2} \left(\p_u \phi_{l} \p_u \phi_{l} - \langle \p_u \phi_{l} \p_u \phi_{l} \rangle_{\Om_H^L}\right) \eps_S \, .
\end{align}
The covariant stress energy tensor, integrated on the spatial sections of $\scri^{+}_R$ is 
\be \label{tuusumtuul}
    t_{uu} := \int_{S^2} T_{uu} \eps_S = \sum_{l = 0}^{+ \infty} t_{uu}^l \, ,
\ee
where we defined the effective two-dimensional stress energy tensor associated to the chiral CFTs with the same quantum number $l$ to be
\be
    t_{uu}^l = :t_{uu}^l:_{0_M} = \int_{S^2} \left(\p_u \phi_{l} \p_u \phi_{l} - \langle \p_u \phi_{l} \p_u \phi_{l} \rangle_{0_M}\right) \eps_S \, . 
\ee
Since for the modes $l \geq L$ the $L$-vacuum is indistinguishable from the Minkowski vacuum we have that
\be
    l \geq L \quad \Rightarrow \quad :t_{uu}^l:_{\Om_H^L} = :t_{uu}^l:_{0_M} = t_{uu}^l
\ee
Therefore, in order to relate the one-sided modular Hamiltonian \eqref{modhamlvacuum} to the real stress energy distribution (integrated on a cross section $S$) we change coordinate and deal with the Schwarzian derivative only when $l < L$. At the end we get
\be
    2 \pi (U - U_i) t_{UU}^l \text{d}U = \frac{2 \pi}{\kappa} (1 - e^{- \kappa(u - u_i)}) t_{uu}^l \text{d}u = \int_{S^2} k_{l} + \mathcal{R}_l
\ee
where 
\be \label{Rl1}
\left\{ 
    \begin{array}{ll}
        \mathcal{R}_l = \frac{2 \pi}{\kappa} (1 - e^{- \kappa(u - u_i)}) \frac{(2l + 1) \kappa^2}{48 \pi} \text{d}u & \mbox{if} \quad l < L \\
        \mathcal{R}_l = [-2 \pi (u - u_i) + \frac{2 \pi}{\kappa} (1 - e^{- \kappa (u- u_i)})] t_{uu}^l \text{d}u & \mbox{if} \quad l \geq L
    \end{array}
    \right. 
\ee
It is clear that the first term in \eqref{Rl1} diverges if $L \rightarrow + \infty$ while the second one goes to zero (if we consider a state $\ket{\Psi} \in \mathscr{H}_{\Om_H^L}$ that have finite energy density). Then, at fixed $L$, we consider the states $\ket{\Psi} \in \mathscr{H}_{\Om_H^L}$
so that 
\be \label{conditionRl}
    \sum_{l=0}^{L-1} \langle \mathcal{R}_l \rangle_\Psi = \frac{2 \pi}{\kappa} (1 - e^{- \kappa(u - u_i)}) \frac{L^2 \kappa^2}{48 \pi} \text{d}u  \gg \sum_{l= L}^{+ \infty} \langle \mathcal{R}_l \rangle_\Psi
\ee
that is a condition always satisfied for $\ket{\Om_H^L}$ since $\sum_{l= L}^{+ \infty} \langle \mathcal{R}_l \rangle_{\Om_H^L} = 0$ and that is true for most of the states if we take $L \gg 1$ (and true in a sense for any state if $L = + \infty$ since the left hand side of \eqref{conditionRl} diverges).\footnote{This condition is immediately satisfied for the states $\ket{\Psi} \in \{ \mathcal{A}^L_i \ket{\Om_H^L} \}$ where the algebra $\mathcal{A}^L_i$ is given by \eqref{4DLsmearing} ($\mathcal{A}_i^L$ is the subalgebra of \eqref{4DLsmearing} restricted to the region $\mathcal{D}_i$ ).} Then, since 
\be
    2 \pi (U - U_i) t_{UU} \text{d}U = \sum_{l = 0}^{+ \infty}  2 \pi (U - U_i) t_{UU}^l \text{d}U 
\ee
we have that 
\begin{align} \label{modhambmassvar}
   \frac{2 \pi}{\kappa} \int_{U_i}^{\Lambda} \kappa (U - U_i) \langle t_{UU} \rangle_\Psi \text{d}U &= \sum_{l = 0}^{+ \infty} \int_{\mathcal{D}_i^\Lambda} \langle k_l \rangle_{\Psi} + \frac{2 \pi}{\kappa} \int_{u_i}^\lambda (1 - e^{- \kappa(u - u_i)}) \frac{L^2 \kappa^2}{48 \pi} \text{d}u
\end{align}
which is of the form \eqref{splittingrealnosch} with $C_{L} = \frac{L^2}{48\pi}$. Therefore in virtue of the lemma we get in the limit $\Lambda \to + \infty$
\be \label{relationbondimass}
    \frac{2 \pi}{\kappa} \Delta M = \Delta \langle K^{\mathcal{A}_i}_{\Om_H^L} \rangle_{\Psi} + \frac{2 \pi}{\kappa} \Delta M_{\Om_H^L} 
\ee
where
\be
    \Delta M_{\Om_H^L} := -(u_2 - u_1) \frac{L^2 \kappa^2}{48 \pi}
\ee
is the stress energy density on $\scri^+_R$ in the state $\ket{\Om_H^L}$, and is therefore the zero-point energy density. This is true as long as \eqref{assumpthor} and \eqref{conditionRl} are valid for $\ket{\Psi}$ (it is all the more true when $L$ is large).

\subsubsection*{Second law from the $L$-vacuum}

If we fix $L \in \mathbb{N}^\ast$ (assuming that it is "large enough" for \eqref{conditionRl} to hold), and and write the fundamental inequality \eqref{eq: total fundamental inequality}
\be \label{mrelvacuum}
     \Delta S^{\text{v.N.}, \mathcal{A}^\mathcal{B}_i}_{\bar \Psi|\bar \Omega^L_H} - \Delta \langle K^{\mathcal{A}_i^{\scri}}_{ \Omega_H^L}\rangle_{\bar \Psi} \geq 0 \,
\ee
where $\ket{\bar \Omega^L} = \ket{\Omega^{\mathcal{H}}} \otimes \ket{\Omega_H^L}$ and $\ket{\bar \Psi} \in \mathscr{H}_{\bar \Omega^L}$. All the steps between \eqref{Rl1} and \eqref{relationbondimass} remain valid as long as $\ket{\bar \Psi}$ also satisfies \eqref{conditionRl}. We see with \eqref{relationbondimass} that the one-sided modular Hamiltonian is related to the difference between the Bondi mass variation in the state $\ket{\bar \Psi}$ and the one in the $L$-vacuum. To make the latter quantity disappear we use the Clausius relation, as we discussed in the paragraph \ref{subsec: GSL HH state}. Indeed the radiation in the $L$-vacuum does not only carry energy but also entropy. Exactly as we argued before, the radiation is thermal so that we cannot extract work from it, hence the energy dissipated should be counted as a heat flux. Then, since in the state $\ket{\Om_H^L}$ the total amount of energy is strictly speaking infinite \footnote{In the $L$-vacuum, the total amount of energy is infinite, as it is the case in the Hartle-Hawking vacuum state or in the more physical Unruh vacuum state, even if the local energy density at $\scri^+_R$ is not. It means that if we calculate the ADM mass in all these states we would get infinity. It is not a problem if we only look at the local observables and local back-reactions, as we are doing here. For instance, the black hole luminosity at $\scri^+_R$, computed in the Unruh vacuum state is a local quantity that has physical significance, even if the ADM mass of the Unruh vacuum is infinite.} and the radiation is in thermal equilibrium, we can apply the Clausius relation to obtain the variation of entropy in the $L$-vacuum (the zero point entropy variation)
\be \label{ClausiusrelLhard}
    \Delta S_{\Om_H^L} = \frac{2 \pi}{\kappa} \Delta M_{\Om_H^L} \, .
\ee
Notice that strictly speaking, the vacuum state $\ket {\Om_H^L}$ is thermal with respect to the flow generated by the boost field $(U - U_i) \p_U$ only if we restrict ourselves to the algebra $\mathcal{A}_i^L$ defined in \eqref{4DLsmearing}. \footnote{This means that the KMS conditions in the vacuum state $\ket{\Om_H^L}$ are satisfied only for any pair of observables $(A,B) \in \mathcal{A}_i^L$ if we take the modular Hamiltonian to be conjugated to the "time" generator $\xi = (U -U_i) \p_U$.} However, in the state $\ket{\Om_H^L}$, the modes $l \geq L$ are not excited with respect to the Minkowski vacuum $\ket{0}_M$, so that they do not carry any energy and entropy, and therefore the thermal reservoir justifying the use of the Clausius relation \eqref{ClausiusrelLhard} is only made of the modes with angular momentum $l \leq L$. Then, by combining \eqref{ClausiusrelLhard}, \eqref{relationbondimass} and \eqref{mrelvacuum}, we get 
\be \label{freeeenrgy}
    \Delta M - T_H \Delta S \leq 0
\ee
where
\be
    \Delta S := \Delta S_{\bar \Psi \lvert \bar{\Om}_H^L}^{\text{v.N.},\mathcal{A}_i^\mathcal{B}} + \Delta S_{\Om_H^L}
\ee
and we obtain that the relevant thermodynamic potential in the $L$-vacuum is again the free energy
\begin{equation}
    \label{eq: potential L vac}
    \mathcal{G}_{\Omega_H^L} = M - T_H S := \mathcal{F} \, .
\end{equation}
Notice that if we take $L = + \infty$, the relation \eqref{freeeenrgy} still makes sense and is perfectly well defined, but $\Delta M$ and $\Delta S$ become infinite: we recover the Hartle-Hawking case \eqref{free energy decreases hh}. As we argued in Section \ref{sec: quantization}, a nice notion of black hole at thermal equilibrium at the Hawking temperature would be to take $L$ arbitrarily large but finite. Then, the only outgoing energy flux that is not compensated by an incoming one would be coming from the modes $l \geq L$, which are actually exponentially suppressed when $l$ is big, so that if we take $L$ to be large enough, we can neglect the contribution of this flux for arbitrarily large time intervals.\footnote{Notice that this equilibrium would nevertheless be very instable, see footnote \ref{foot: unstable equilib}.}


\subsection{Soft regularization: the $\kappa_l$-vacuum}
\label{subsec: GSL soft}

We consider finally the $\kappa_l$-vacuum state $\ket{\Omega^{\{ \kappa_l\} }_H}$ defined using the notion of time \eqref{ulmsoftdef} and the positive frequency modes \eqref{eq: positive space kappa vacuum}. This state admits the tensor product decomposition \eqref{vallmten} with $U^{(l,m)} = U^l = e^{\kappa_l u} $ on $\scri^+_R$. Then, we decompose the field operator $\phi$ in a sum of scalar fields $\phi_l$ corresponding to the different sectors $l$
\begin{equation}
    \label{defphil}
    \phi = \sum_l \phi_l \Longrightarrow \phi_l = \sum_{m = -l}^{l} \int_{0}^{+ \infty} \frac{d \Om}{\sqrt{4 \pi \Om}} \left(Y_m^l \hat a_{\Om l m} e^{- i \Om U^{(l,m)}} + \bar{Y}_m^l \hat a_{\Om l m}^\dag e^{i \Om U^{(l,m)}}\right)
\end{equation}
so that the $l$-stress-energy tensor (renormalized wrt the vacuum $\ket{\Omega^{U^l}_H}$ attached to each sector $l$, see \eqref{vallmten}) reads
\begin{equation}
    \label{eq: lstresstensor}
    :T_{U^l U^l}^{l}:_{\Omega^{U^l}} = \partial_{U^l} \phi_l \partial_{U^l} \phi_l - \langle \partial_{U^l} \phi_l \partial_{U^l} \phi_l \rangle_{\Omega^{U^l}} \, .
\end{equation}
Like in the previous paragraph on the $L$-vacuum, we denote $t_{uu}$ the component of the stress-energy tensor integrated over a spatial cross section of $\scri_R$ i.e. for instance
\begin{equation}
    \label{eq: integrated stress}
    :t_{U^l U^l}^l:_{\Omega^{U^l}_H} = \int_S :T_{U^l U^l}^l:_{\Omega^{U^l}_H} \eps_S \, .
\end{equation}

\subsubsection*{Modular Hamiltonian of the $\kappa_l$-vacuum}

From the general results of Appendix \ref{app: modular boost energy}, we can compute the one-sided modular Hamiltonian of the state $\ket{\Omega^{U^l}_H}$ with respect to the algebra $\mathcal{A}_i^\scri$
\begin{equation}
    \label{modhamUlvac}
    K^{\mathcal{A}_i^\scri}_{\Omega^{U^l}_H} = 2\pi \int_{\mathcal{D}_i^\scri}(U^l -U^l_i) :T_{U^l U^l}^l:_{\Omega^{U^l}} \text{d}U^l \wedge \epsilon_S \, .
\end{equation}
The relation between the covariant and normal-ordered stress-energy tensor, itegrated over $S$, reads
\be \label{splittingrealnosch2new}
    t_{U^l U^l}^l = \, :t_{U^l U^l}^l:_{\Omega^{U^l}_H} +\,  \frac{1}{(\kappa_l U_l)^2} \frac{(2 l + 1 ) \kappa_l^2}{48 \pi} \, ,
\ee
that is in the form \eqref{splittingrealnosch} with $C_l = \frac{(2l+1)}{48\pi}$. Therefore we deduce that 
\be \label{realstressenergysoftlnew}
    \frac{2 \pi}{\kappa_l}  \kappa_l (U^l - U_i^l) t_{U^l U^l}^l \text{d}U^l  = 2 \pi  (U^l - U_i^l) :t_{U^l U^l}^l:_{\Omega^{U^l}_H} \text{d}U^l   + \frac{2 \pi}{\kappa_l} (1 - e^{- \kappa_l (u - u_i)}) \frac{(2 l + 1 ) \kappa_l^2}{48 \pi} \text{d}u \, .
\ee
The integration of \eqref{realstressenergysoftlnew} over the whole spacetime domain $\mathcal{D}_i^\Lambda$ (it remains to perform the integral over the null time direction) gives
\begin{equation}
    \label{eq: integration over domain}
    K^{\mathcal{A}_i^\scri}_{\Omega^{U^l}_H} = \frac{2\pi}{\kappa_l} \left( E^l + \int_{u_i}^{\lambda} \left(e^{-\kappa_l (u - u_i)} - 1\right) \frac{(2l+1) \kappa_l^2}{48\pi} \text{d}u \right) \, ,
\end{equation}
that is well defined although the two terms in the rhs of \eqref{eq: integration over domain} are not, so that we need to introduce an infrared regulator $\Lambda := \kappa^{-1} \ln \lambda$ to manipulate them individually. We also have introduced $E^l$, the \emph{energy of mode l} 
\begin{equation}
    \label{deftotalenergyl}
    E^l := \int_{\mathcal{D}_i^\scri}  \kappa_l (U^l - U_i^l) T_{U^l U^l}^l \text{d}U^l \wedge \eps_S = \int_{\mathcal{D}_i^\scri} \left( 1 - e^{-\kappa_l(u-u_i)} \right) T_{uu}^l \text{d}u \wedge \eps_S \, . 
\end{equation}
that is also regularized with an infrared cutoff $\Lambda$ as in the proof of the lemma \eqref{splittingrealnosch}. Since in the end one will only look at differences of quantities like \eqref{deftotalenergyl}, the presence of such cutoffs is only temporary and the final results will be of course regularization-independent. Therefore, for the sake of clarity, the presence of the infrared cutoff $\Lambda$ in \eqref{deftotalenergyl} is kept implicit. At this stage, it is interesting to introduce a new quantity $\tilde{\m}_l$ is order to relate our analysis to usual thermodynamics. We set 
\be
\label{eq: def chemical potential l}
    \tilde{\m}_l := 1 - \frac{\kappa}{\kappa_l}
\ee
and we notice that since $\kappa \geq \kappa_l$, $ \tilde{\m}_l$ is negative and vanishes if we have a perfect transmission, i.e $\tilde{\m}_l = 0$ when $\kappa_l = \kappa$. As we shall show latter in the proof, $\tilde \mu_l$ can be interpreted as a \emph{chemical potential} for the two-dimensional effective CFT underpinning the field at angular momentum mode $l$. Then we can rewrite \eqref{eq: integration over domain} as
\begin{equation}
    \label{eq: integration domain with chemical}
    K^{\mathcal{A}_i^\scri}_{\Omega^{U^l}_H} = \frac{2\pi}{\kappa} (1 - \tilde{\mu}_l) \left( E^l + \int_{u_i}^{\lambda} \left(e^{-\kappa_l (u - u_i)} - 1\right) \frac{(2l+1) \kappa_l^2}{48\pi} \text{d}u \right) \, .
\end{equation}
so that the one-sided modular Hamiltonian of the $\kappa_l$-vacuum is obtained only via a sum over the one-sided modular Hamiltonians associated to each sector $l$
\begin{equation}
    \label{eq: kappalvacmodham}
    K^{\mathcal{A}_i^\scri}_{\Om^{\{ \kappa_l \}}_H} := \sum_l  K^{\mathcal{A}_i^\scri}_{\Omega^{U^l}_H} = \sum_l \left[\frac{2\pi}{\kappa} (1 - \tilde{\mu}_l) \left( E^l + \int_{u_i}^{\lambda} \left(e^{-\kappa_l (u - u_i)} - 1\right) \frac{(2l+1) \kappa_l^2}{48\pi} \text{d}u \right) \right] \, .
\end{equation}
Of course, we choose the sequence $\{ \kappa_l \}_{l \geq 0}$ so that\footnote{Notice that $\sum_{l \geq 0} (2l + 1) \kappa_l < + \infty$ also implies that $\sum_{l \geq 0} (2l + 1) \kappa_l^2 < + \infty$}
\be
    \sum_{l \geq 0} (2l + 1) \kappa_l < + \infty
\ee
in order to get finite quantities.  \footnote{\label{foot: proposal kappa l}Since the potential barrier has a height that is of order $l(l+1)$, it would be natural to choose $\kappa_l \sim e^{- \alpha l(l + 1)}$ with $\a > 0$ a constant, to try to model a real black hole. Such a choice of $\kappa_l$ ensures convergence.}

\subsubsection*{Monotonicity of relative entropy}

Now, we look at the relative entropy between a state $\ket{\bar \Psi} \in \mathscr{H}_{\bar \Om^{\{ \kappa_l \}}_H}$ and the total vacuum $\ket{\bar \Omega^{\{ \kappa_l \}}_H} = \ket{\Omega^\mathcal{H}} \otimes \ket{\Om^{\{ \kappa_l \}}_H}$ with respect to the algebra $\mathcal{A}_i^\mathcal{B}$. Then, from the fundamental inequality \eqref{eq: total fundamental inequality} we can write
\be \label{monotoagainomkappalnew}
   \Delta S^{\text{v.N.}, \mathcal{A}_i^\mathcal{B}}_{\bar \Psi \lvert \bar \Om^{\{ \kappa_l \}}_H} - \Delta \langle K^{\mathcal{A}_i^{\scri}}_{\Om^{\{ \kappa_l \}}_H} \rangle_{\bar \Psi} \geq 0 \, .
\ee
Given \eqref{eq: kappalvacmodham} we have
\begin{equation}
    \label{eq: deltamodhamkappal}
    \Delta \langle K^{\mathcal{A}_i^\scri}_{\Om^{\{ \kappa_l \}}_H} \rangle_{\bar \Psi} = \frac{2\pi}{\kappa} \sum_l \left(\Delta E^l - \Delta E^l_{\Omega^{U^l}_H} \right) - \frac{2\pi}{\kappa} \sum_l \tilde{\mu_l}\left(\Delta E^l - \Delta E^l_{\Omega^{U^l}_H} \right) \, ,
\end{equation}
where $\Delta E^l_{\Omega^{U^l}_H}$ is the \emph{energy-variation of the mode $l$ in the vacuum} $\ket{\Omega^{U^l}_H}$ attached to the angular mode $l$. Following \eqref{eq: bond mass vacuum variation}, the latter reads
\be \label{vacuumvariationsnew}
    \Delta E^l_{\Om^{U^l}_H} = -(u_2 - u_1) \frac{(2l + 1) \kappa_l^2}{48 \pi}\, ,
\ee
where all the quantities defined above are finite due to the lemma \eqref{splittingrealnosch}.\footnote{In particular, the dependence on the cutoff $\Lambda$ has been removed at this stage.} Now, we use an assumption similar to \eqref{assumpthor} on the state $\ket{\bar \Psi}$
\be \label{assumfoanyl}
    \forall l, \qquad \kappa_l^{-1} \p_u \langle T_{uu}^l \rangle_{\bar \Psi} \ll \langle T_{uu}^l \rangle_{\bar \Psi} 
\ee
so that we can neglect the exponential term in \eqref{deftotalenergyl} in front of the constant one, 
\footnote{Indeed, notice that by integrating by part the second term in the rhs of \eqref{eq: integration domain with chemical}, we have that 
\begin{align}
    \int_{u_i}^{\lambda} \int_S \left(1 - e^{- \kappa_l (u - u_i)}\right) \langle T_{uu}^l \rangle_{\bar \Psi} \text{d}u \wedge \eps_S =& \int_{u_i}^{\lambda} \int_S \langle T_{uu}^l \rangle_{\bar \Psi} \text{d}u \wedge \eps_S + \int_{u_i}^{\lambda} \int_S \kappa_l^{-1} \p_u \langle T_{uu}^l \rangle_{\bar \Psi} \text{d}u \wedge \eps_S \\ \nonumber
    &- \kappa^{-1}_l \int_S \langle T_{uu}^l \rangle_{\bar \Psi} \Big|_{\lambda} + O\left(\int_{u_i}^{\lambda} \int_S \kappa_l^{-1} \p_u \langle T_{uu}^l \rangle_{\bar \Psi} \text{d}u \wedge \eps_S\right) \, ,
\end{align}
if we take $\lambda - u_i \gg \kappa_l^{-1}$, which we can always do by taking $\lambda$ sufficiently large. Then, by subtracting the above result for the two regions $\mathcal{D}_1^\lambda$ and $\mathcal{D}_2^\lambda$, we get 
\be \label{derivapproxnew}
    \Delta \left( \int_{\mathcal{D}_i^\lambda} \left(1 - e^{- \kappa_l(u - u_i)} \right) \langle T_{uu}^l \rangle_{\bar \Psi} \right) \text{d}u \wedge \eps_S = \Delta \left( \int_{\mathcal{D}_i^\lambda} \left(\langle T_{uu}^l \rangle_{\bar \Psi} + \kappa_l^{-1} \p_u \langle T_{uu}^l \rangle_{\bar \Psi} + O\left(\kappa_l^{-1} \p_u \langle T_{uu}^l \rangle_{\bar \Psi}\right)\right) \text{d}u \wedge \eps_S \right)
\ee
and taking the limit $\lambda \rightarrow + \infty$, the lhs of \eqref{derivapproxnew} becomes $\Delta E_l$ if we assume \eqref{assumfoanyl}.} 
so that we can relate the variation of total energy in the state $\ket{\bar \Psi}$ ie $\Delta E = \sum_l \Delta E^l$ to the Bondi mass via \eqref{diffbondimass}
\be
    \Delta M = \sum_l \Delta \langle E^l \rangle_{\bar \Psi} \, .
\ee

Then, as we did in the previous subsections, since the energy variations in the vacuum state appears in \eqref{eq: deltamodhamkappal} we can use the Clausius relation to relate it to the entropy variation of the radiation in the state. Note that in the state $\ket{\Om^{U^l}_H}$, the modes with angular momentum $l$ are in thermal equilibrium at temperature $T_l = \frac{\kappa_l}{2 \pi}$. Modes of the quantum field with different angular momentum $l$ are therefore emitted from distinct reservoirs with local temperature $T_l = \frac{\kappa_l}{2 \pi}$. The Clausius relation applied on each reservoir $l$ tells us that 
\be
    \frac{\Delta E^l_{\Om^{U^l}_H}}{T_l} = \frac{\Delta E^l_{\Om^{U^l}_H}}{T_H} - \frac{\tilde{\m}_l}{T_H} \Delta E^l_{\Om^{U^l}_H} = \Delta S^l_{\Om^{U^l}_H} 
\ee
and since the vacuum state $\ket{\Om^{\{ \kappa_l \}}_H}$ is in a factorized state of the form \eqref{vallmten}, one sums over $l$ to get
\be
    \frac{\Delta M_{\Om^{\{ \kappa_l \}}_H}}{T_H} - \sum_l \frac{\tilde{\m}_l}{T_H} \Delta E^l_{\Om^{U^l}_H} = \sum_l \Delta S^l_{\Om^{U^l}_H} = \Delta S_{\Om^{\{ \kappa_l \}}_H}
\ee
so that \eqref{eq: deltamodhamkappal} becomes 
\be
    \Delta \langle K^{\mathcal{A}_i^\scri}_{\Om^{\{ \kappa_l \}}_H} \rangle_{\bar \Psi} =  \frac{\Delta M}{T_H} - \sum_{l = 0}^{+ \infty} \frac{\tilde{\m}_l}{T_H} \langle \Delta E^l \rangle_{\bar \Psi}  - \Delta S_{\Om^{\{ \kappa_l \}}_H}  \, .
\ee
Eq. \eqref{monotoagainomkappalnew} can therefore be written as 
\be \label{grandpotentialonlylnew}
    \frac{\Delta M}{T_H} - \sum_{l = 0}^{+ \infty} \frac{\tilde{\m}_l}{T_H} \langle \Delta E^l \rangle_{\bar{\Psi}} - \Delta S_{\bar{\Psi}} \leq 0
\ee
with 
\be
    \Delta S_{\bar{\Psi}} = \Delta S^{\text{v.N.}, \mathcal{A}_i^\mathcal{B}}_{\bar{\Psi} \lvert \bar{\Om}^{\{ \kappa_l \}}_H} + \Delta S_{\Om^{\{ \kappa_l \}}_H}
\ee
being the total entropy radiated away. Consequently, within the soft regularization, the free energy is no longer the adequate thermodynamic potential. Instead emerges the grand potential $\mathcal{G}_{\Omega^{\{\kappa_l\}}}$
\be \label{secondlawnewthermod}
    \Delta \mathcal{G}_{\Omega^{\{ \kappa_l \}}_H} := \Delta M - \sum_{l = 0}^{+ \infty} \tilde{\m}_l\langle \Delta E^l \rangle_{\bar \Psi}  - T_H\Delta S_{\bar \Psi} \leq 0 \, ,
\ee
which gives us information on the spontaneous evolution of our physical system. The change of thermodynamic potential, wrt the hard regularization, is tied to the fact that the vacuum state $\ket{\Om_H^{\{ \kappa_l \}}}$ is not "thermal" anymore with respect to the usual Hamiltonian. Indeed, we have now an infinite number of reservoirs at different temperatures. Having a bunch of reservoirs at hand allows in principle for work extraction from the quantum fields in the vacuum state $\ket{\Om_H^{\{ \kappa_l \}}}$. These considerations are left for the second part of this work \cite{ARBMVsoon}.

\subsubsection*{An enlightening rewriting}

Before coming to the last part of the proof in which the black hole area enters the game, we want to draw the reader's attention to the fact that we can rewrite 
\be \label{Eldef}
    E_l - \langle E_l \rangle_{\Om_H^{\{ \kappa_l \}}} = \sum_{m=-l}^{m = +l} \sum_{\om > 0} \om \left(N_{\om l m} - \langle N_{\om l m} \rangle_{\Om_H^{\{ \kappa_l \}}}\right)
\ee
since it corresponds to the contribution of the $l$-modes to the one-sided boost energy on top the vacuum distribution. Therefore, since any vector $\ket{\bar \Psi} \in \mathscr{H}_{\bar \Om_H^{\{ \kappa_l \}}}$ can be obtained through a finite excitation on top of the $\kappa_l$-vacuum, the operator in \eqref{Eldef} is well defined on $\mathscr{H}_{\bar \Om_H^{\{ \kappa_l \}}}$ (although unbounded). Alternatively, we can write \eqref{Eldef} as 
\be
    E^l - \langle E^l \rangle_{\Om_H^{\{ \kappa_l \}}} = \sum_{m=-l}^{m = +l} \int_{0}^{+ \infty} \om \bar{n}_{\om l m} d \om, \qquad \bar{n}_{\om l m} = \underset{\Delta \om \rightarrow 0}{\lim} \sum_{\om, \om + \Delta \om } \frac{N_{\om l m} - \langle N_{\om l m} \rangle_{\Om_H^{\{ \kappa_l \}}}}{\Delta \om}
\ee
where $\bar{n}_{\om l m}$ is the excess density of modes compared to the vacuum $\ket{\Om_H^{\{ \kappa_l \}}}$. Of course, we should not forget that $\om$ is here the frequency conjugated to the time \eqref{modaffinetimeu0} associated to the region $u \geq u_0$. Likewise, we can decompose the equation \eqref{vacuumvariationsnew} so that
\be
    \Delta E^l_{\Om_H^{\{ \kappa_l \}}} = -(u_2 - u_1) (2l + 1) \int_{0}^{+ \infty} \frac{1}{2 \pi} \frac{\om}{e^{\frac{2 \pi  \om}{\kappa_l}} - 1} d \om = -(u_2 - u_1) \sum_{m = -l}^{m = l} \int_0^{\infty} \om \langle N_{\om l m} \rangle_{\Om_H^{\{ \kappa_l \}}} d \om
\ee
where of course $\langle N_{\om l m} \rangle_{\Om_H^{\{ \kappa_l \}}}$ is the density of particles in the state $\ket{\Om_H^{\{ \kappa_l \}}}$ on $\scri_R^+$, given by a Bose-Einstein distribution at the temperature $T_{l} = \frac{\kappa_l}{2 \pi}$.
Now, we can set 
\be
\label{def: mu omega l}
    \m_{\om l} := \om \tilde{\m}_l = \omega \left( 1 - \frac{\kappa}{\kappa_l}\right)
\ee
so that \eqref{grandpotentialonlylnew} can now be written
\be \label{grandpotential2var}
    \frac{\Delta M}{T_H} - \sum_{l = 0}^{+ \infty} \sum_{m = -l }^{m = +l} \int_{0}^{+ \infty} \frac{\m_{\om l}}{T_H} \langle \Delta n_{\om l m} \rangle_{\bar \Psi} d\om - \Delta S_{\bar{\Psi}} \leq 0
\ee
where 
\be
     \Delta n_{\om l m} := \Delta \bar{n}_{\om l m} - (u_2 - u_1) \langle N_{\om l m} \rangle_{\Om_H^{\{ \kappa_l \}}} 
\ee
is the spectral density operator. By analogy to ordinary laws of thermodynamics, the quantity $\m_{\om l}$ is indeed a \textit{chemical potential}, and has been introduced first in \cite{ARB24}. In addition, we recover exactly the same relation as in \cite{ARB24} between the effective temperature and the chemical potential, namely \eqref{def: mu omega l}. 

\subsection{Complete proof}
\label{subsec: open horizon}

\begin{figure}[ht]
    \centering
    \includegraphics[width=0.5\textwidth]{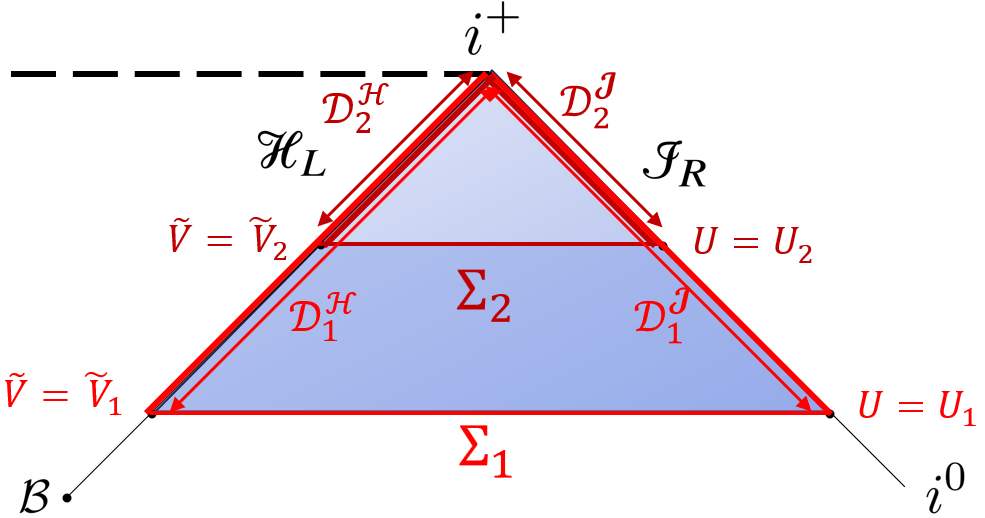}
    \caption{The two spacelike hypersurfaces $\Sigma_1$ and $\Sigma_1$  we consider in this proof, as well as the regions on the horizon and null infinity on which the algebras are defined.}
    \label{fig: setup GSL complete}
\end{figure}

Consider now that the two spacelike hypersurfaces $\Sigma_1$ and $\Sigma_2$ starts at different cuts $S_1$ and $S_2$ on the horizon $\mathcal{H}_L^+$, say, at $\tilde V = \tilde V_1$ and $\tilde V = \tilde V_2$ with $\tilde V_2 > \tilde V_1 > 0$, and ends at two different cuts $\mathcal{C}_1$ and $\mathcal{C}_2$ at $\scri_R^+$, say, at $U = U_1$ and $U = U_2$ with $U_2 > U_1 > 0$ (see Figure \ref{fig: setup GSL complete}).\footnote{Strictly speaking, in order to be spacelike, we should add to these regions the part between the bifurcation surface $\mathcal{B}$ and the cut $S_i$ together with the part between $\iota^0$ and the cut $\mathcal{C}_i$, see footnote \ref{foot: part to add spacelike}.}
Pushing the $\Sigma_i$ to infinity we observe that they correspond to the domains
\begin{equation}
    \label{eq: def domain total 2}
    \mathcal{D}_i^{\text{tot}} = \underbrace{\left( (\tilde V_i, +\infty) \times S^2_{\mathcal{H}}\right)}_{:= \mathcal{D}_i^{\mathcal{H}}} \cup \underbrace{\left( (U_i, +\infty) \times S^2_{\scri}\right)}_{:= \mathcal{D}_{i}^{\scri}} \, ,
\end{equation}
to which we attach the algebra of observables $\mathcal{A}^{\text{tot}}_i = \mathcal{A}^{\mathcal{H}}_i \otimes \mathcal{A}_i^{\scri}$. We give explicit details on how to complete our proof in this more general setup only in the case of the $L$-vacuum. We consider as reference state the tensor product state between the Hartle-Hawking state on the horizon\footnote{Recall that the radiation emitted by a black hole is \emph{exactly} thermal on the horizon.} $\ket{\Omega_{H}^{\mathcal{H}_{L}^{+}}}$ and the $L$-vacuum at future null infinity
\begin{equation}
    \label{totalvacstate}
    \ket{\Omega_L^{\text{tot}}} = \ket{\Omega_{H}^{\mathcal{H}_{L}^{+}}} \otimes \ket{\Omega_H^L} \, .
\end{equation}
Consequently the total one-sided modular Hamiltonian is the sum of contributions from the horizon and null infinity
\begin{equation}
    \label{totalmodhamil}
    K^{\text{tot}, \mathcal{A}_{i}^{\text{tot}}}_{\Omega^{\text{tot}}_{L}} := K^{\mathcal{A}_{i}^{\mathcal{H}}}_{\Omega^\mathcal{H}_H} + K^{\mathcal{A}_i^\scri}_{\Omega_H^L}
\end{equation}
with
\begin{equation}
    \label{horizonmodhamil}
    K^{\mathcal{A}_{i}^\mathcal{H}}_{\Omega^\mathcal{H}_H} = 2\pi \int_{\mathcal{D}_i^\mathcal{H}}(\tilde V-\tilde V_i):T_{\tilde V \tilde V}:_{\Omega_{H}^\mathcal{H}} \text{d}\tilde V \wedge \eps_S = 2\pi \int_{\mathcal{D}_i^\mathcal{H}}(\tilde V-\tilde V_i) T_{\tilde V \tilde V} \text{d}\tilde V \wedge \eps_S \, .
\end{equation}
From eq. \eqref{differenceareasde} i.e. from the dynamics of the horizon we relate the variation of area between $\tilde{V}_1$ and $\tilde{V}_2$ to the change of vev of modular Hamiltonian via
\begin{equation}
    \label{changeareamodhamil}
    \Delta\left(\frac{A}{4G}\right) = \Delta \langle K^{\mathcal{A}_{i}^\mathcal{H}}_{\Omega^{\mathcal{H}}_H}  \rangle_{\bar \Psi} \, .
\end{equation}
The arbitrary state $\ket{\bar \Psi} \in \mathscr{H}_{\Omega_L^{\text{tot}}}$ is taken to be such that hypotheses \eqref{assumpthor} and \eqref{conditionRl} are both satisfied, which requires to take $L$ to be large enough, but finite. Finally the monotonicity of the relative entropy between such a state $\ket{\bar \Psi}$ and our reference state \eqref{totalvacstate} straightforwardly leads to the decreasing of the \emph{generalized free energy}
\begin{equation}
    \label{decrease free energy between D_i}
    \Delta \mathcal{F}_{\text{gen}} \leq 0
\end{equation}
with 
\begin{equation}
    \label{def gen free energy}
    \mathcal{F}_{\text{gen}} = M - T_{H}\left( S_i + \frac{A}{4G} \right) \, ,
\end{equation}
where
\begin{equation}
    S_i = S^{\text{v.N.},\mathcal{A}_i^{\text{tot}}}_{\bar \Psi \lvert \Omega^{\text{tot}}_{L}} + S_{\Om^L_H}
\end{equation}
is the total entropy variation between the two open slices $\Sigma_i$.\footnote{In the Hartle-Hawking vacuum (regularized here to the $L$-vacuum with large $L$), there is no entropy flux on the horizon while there is an entropy flux at $\scri^+_R$. The discussion is similar to the energy flux: the increase in entropy due to  the outgoing radiation is exactly compensated by the entropy of the ingoing radiation falling into the black hole.} We see appearing the generalized entropy
\begin{equation}
    \label{genentropy}
    S_{i, \text{gen}} = S_i + \frac{A}{4G} \, .
\end{equation}
First established between the two regions $\mathcal{D}_i^{\text{tot}}$, we propagate the result \eqref{decrease free energy between D_i} to the arbitrary hypersurfaces $\Sigma_i$ by unitarity as usual. We have therefore shown that in an eternal Schwarzschild black hole background, the quantity which decreases between two arbitrary hypersurfaces during any physical process is, in the hard regularized Hartle-Hawking state, the generalized free energy \eqref{def gen free energy}. Within the soft regularization scheme we would have gotten the generalized version of the grand potential \eqref{secondlawnewthermod} with the chemical potentials $\mu_{\omega l}$.


\section{Outlook and perspectives}
\label{sec: conclusion}

We showed that from the point of view of asymptotic observers the spontaneous evolution of quantum fields in a perturbed Schwarzschild black hole background was equivalent to the monotonicity of a thermodynamic potential built upon the relevant geometric quantity at future null infinity, namely the Bondi mass, and eventually additional non geometric terms to which we provided a thermodynamic interpretation. Depending on the manner the Hartle-Hawking state, unsuitable for such a study due to its divergent net energy flux at $\scri_R^+$, is regularized, the dual GSL states the decrease of either the free energy or the grand potential. Non geometric corrections are encoded in a set of chemical potentials modeling the back-reaction process due to the potential barrier. When the hypersurfaces are opened both at infinity and on the horizon, geometric quantities associated to both semiclassical dynamics (area and mass) enters the, now generalized, thermodynamic potential. The monotonicity of these thermodynamic potentials has been proven in this setting within the semi-classical regimes for a large class of states representing local excitations of the field on top of the Hartle-Hawking vacuum state and/or its regularizations.

The first natural extension of our work is to consider the Unruh state, which is the adapted state to consider in order to model accurately the state of a black hole after a collapse. While the different classes of vacuum states defined on the algebra of observables at null infinity were modeling some features of the Unruh vacuum, such as the fact that modes with different angular momentum are back-scattered differently by the black hole background, none of them describes accurately the late time Hawking radiation. Indeed, the dependence in the frequency of the transmission coefficients prevents us from describing the Unruh vacuum as the natural vacuum state associated to a positive frequency decomposition attached with a time parameter at maximally extended null infinity. Instead, as proposed in \cite{ARB24}, one should study a transmission process between the white hole horizon where the radiation is exactly thermal at Hawking temperature, and future null infinity. Due to the potential barrier, not all the angular modes will be observable by an asymptotic observer, hence chemical potentials are expected to appear in the relevant thermodynamic potential, exactly as what happened for late time  boundary conditions compatible with the $\kappa_l$-vacua (see \eqref{grandpotential2var}). In \cite{ARB24}, preliminary computations showed that the latter depend precisely on the transmission coefficients. The more algebraic setup presented in this work can be used to put this proof on more solid grounds. Additionally, treating the Unruh state allows to deal with rotating black holes and therefore to study the dual GSL also in that context.

A second extension is based on several observations we have made along this work. On the one hand, as stated in the Introduction, finding the free energy or the grand potential in the dual GSL shows that, from the point of view of usual thermodynamics, there exists a deeper connection between quantization of a field on a non-expanding null hypersurface and the theory of of open quantum systems. One further hint towards that direction was the assumption \eqref{assumpthor}, comparing a dynamical time $\tau_D = \frac{\langle T_{uu}\rangle_\Psi}{\partial_u \langle T_{uu}\rangle_\Psi}$ related to the characteristic variation of the stress energy tensor, to what can be interpreted as a correlation time $\tau_B = \kappa^{-1} \sim \frac{\hbar}{T_H}$. In the theory of open quantum systems weakly interacting with a thermal bath, the typical scale at which one can rely on Markovian dynamics is tied to the inverse temperature \cite{breuer2002theory}.\footnote{To be precise, the Markovian approximation also depends on additional timescales set by the bath spectral density, including possible ultraviolet cutoffs. However, for thermal equilibrium baths, the inverse temperature defines a universal thermal correlation timescale.} This suggests a close interplay between the relaxation of quantum fields toward a vacuum state at null infinity (e.g., the Hartle–Hawking state) and the equilibration of a small system, regarded as a subpart of a larger bath, toward a (generalized) Gibbs state. It would be interesting to unveil this analogy between black hole physics and Markovian systems, explaining in particular why, while being a very strong hypothesis, Markovianity is immediately achieved once we place ourself on a null hypersurface. Lastly, the correlation with thermodynamics of open quantum systems can be push further by going back to the expression \eqref{grandpotential2var}. The non-geometric terms depending on the chemical potential are related to work terms. This seems to tell us that, once correctly described by an appropriate quantum state at infinity,\footnote{By this we mean not the Hartle-Hawking state.} work could in principle be extracted from the non-thermal radiation emitted by a non-rotating black hole. 

We leave the treatment of the Unruh state, the connection with the thermodynamics of open systems and the construction of a machine which can extract work from the radiation emitted by a non-rotating black hole for the companion paper \cite{ARBMVsoon}.


\section*{Acknowledgments}
ARB would like to thank Mohamed Boubakour, Luca Ciambelli, Cyril Elouard, Anthony Speranza, Simone Speziale and Aron Wall for stimulating discussions on this topic. MV thanks Cyril Elouard for discussions during his visit in Nancy. ARB is also grateful to the Julian Schwinger Foundation for financial support at the 2025 Peyresq Spacetime Meeting, where useful discussions influenced this work. The work of ARB is funded by the European Union. Views and opinions expressed are however those of the author(s) only and do not necessarily reflect those of the European Union or the European Research Council. Neither the European Union nor the granting authority can be held responsible for them. This work is supported by ERC grant QARNOT, project number 101163469. The work of MV is supported by the Fonds de la Recherche Scientifique -- FNRS under the Grant No. T.0047.24. We thank each other institutions for hosting and financial support during visits at the last stages of this work.

\appendix


\section{Unruh effect on a non-expanding null hypersurface}
\label{app: unruh}

In this Appendix we present a comprehensive proof of the Unruh effect from a direct quantization of a free massless field on a non-expanding null hypersurface, completing and making explicit some details of standard proofs e.g. \cite{Unruh:1976db, Bisognano:1975ih, kay1991theorems, sewell1982quantum, hislop1982modular, wald1994quantum}. This effect was the key point in Section \ref{sec: GSL proof} to ensure that the vacuum state $\ket{\Omega}$, defined through a positive frequency decomposition induced by a choice of time, was indeed thermal when restricted to the subalgebras of observables we considered.

Let $\mathcal{N}$ be a non-expanding null hypersurface parametrized by a coordinate $U \in (-\infty, +\infty)$ (e.g. $\scri_R$). Its normal vector is therefore $\partial_U$. On this background we quantize a massless scalar field.\footnote{Our proof extends immediately to fermionic fields, by simply changing the commutation relations of the ladder operators into their fermionic counterpart.} The name of the game being to restrict the projected  bulk vacuum state to a subregion of $\mathcal{N}$, we start by analyzing a restriction to $U > 0$. Our proof is divided into two parts. First we show that such a restriction generates a Bose-Einstein distribution\footnote{Or Fermi-Dirac for fermionic fields.} of particles for the average value of the number operator. Then, we explicitly show that the projected vacuum state on $\mathcal{N}$ can be written as a thermal entanglement between the eigenmodes of the field in the two regions, above and below the cut $U=0$. Note that in general, especially when using the Unruh effect in proofs of the GSL the vacuum is invariant (in particular) under translations on $\mathcal{N}$, see Section \ref{sec: quantization}, making the precise locus of the cut in $U$ irrelevant. Therefore, we take $U=0$ for simplicity.

We give all the details of the proof in the two-dimensional case while we only display the end result for the four-dimensional extension.

\subsubsection*{The global vacuum state}

Consider a Klein-Gordon field $\hat{\phi}(U)$ on $\mathcal{N}$, which we canonically quantize using the (correctly normalized)\footnote{With respect to the scalar product \eqref{symplecticformsu}.} modes $\frac{e^{-i\Omega U}}{\sqrt{4\pi \Omega}}$ with $\Omega > 0$. This gives
\begin{equation}
    \label{eq: scalar quantized U}
    \hat{\phi}(U) = \int_{0}^{+\infty}\, \frac{\text{d}\Omega}{\sqrt{4\pi \Omega}} \,\left(\hat{a}_{\Omega}e^{-i \Omega U} + \hat{a}_{\Omega}^{\dagger}e^{+i\Omega U} \right) \;
\end{equation}
which, together with its canonical momenta $\hat{\pi}(U) \equiv \partial_{U}\hat{\phi}$, satisfies the canonical equal-time commutation relations
\begin{equation}
    \label{eq: cano commut relations}
    [\hat{\phi}(U), \hat{\pi}(U')] = \frac{i}{2}\delta(U - U') \, ,
\end{equation}
given that
\begin{equation}
    \label{eq: Omega commut crea op}
    \left[\hat{a}_{\Omega}, \hat{a}_{\Omega'}^{\dagger}\right] = \delta(\Omega-\Omega') \, .
\end{equation}
From this QFT construction we get the one-particle Hilbert space associated with the coordinate $U$ whose the vacuum state $\ket{\Omega^U}$ is the state such that for all frequencies $\Omega > 0$ we have $\hat{a}_{\Omega}\ket{\Omega^U} = 0$.

\subsubsection*{Quantization in the region $U > 0$}

We restrict our attention to the upper half of $\mathcal{N}$, namely the region $U > 0$ that we shall denote region $I$. There, $U$ is no longer a good coordinate for quantization as it does not range in the whole set of real numbers. We consider instead the usual change of coordinates
\begin{equation}
    \label{eq log relation}
    u = \ln U
\end{equation}
which defines $u \in (-\infty, + \infty)$ as we go along region $I$. Our objective is now to decompose the restriction of $\hat{\phi}$ in region $I$, denoted $\hat{\phi}_{+}$, but into positive-frequency modes for the coordinate $u$ i.e. in the basis $\frac{e^{-i \omega u}}{\sqrt{4\pi \omega}}$. This leads to
\begin{equation}
    \label{eq: quantif scalar half space}
    \hat{\phi}_{+}(u) = \int^{+\infty}_{0}\, \frac{\text{d}\omega}{\sqrt{4\pi \omega}} \left(\hat{a}_{\omega, I}e^{-i \omega u} + \hat{a}_{\omega, I}^{\dagger}e^{+i\omega u} \right) \, .
\end{equation}
This field should coincide with the restriction of \eqref{eq: scalar quantized U} to region $I$; therefore the positive modes $\Omega$ should be decomposed in the basis of modes $\omega$. Hence, as nothing prevent negative $\omega$ modes to appear, we need the following relations to hold when $U > 0$
\begin{align}
    \label{eq: big A et a}
    \hat{A}_I &= \int_{0}^{+\infty} \frac{\text{d}\Omega}{\sqrt{4\pi \Omega}}\hat{a}_{\Omega}e^{-i\Omega U} \equiv \int_{-\infty}^{+\infty}\frac{\text{d}\omega}{\sqrt{4\pi |\omega|}}\tilde{a}_{\omega, I}e^{-i\omega u} \, ,
    \\
    \label{eq: big A et a dag}
    \hat{A}^{\dagger}_I &= \int_{0}^{+\infty} \frac{\text{d}\Omega}{\sqrt{4\pi \Omega}}\hat{a}_{\Omega}^{\dagger}e^{i\Omega U} \equiv \int_{-\infty}^{+\infty}\frac{\text{d}\omega}{\sqrt{4\pi |\omega|}}\tilde{a}_{\omega, I}^{\dagger}e^{i\omega u} \, ,
\end{align}
so that of course $\hat{\phi}_+ = \hat{A} + \hat{A}^{\dagger}$ and the $\tilde{a}_{\omega}$ are implicitly defined by 
\begin{equation}
    \label{eq: aux ladder from KG}
    \tilde{a}_{\omega, I} = \left(\hat{A}, \frac{e^{-i \omega u}}{\sqrt{4 \pi |\omega|}} \right)_{KG} \quad \text{and} \quad \tilde{a}_{\omega, I}^{\dagger} = - \left(\hat{A}^{\dagger}, \frac{e^{i \omega u}}{\sqrt{4 \pi |\omega|}} \right)_{KG}
\end{equation}
where we recall for completeness the Klein-Gordon product in region $I$
\begin{equation}
    \label{eq: KG product on N}
    (f,g)_{KG} = i \int_{-\infty}^{+\infty} \left(\bar{f}(u) \partial_{u}g(u) - g(u)\partial_{u}\bar{f}(u) \right) \text{d}u
\end{equation}
where the bar denotes complex conjugation. In \eqref{eq: big A et a} and \eqref{eq: big A et a dag}, due to the integral over all real frequencies, the operators $\tilde{a}_{\omega,I}$ are not the ladder operators of the field in coordinate $u$. They are just convenient variables for our purposes and the actual ladder operators will turn out to be linear combinations of the $\hat{a}_\Omega$ and $\hat{a}_\Omega^\dag$. We start by considering $\hat{A}_ I$, from \eqref{eq: big A et a} and \eqref{eq: aux ladder from KG} we get
\begin{equation}
    \label{eq: A in terms of F}
    \hat{A}_I(u) = \int_{-\infty}^{+\infty}\int_{0}^{+\infty} \frac{\text{d}\omega}{\sqrt{4\pi |\omega|}} \, \frac{\text{d}\Omega}{2\pi} \hat{a}_{\Omega}\sqrt{\frac{\omega}{\Omega}} F(\Omega, \omega)e^{-i\omega u} \,
\end{equation}
where the function $F(\Omega, \omega)$ reads for positive $\Omega$
\begin{equation}
    \label{eq: def F}
    F(\Omega, \omega) = \int_{-\infty}^{+\infty} e^{i\Omega U}e^{-i\omega u} \text{d}u = \int_{0}^{+\infty} e^{i\Omega U} U^{-i\omega - 1} \text{d}U \, .
\end{equation}
Similarly we get for the Hermitian conjugate $\hat{A}^{\dagger}$
\begin{equation}
    \label{eq: A dag as F bar}
    \hat{A}^{\dagger}_I(u) = \int_{-\infty}^{+\infty}\int_{0}^{+\infty} \frac{\text{d}\omega}{\sqrt{4\pi |\omega|}} \, \frac{\text{d}\Omega}{2\pi}\hat{a}_{\Omega}^{\dagger}\sqrt{\frac{\omega}{\Omega}} \bar{F}(\Omega, \omega)e^{i\omega u} \,
\end{equation}
where appears $\bar{F}$, the complex conjugate of $F$. It turns out that the following relation holds
\begin{equation}
    \label{eq: relation F bar F}
    F(\Omega, \omega) = F(-\Omega, \omega) e^{\pi \omega} \, .
\end{equation}
To show \eqref{eq: relation F bar F}, we can complexify the coordinate $U$ and integrate on the upper half complex plane where $\mathfrak{Im}(U) > 0$ between the half-circles of radius $R \gg 1$ and $r \ll 1$. In particular, if we set $U' = -U$, we have
\begin{align}
    \label{eq: change var in F}
    F(\Omega, \omega) &= \lim_{r\to0 ; R \to +\infty} \int_{r}^{R}\text{d}U e^{i\Omega U} U^{-i\omega - 1} \\
    \nonumber
    &= e^{\pi \omega} \lim_{r\to0 ; R \to +\infty} \int_{-r}^{-R}\text{d}U e^{-i\Omega U} U^{-i\omega - 1} \, ,
\end{align}
where we have used $-1 = e^{+i\pi}$.\footnote{We have to write $-1 = e^{+i \pi}$ because we have to take the upper half circle of radius $r \ll 1$ in order to have no residue. Therefore, the cut of the logarithm must be chosen in the upper half plane and so we have $-1 = e^{+i \pi}$ and not $-1 = e^{-i \pi}$.} The integral on the circle of radius $R \to + \infty$ vanishes because of the real exponential and the condition $\Omega > 0$ while the integral on the circle $r \to 0$ vanishes since the contribution in the integrand is proportional to $r^{i \omega}$ that vanishes in that limit.\footnote{When $r \rightarrow 0$ we can take $\omega' = \omega' - i \epsilon$, $\epsilon > 0$ and then taking the limit $\epsilon \to 0$ we get the advertised result.} Therefore we deduce \eqref{eq: relation F bar F} from \eqref{eq: change var in F}. Similarly for the complex conjugate we get
\begin{equation}
    \label{eq: relation bar F}
    \bar{F}(\Omega, \omega) := F(-\Omega, - \omega) = F(\Omega, -\omega) e^{\pi \omega} \, .
\end{equation}
Now starting from the decomposition $\hat{\phi}_+ =\hat{A} + \hat{A}^{\dagger}$, using the expressions \eqref{eq: A in terms of F} and \eqref{eq: A dag as F bar} and the relations \eqref{eq: relation F bar F} and \eqref{eq: relation bar F} we can decompose the field only in terms of positive frequencies in $\omega$
\begin{equation}
    \label{eq: phi in omega modes}
    \hat{\phi}_+(u) = \int_{\omega = 0}^{+\infty}\frac{\text{d}\omega}{\sqrt{4\pi \omega}} \left(\hat{a}_{\omega, I}e^{-i\omega u} + \hat{a}^{\dagger}_{\omega, I}e^{i\omega u} \right) \, ,
\end{equation}
where we the ``true'' ladder operators for $\omega > 0$ read
\begin{align}
    \label{eq: omega annihilation}
    \hat{a}_{\omega, I} &\equiv \frac{1}{2\pi}\int_{\Omega = 0}^{+\infty} \sqrt{\frac{\omega}{\Omega}} \text{d}\Omega \, F(\Omega, \omega) \left(\hat{a}_{\Omega} + e^{-\pi \omega}\hat{a}^{\dagger}_{\Omega}\right) \\
    \label{eq: omega creation}
    \hat{a}^{\dagger}_{\omega, I} &\equiv \frac{1}{2\pi}\int_{\Omega = 0}^{+\infty} \sqrt{\frac{\omega}{\Omega}} \text{d}\Omega \, \bar{F}(\Omega, \omega) \left(\hat{a}^{\dagger}_{\Omega} + e^{-\pi \omega}\hat{a}_{\Omega}\right) \, .
\end{align}
This concludes our objective of expressing our field in positive frequencies of the accelerated observer at coordinate $u$.\footnote{Note \emph{en passant} that the vacuum state in region $I$ is the state $\ket{\Omega^u}$ annihilated by all positive-modes $\hat{a}_{\omega}$ for $\omega > 0$.}

\subsubsection*{Commutation relations and number operator}

Being genuine ladder operators, the commutators between \eqref{eq: omega annihilation} and \eqref{eq: omega creation} match the canonical ones. We indeed find
\begin{equation}
    \label{eq: zero commut a omega}
    \left[\hat{a}_{\omega, I}, \hat{a}_{\omega', I} \right] = \left[\hat{a}_{\omega, I}^{\dagger}, \hat{a}_{\omega', I}^{\dagger} \right] = 0 \, ,
\end{equation}
together with
\begin{equation}
    \label{eq: non zero commut}
    \left[\hat{a}_{\omega, I}, \hat{a}_{\omega', I}^{\dagger} \right] = \delta(\omega - \omega') \, .
\end{equation}
Using the definitions \eqref{eq: omega annihilation} and \eqref{eq: omega creation} together with the commutators \eqref{eq: Omega commut crea op} we get
\begin{align}
    \label{eq: commut non zero omega}
    \left[\hat{a}_{\omega, I}, \hat{a}_{\omega', I}^{\dagger} \right] &= \frac{1}{4\pi^2}\int_0^{+\infty} \text{d}\Omega \int_{0}^{+\infty} \text{d}\Omega' \sqrt{\frac{\omega \omega'}{\Omega \Omega'}} F(\Omega, \omega) \bar{F}(\Omega, \omega) \left( \left[\hat{a}_{\Omega}, \hat{a}^{\dagger}_{\Omega'}\right] - e^{-\pi(\omega - \omega')} \left[\hat{a}_{\Omega'}, \hat{a}^{\dagger}_{\Omega} \right] \right) \\
    \nonumber
    &= \frac{\sqrt{\omega \omega'}}{4\pi^2} \int_{0}^{+\infty} \frac{\text{d}\Omega}{\Omega} F(\Omega, \omega) \bar{F}(\Omega, \omega') \left(1 - e^{-\pi(\omega + \omega')} \right) \\
    \nonumber
    &= \frac{\sqrt{\omega \omega'}}{4\pi^2}\left(1 - e^{-\pi(\omega + \omega')} \right) \int_{0}^{+\infty} \frac{\text{d}\Omega}{\Omega} \left[\int_{0}^{+\infty} \text{d}U e^{i\Omega U} U^{-i\omega -1}\right]\left[\int_{0}^{+\infty} \text{d}V e^{-i\Omega V} V^{i\tilde \omega -1}\right]
\end{align}
where to get the second line we used the definition of $F$ in \eqref{eq: def F} and of its complex conjugate. Applying successively the changes of variables $(U,V) = (e^u, e^v)$ and $(u,v) = (x+y, x-y)$ we find
\begin{equation}
    \label{eq: commut as function x and y}
    \left[\hat{a}_{\omega, I}, \hat{a}_{\omega', I}^{\dagger} \right] = \frac{\sqrt{\omega \omega'}}{4\pi^2}\left(1 - e^{-\pi(\omega + \omega')} \right) \int_{0}^{+\infty} \frac{\text{d}\Omega}{\Omega} \int_{-\infty}^{+\infty}\text{d}x \int_{-\infty}^{+\infty}\text{d}y \, e^{-i\Omega e^{x}(e^{y}-e^{-y})} e^{i(\omega -\omega')x} e^{i(\omega + \omega')y}
\end{equation}
and we observe that the integral 
\begin{equation}
    \label{eq: I integral}
    I(x, \omega, \omega') := \int_{0}^{+\infty} \frac{\text{d}\Omega}{\Omega}\int_{-\infty}^{+\infty}\text{d}y \, e^{-i\Omega e^{x}(e^{y}-e^{-y})} e^{i(\omega + \omega')y}
\end{equation}
is independent of $x$.\footnote{One has to set $\tilde{\Omega} = \Omega e^x$.} Therefore it follows that
\begin{align}
    \label{eq: commut ladder step 4}
    \left[\hat{a}_{\omega, I}, \hat{a}_{\omega', I}^{\dagger}\right] &= \frac{\sqrt{\omega \omega'}}{4\pi^2}\left(1 - e^{-\pi(\omega + \omega')} \right) I(\omega, \omega') \int_{-\infty}^{+\infty}\text{d}x e^{i(\omega - \omega')x} \\
    \nonumber
    &= \frac{\omega}{2\pi}\left(1 - e^{-\pi(\omega + \omega')} \right) I(\omega, \omega) \delta(\omega - \omega') \, .
\end{align}
Ultralocality in $\omega$ is therefore expected with the appearance of the delta function. Using \eqref{eq: commut ladder step 4} in a previous step of the computation, namely the first line of \eqref{eq: commut non zero omega} gives
\begin{equation}
    \label{eq: commut ladder step 5}
    \left[\hat{a}_{\omega, I}, \hat{a}_{\omega', I}^{\dagger}\right] = \frac{\omega}{4\pi^2} \left(1 - e^{-2\pi\omega} \right) \int_{0}^{+\infty} \frac{\text{d}\Omega}{\Omega} |F(\Omega, \omega)|^2 \delta(\omega - \omega') \, = \delta(\omega - \omega') \, ,
\end{equation}
in virtue of \eqref{eq: non zero commut}. Therefore we find the relation
\begin{equation}
    \label{eq: bose einstein 1}
    \frac{1}{4\pi^2}\int_{0}^{+\infty}\frac{\text{d}\Omega}{\Omega}\omega |F(\Omega, \omega)|^2 = \frac{1}{1-e^{-2\pi \omega}} \, .
\end{equation}
We emphasize that this result crucially depends on the $\frac{1}{2}$ factor appearing in \eqref{eq: cano commut relations}, therefore on the fact that we are dealing with a null hypersurface. We are now equipped to compute the number of particles detected by an accelerated observer (with coordinate $u$) at frequency $\omega$, while the vacuum state was described by the coordinate $U$ and modes $\Omega$. The vev of the number operator $\hat{N}_{\omega, I}$ in the vacuum $\ket{\Omega^U}$ is
\begin{align}
\label{eq: number operator}
    N_{\omega, I} &= \bra{\Omega^U} \hat{N}_{\omega, I} \ket{\Omega^U} = \bra{\Omega^U} \hat{a}^\dagger_{\omega, I} \hat{a}_{\omega, I} \ket{\Omega^U} = \frac{1}{4 \pi^2} \int_0^{+ \infty} \frac{\text{d}\Omega}{\Omega} \omega \lvert F(\Omega, \omega) \rvert^2 e^{- 2 \pi \omega} \nonumber \\
    &= \frac{1}{e^{2 \pi \omega} - 1}
\end{align}
which gives, as expected, a Bose-Einstein distribution. We recovered the first step towards the Unruh effect but in the more general set-up of an arbitrary null hypersurface. Note that the assumption of non-expansion is useless here, as a non-vanishing expansion only changes the volume form of the space of generators, which does not exist when the bulk is two-dimensional.

However, it is not enough to compute the average number of particles at a given frequency to find the exact expression of the vacuum. The latter will be thermal if one can show a thermal entanglement between the region $U > 0$ and the region $U < 0$. We therefore turn our attention to this last point.

\subsubsection*{Thermal entanglement of the vacuum state}

The quintessence of the Unruh effect comes from the fact that we cut the spacetime in two regions and observe the vacuum state from only one of the two, by tracing out the degrees of freedom of the other region. In order to get a thermal state upon tracing, the density matrix of the vacuum state $\ket{\Omega^U}$ should be written as a sum of entangled states between the two regions, with exponential coefficients.\footnote{This is equivalent in the well-known Rindler framework to a decomposition of the vacuum density matrix in the Rindler basis.} This is what we shall show in this paragraph. 

The region $U > 0$ has been taken into account in the last paragraph so we should now focus on the complementary region $U < 0$. Quantities related to the region $U < 0$ will be denoted with a subscript $II$. Again to get a good coordinate for quantization we can choose
\begin{equation}
    \label{eq: def coord v}
    u_- = \ln(-U) \, .
\end{equation}
Even though it ranges in $(-\infty, +\infty)$ as it should, one should remark that the cut $U = 0$ is located at $u_- = -\infty$, therefore $u_-$ is past-directed and so the positive frequency modes for $u_-$ (of frequency $\tilde \omega > 0$) are written as $\frac{e^{i\tilde\omega u_-}}{\sqrt{4\pi \tilde \omega}}$ with a plus sign. The counterpart of the relations \eqref{eq: big A et a} and \eqref{eq: big A et a dag} are
\begin{align}
    \label{eq: big A et a neg case}
    \hat{A}_{II} &= \int_{0}^{+\infty} \frac{\text{d}\Omega}{\sqrt{4\pi \Omega}}\hat{a}_{\Omega}e^{-i\Omega U} \equiv \int_{-\infty}^{+\infty}\frac{\text{d}\tilde \omega}{\sqrt{4\pi \tilde{ |\omega|}}}\tilde{a}_{\tilde \omega, II}e^{i\tilde \omega u_-} \, ,
    \\
    \label{eq: big A et a dag neg case}
    \hat{A}^{\dagger}_{II} &= \int_{0}^{+\infty} \frac{\text{d}\Omega}{\sqrt{4\pi \Omega}}\hat{a}_{\Omega}^{\dagger}e^{i\Omega U} \equiv \int_{-\infty}^{+\infty}\frac{\text{d}\tilde\omega}{\sqrt{4\pi \tilde{|\omega|}}}\tilde{a}_{\tilde \omega, II}^{\dagger}e^{-i\tilde \omega u_-} \, ,
\end{align}
and this defines the ``convenient'' operators we use in the region $II$
\begin{equation}
    \label{eq: aux ladder from KG neg space}
    \tilde{a}_{\tilde \omega, II} = - \left(\hat{A}_{II}, \frac{e^{i \tilde \omega u_-}}{\sqrt{4 \pi \tilde{|\omega|}}} \right)_{KG} \quad \text{and} \quad \tilde{a}_{\tilde \omega, II}^{\dagger} =  \left(\hat{A}^{\dagger}_{II}, \frac{e^{-i \tilde \omega u_-}}{\sqrt{4 \pi \tilde{|\omega|}}} \right)_{KG} \, .
\end{equation}
The equivalent of $F$ in \eqref{eq: A in terms of F} is here given by $G$ such that
\begin{equation}
    \label{eq: def G}
    G(\Omega, \tilde \omega) = \int_{-\infty}^{+\infty}e^{i\Omega U}e^{i\tilde \omega u_-} \text{d}u_- = \int_{0}^{+\infty} e^{-i\Omega U} U^{i\tilde \omega - 1} \text{d}U = F(-\Omega, -\tilde \omega) = \bar{F}(\Omega,\tilde \omega) \, ,
\end{equation}
where we have used \eqref{eq: def F} and \eqref{eq: relation bar F}. We also have the following relation on the complex conjugate of $G$
\begin{equation}
    \label{eq: relation bar G et F}
    \bar{G}(\Omega, \omega) = G(-\Omega, -\omega) = F(\Omega, \omega)
\end{equation}
using \eqref{eq: def G}. From \eqref{eq: def G} and \eqref{eq: relation bar G et F} we can deduce the expression of the genuine ladder operators for the coordinate $u_-$. Calling them $\hat{a}_{\omega, II}$ and $\hat{a}^{\dagger}_{\omega, II}$ we find
\begin{align}
    \label{eq: negative annihilation}
    \hat{a}_{\omega, II} &= \frac{1}{2\pi} \int_{0}^{+\infty} \sqrt{\frac{\omega}{\Omega}}\text{d}\Omega \bar{F}(\Omega, \omega) \left(\hat{a}_{\Omega} + e^{-\pi \omega}\hat{a}_{\Omega}^{\dagger}\right) \\
    \label{eq: negative creation}
    \hat{a}_{\omega, II}^{\dagger} &= \frac{1}{2\pi} \int_{0}^{+\infty} \sqrt{\frac{\omega}{\Omega}}\text{d}\Omega F(\Omega, \omega) \left(\hat{a}_{\Omega}^{\dagger} + e^{-\pi  \omega}\hat{a}_{\Omega}\right) \, .
\end{align}
These operators also satisfy the canonical commutation relations \eqref{eq: zero commut a omega} and \eqref{eq: non zero commut}, but we should emphasize the fact that the couple $\left(\hat{a}_{\omega, II}, \hat{a}_{\omega, II}^{\dagger}\right)$ acts on the region $U < 0$ while $\left(\hat{a}_{\omega, I}, \hat{a}_{\omega, I}^{\dagger}\right)$ was acting on the region $U > 0$. The two couples of operators do not act on the same Hilbert space and therefore they necessarily commute with each other.

Recalling that the global vacuum state on $\mathcal{N}$ is by definition annihilated by all $\hat{a}_{\Omega}$ for positive $\Omega$, we deduce from \eqref{eq: omega annihilation}, \eqref{eq: omega creation}, \eqref{eq: negative annihilation} and \eqref{eq: negative creation} that the following combination of the restricted operators also annihilate $\ket{\Omega^U}$
\begin{align}
    \label{eq: first anihil two regions}
    \left(\hat{a}_{\omega, II} - e^{-\pi \omega}\hat{a}^{\dagger}_{\omega, I} \right) \ket{\Omega^U} &= 0 \\
    \label{eq: second anihil two regions}
    \left(\hat{a}_{\omega, I} - e^{-\pi \omega}\hat{a}^{\dagger}_{\omega, II} \right) \ket{\Omega^U} &= 0 \, .
\end{align}
These relations imply that the number operators $\hat{N}_{\omega, I}$ and $\hat{N}_{\omega, II}$ in each of the two regions are necessarily identical since one can deduce from \eqref{eq: first anihil two regions} and \eqref{eq: second anihil two regions}
\begin{equation}
    \label{eq: equality number operators}
    \hat{a}_{\omega, II}^{\dagger} \hat{a}_{\omega, II} \ket{\Omega^U} = \hat{a}_{\omega, I}^{\dagger} \hat{a}_{\omega, I} \ket{\Omega^U} \,
\end{equation}
meaning that one can choose a basis of the total Hilbert space $\mathscr{H}_{\Omega}$ made of vectors representing $N$-particle states in the two regions, i.e. states of the form $\ket{N_{\omega, I}, N_{\omega, II}} = \ket{N_{\omega, I}} \otimes \ket{N_{\omega, II}}$ where as we said $N_{\omega, I} = N_{\omega, II}$. In this basis the vacuum state can be written as 
\begin{equation}
    \label{eq: N vac in double basis}
    \ket{\Omega^U} = \prod_{\omega = 0}^{+\infty} Z_\omega \sum_{N_{\omega, I} = N_{\omega, II}} A_{N_{\omega}}(\omega) \ket{N_{\omega, I}} \otimes \ket{N_{\omega, II}}
\end{equation}
with $A_{N_{\omega}}(\omega)$ coefficients yet to be determined and $Z_\omega$ a ($\omega$-dependent) normalization factor. The thermal character (or not) of the restricted vacuum state is all contained in these coefficients. Acting on \eqref{eq: N vac in double basis} with a particular frequency $\omega$ we get a recursion relation for the $A_{N_{\omega}}$
\begin{equation}
    \label{eq: recursion}
    A_{N_\omega} = e^{- \pi \omega} A_{N_\omega - 1}
\end{equation}
from which we deduce $A_{N_\omega} = e^{- \pi N_\omega \omega} A_0$ by iteration. Knowing $A_{N_{\omega}}$ we get from the requirement that $\braket{0|0}_U = 1$ the expression of the normalization factor ($A_0 = 1$ necessarily)
\begin{equation}
    \label{eq: normalization 2D}
    Z_\omega = \sqrt{1 - e^{-2\pi \omega}} \, .
\end{equation}
Hence the density matrix of the $\mathcal{N}$ vacuum reads
\begin{equation}
    \label{eq: vac density matrix}
    \ket{\Omega^U}\bra{\Omega^U} = \prod_{\omega = 0}^{+\infty} \left(1 - e^{-2\pi \omega}\right) \sum_{N_{\omega, I} = N_{\omega, II}} e^{-\pi \omega N_{\omega}}\ket{N_{\omega, I}} \otimes \ket{N_{\omega, II}}\bra{N_{\omega, I}}\otimes \bra{N_{\omega, II}} \, ,
\end{equation}
displaying a thermal entanglement between the two sub-regions. Any partial tracing over one of the two Hilbert spaces, i.e. any restriction of the vacuum state to one of the two sub-regions, will lead to a thermal state. This is the Unruh effect in two dimensions.

\subsubsection*{Four-dimensional proof}

We complement the coordinate $U$ by two angular coordinates collectively denoted by $x^A$ with $A = 1,2$ and expand the field as
\begin{equation}
    \label{eq: phi in modes and harmonics}
    \hat{\phi}(U, x^A) = \sum_{l = 0}^{+\infty} \sum_{m = -l}^{+l} \int_{0}^{+\infty} \frac{\text{d}\Omega}{\sqrt{4\pi \Omega}}\left(\hat{a}_{\Omega l m} Y^{l}_{m}(x^{A}) e^{-i\Omega U} + \hat{a}_{\Omega l m}^{\dagger}\bar{Y}^{l}_{m}(x^{A})e^{+i\Omega U} \right) \, ,
\end{equation}
where the modes are correctly normalized as indeed
\begin{equation}
    \label{eq: 4D normalization}
    \left(Y^{l}_{m}(x^A) \frac{e^{-i\Omega U}}{\sqrt{4\pi \Omega}} , Y^{l'}_{m'}(x^A) \frac{e^{-i\Omega' U}}{\sqrt{4\pi \Omega'}} \right)_{KG} = \delta_{ll'}\delta_{mm'}\delta(\Omega - \Omega') \, ,
\end{equation}
with the Klein-Gordon scalar product which reads now
\begin{equation}
    \label{eq: 4D KG product}
    \left(f(U,x^A), g(U,x^A)\right)_{KG} = i\int_{-\infty}^{+\infty}  \int_{S^2}\text{d}U \w \eps_S \left( \partial_{U}\bar{f} g - \partial_{U}g \bar{f} \right) \, .
\end{equation}
We see appearing the volume form of the two-dimensional spatial sections of $\mathcal{N}$, $\eps_S$, which we take to be metrically a unit $2$-sphere $S^2$.\footnote{This choice is convenient as it allows to decompose the field in the well-known basis of spherical harmonics. Actually it is only necessary to ask the space of generators to be topologically a $2$-sphere. Indeed, if the space of generators is not a metric two sphere, one just shifts $Y_m^l \rightarrow \frac{Y_m^l}{\sqrt{q}}$ where $q$ is the conformal factor so that $\eps_S = q \eps^0_S$ with $\eps_S^0$ being the volume form of a unit metric two sphere. Now new modes are normalized and \eqref{eq: 4D normalization} is satisfied. This small modification allows one to quantize on a Kerr black hole horizon or at $\scri^+$ in a non-Bondi frame using the framework presented here.} The latter is meant to change when the null hypersurface at hand possesses a non zero expansion. Consequently, nothing guarantees the normalization condition \eqref{eq: 4D normalization} to be valid everywhere on $\mathcal{N}$. As our proof requires the mode decomposition \eqref{eq: phi in modes and harmonics} to be valid everywhere, we are led to assume that $\mathcal{N}$ is a non-expanding null-hypersurface.

After restricting the field to region $I$ and to region $II$, we get the combinations of operators of region $I$ and of region $II$ that annihilate the vacuum
\begin{align}
    \label{eq: 4d first anihil two regions}
    \left(\hat{a}_{\omega l m, II} - (-1)^{m} e^{-\pi \omega}\hat{a}^{\dagger}_{\omega l -m, I} \right) \ket{\Omega^U} &= 0 \\
    \label{eq: 4d second anihil two regions}
    \left(\hat{a}_{\omega l m, I} - (-1)^{m} e^{-\pi \omega}\hat{a}^{\dagger}_{\omega l -m, II} \right) \ket{\Omega^U} &= 0 \, .
\end{align}
These two equations leads, using similar techniques than in the two-dimensional case, to an expression of the $\mathcal{N}$-vacuum density matrix
\begin{equation}
    \label{eq: 4d vac density matrix}
    \ket{\Omega^U}\bra{\Omega^U} = \prod_{\omega lm} \left(1 - e^{-2\pi \omega}\right)\sum_{N_{\omega l m, I} = N_{\omega l -m, II}} e^{-\pi \omega N_{\omega}}\ket{N_{\omega l m, I}} \otimes \ket{N_{\omega l -m, II}}\bra{N_{\omega l m, I}} \otimes \bra{N_{\omega l -m, II}} \, ,
\end{equation}
betraying the appearance of a thermal state upon partial tracing. All these computations remain valid upon replacing $U$ by $U^{(l,m)}$ defined in \eqref{ulmsoftdef}. This concludes our proof of the Unruh effect on any arbitrary non-expanding null hypersurface. We shed the light once again on the importance of both assumptions, while being on a null-hypersurface is used to prove the Bose-Einstein distribution of the number of particles in both sub-regions, the non-expanding character of $\mathcal{N}$ is essential in four-dimensions to be able to decompose the field in a basis of (everywhere) correctly normalized modes.


\section{The spacetime approach to QFT in curved spacetimes}
\label{app: QFT curved}

\subsection{Generalities}
\label{appsub: generalities spacetime approach}

In Minkowski spacetime, Klein-Gordon fields form an infinite collection of harmonic oscillators, some with positive frequency while others with negative frequency. The classical field, i.e. the solution of the Klein-Gordon equation, is expanded in a basis of positive frequency modes (e.g. $~e^{-i\omega t}, \omega > 0$ with $t$ the Poincaré global time) and the Fourier coefficients $a_\omega, a^\ast_\omega$ are promoted to quantum operators satisfying the canonical commutation relations. This gives the quantum version of the field from which the notion of vacuum state, one-particle Hilbert-space and Fock space are built as usual. Hence, in a nutshell, quantizing a classical field amounts first to define a space of positive frequency modes from which the one-particle Hilbert-space arises and then to construct the associated Fock space. What renders Minkowski spacetime peculiar w.r.t. other spacetimes (in particular curved ones), is the existence of a global notion of time $t$. The field is then basically quantized on a $t = \text{cst}$ hypersurface.

In curved spacetimes there does not exist generically\footnote{Especially for non stationary spacetimes.} a unique notion of time, even locally. One then selects one time, among others, and starts by considering the space of initial data of a configuration variable (the field) and its conjugate momenta (namely the derivative of the field w.r.t. the selected notion of time). This ``initial" surface can be either spacelike or null and this leads to the quantization of fields on spacelike or null hypersurfaces. It is actually more convenient to step back and look at a more covariant manner to quantize the theory, without specifying any particular initial surface data nor any notion of equal time. The basic idea is to focus on the solutions of the equations of motion themselves. The fundamental commutation relations are then encoded by the symplectic form of the theory, defined on the solution space. The symplectic form plays therefore a fundamental role in the quantization of fields in curved spacetimes. Once an invertible symplectic form is defined on the phase space,\footnote{Which, as we just said, is in one to one correspondence with the space of solutions of the equations of motion.} the latter is decomposed in positive and negative frequency modes with respect to the symplectic inner product and finally the Fock space of the theory is built upon the one-particle Hilbert space coming from the positive frequency modes.

Given our knowledge on the spacelike case, we present in this Appendix the quantization of fields on a non-expanding null-hypersurface, as complement to Section \ref{sec: quantization}. Care has to be taken when translating the spacelike steps to this situation as e.g. the naive definition of the symplectic form yields a (undesirable) degenerate quantity. After imposing therefore a slight prescription, we obtain the one-particle Hilbert space and the Fock-space. As a genuine feature of QFT where infinitely many degrees of freedom are considered, the constructed Hilbert spaces crucially depend on the choice of positive frequency modes.\footnote{The general statement, tamed in the Stone-von Neumann theorem \cite{rosenberg2004selective}, is that for a finite number of degrees of freedom, there is only one unitary equivalent class of Hilbert-spaces one can construct with the canonical procedure. Therefore, physics is undistinguishable between two constructions.} We can thus construct non-unitarily related Hilbert spaces. This observation is at the heart of the Unruh effect proved in Appendix \ref{app: unruh}. Following Wald \cite{wald1994quantum}, we explain how the unavoidable UV and IR divergences of QFT are a guide towards the reduction of the indeterminacy in the choice of Hilbert-space. 

Hereafter, we use the abstract index notation when necessary. Most of the material presented here is a translation of well-known material to the case of null-hypersurfaces. The interested reader can look at reviews like e.g. \cite{wald1994quantum,Jacobson:2003vx}.

\subsection{Quantization on a non-expanding null hypersurface}
\label{subsec: QFT null like}

Consider $\phi$ a massless free scalar field and a non-expanding null hypersurface. The assumption of non-expansion is important both for applications of the present formalism at $\scri_R$ and for reasons related to the change of spatial volume form, see \eqref{eq: pi smearing} below. 

The process of quantization requires a non-degenerate symplectic structure. The latter is naturally defined on the theory's phase space and then extended to the whole set of solutions of the equations of motion. In the spacelike case, the phase space $\Gamma$ of the Klein-Gordon theory is defined to be the space of initial data $(\phi, \pi)$ with $\pi$ the conjugate momenta of the field $\phi$, which are smooth and of compact support on the spacelike Cauchy surface $\Sigma$ i.e. $\phi, \pi \in \mathcal{C}^{\infty}_{0}(\Sigma)$. While, at least locally, a unique family of spacelike Cauchy surfaces is enough in that case, when dealing with null initial value surface two families of hypersurfaces are needed. Indeed, the Klein-Gordon equation generically separates between the left and right moving modes. The spacelike initial value problem becomes a null characteristic initial value problem where initial data live e.g. at future null infinity and on the future light cone emanating from the origin (see e.g. the general discussion of \cite{Kroon:2016ink} and the explicit case of the scalar field recently developed in \cite{Bekaert:2024tkv}).\footnote{This observation comes from the fact that a Cauchy surface is either a one-sheeted spacelike hypersurface or a two-sheeted null hypersurface. A relation can be made between the characteristic initial value problem and the equivalent Cauchy problem in a neighborhood of the codimension-$2$ surface in which the two null hypersurfaces intersect, see e.g. \cite{Kroon:2016ink}.} In our work however, we stick onto one of the two null hypersurfaces namely either only the horizon or only right maximally extended null infinity $\scri_{R}$, therefore we describe only half of the modes. 

The bulk manifold $\mathcal{M}$ is charted by the coordinates $(U,V, x^A)$ and $\scri_R$ is a non-expanding null hypersurface located at $V = +\infty$. There lives only the $U$-modes of the Klein-Gordon equation, namely fields of the form $\phi = \phi(U,x^A)$ with conjugate momenta $\pi = \pi(U,x^A) := \partial_U \phi$. The description hereafter can be repeated to treat the $V$-modes on the complementary null hypersurface at $U = +\infty$. The space of solutions of the equations of motion, $\mathscr{S}$, is therefore made of any function independent of the coordinate $V$ i.e. and can be restricted to the space of Schwartz functions $f(U, x^A) \in \mathcal{T}(\mathbb{R} \times S^2)$. This reinterpretation of the solution space in terms of test functions leads on the one hand for a smooth transition towards the algebraic approach described in Appendix \ref{app: AQFT}, while on the other hand it allows for a rigorous definition of the quantum field operator as an operator-valued distribution. Consider all possible smearing of the scalar field $\phi \in \mathscr{S}$ by elements
$f \in \mathcal{T}(\mathbb{R} \times S^2)$
\begin{equation}
    \label{eq: null observables}
    \phi(f) = \int_{\scri_R} f \phi \, \text{d}U \wedge \epsilon_{S} \, ,
\end{equation}
where the integral is over the support of $f$ and with $\eps_{S}$ the volume form of the spatial cross-sections $S$ of $\scri_R$. 
We get from the classical theory the following symplectic product
\begin{equation}
    \label{eq: null symplec product real}
    \forall \phi \in \mathscr{S}, \forall f \in \mathcal{T}(\mathbb{R} \times S^2); \qquad \Omega(f,\phi) = \int_{\scri_R}\left(f \partial_U \phi - \phi \partial_U f \right) \, \text{d}U \wedge \epsilon_{S} \, .
\end{equation}
Contrarily to the spacelike case, the null symplectic product \eqref{eq: null symplec product real} is generically degenerate due to the zero-modes in $U$ of $\phi$. Hence, one can quantize the theory by using the operators 
\begin{equation}
    \label{eq: pi smearing}
     \Om(f, \phi) := 2 \pi(f)
\end{equation}
instead of \eqref{eq: null observables}. \footnote{Note that the restriction to operators \eqref{eq: pi smearing} prevents us from representing the soft sector of the theory, namely the difference between the value of the field at infinite past and late time. As in this work we only focus on local observables, this issue does not matter.} From \eqref{eq: pi smearing}, we deduce the commutation relations
\begin{equation}
    \label{eq: null commut relations}
    \forall (f, g)\in \mathcal{T}(\mathbb{R} \times S^2), \;  \left[\hat{\pi}(f), \hat{\pi}(g)\right] = -\frac{i}{4}\Omega(f,g) \, ,
\end{equation}
which indeed leads to the usual formula
\begin{equation}
    \label{eq: usual null commut relation}
    \left[\hat{\phi}(U, x^A_1), \hat{\pi}(U', x^A_2)\right] = \frac{i}{2}\delta(U - U') \delta^{(2)}(x^A_1 - x^A_2) \, .
\end{equation}
Note that, as we stated in the main content of the article, that one-half factor in \eqref{eq: usual null commut relation}, or equivalently the one-fourth factor in \eqref{eq: null commut relations} depicts the fact that our construction discards half of the initial conditions, namely the ones living on the complementary null hypersurface (here $\scri_R =  \scri^{+}_{I} \cup \scri^{-}_{II}$).

To define the space of positive frequency solutions we consider the complexified solution space $\mathscr{S}_{0}^{\mathbb{C}}$ of Schwartz functions on $\scri_R$ and define on it the complex symplectic form which is called the \emph{Klein-Gordon product}
\begin{equation}
    \label{eq: complex null symplectic product}
    \forall(f,g) \in \left(\mathscr{S}_{0}^{\mathbb{C}}\right)^2, \; \Omega^{\mathbb{C}}(f,g) = i\int_{\scri_R}\left(\bar{f}\partial_U g - g\partial_U \bar{f} \right)\text{d}U \wedge \epsilon_S = 2i \int_{\scri_R} \bar{f}\partial_U g \, \text{d}U \wedge \epsilon_S \, .
\end{equation}
It is antilinear in the first argument and linear in the second one. Also it satisfies $\Om^{\mathbb{C}}(f, g) = \overline{\Om^{\mathbb{C}}(g, f)}$ therefore it is almost an inner product. To make it a genuine inner product we consider a projector $K$ which splits the solution space into the direct sum of positive and negative frequency modes $\mathscr{S}_0^{\mathbb{C}} = \mathscr{S}^{>0}_K \oplus \mathscr{S}^{<0}_K$, with $\mathscr{S}^{<0}_K = \overline{\mathscr{S}^{>0}_K}$. In that case we have that $\forall (f, g) \in \mathscr{S}_0^{\mathbb{C}}$ 
\be
    \Om^{\mathbb{C}}(Kf, K f) > 0, \qquad \Om^{\mathbb{C}}(\overline{Kf}, \overline{K f}) < 0, \qquad \Om^{\mathbb{C}}(Kf, \overline{Kg}) = 0
\ee
where $\overline{K}$ the projector onto $\mathscr{S}^{<0}_K$. The one-particle Hilbert-space $\mathscr{H}^{K}_1$ is then the Cauchy completion of $\mathscr{S}_K^{>0}$. The vacuum state $\ket{\Omega^K}$ is the unique state annihilated by all operators
\be \label{annihilationop}
    a(Kf) = \Om^{\mathbb{C}}(Kf, \phi) = \langle f, \phi \rangle_K \, , \, f \in \mathscr{S}^{>0}_{K} \, .
\ee
and the (symmetrized) Fock space reads
\begin{equation}
    \label{eq: sym fock space}
    \mathscr{F}_{\phi}^{K} := \mathscr{F}_{s}(\mathscr{H}^{K}_1) = \bigoplus_{n = 0}^{\infty} \left(\text{sym} \bigotimes_{k=0}^{n} \mathscr{H}^{1}_K \right) \,,
\end{equation}
where\footnote{When $n=0$ and $k=0$ the sum reduces to $\mathbb{C}$.} ``sym'' denotes the symmetrization.\footnote{The symmetrized version of the Fock space has been chosen in agreement with the spin statistic theorem, which would have required to take the antisymmetrized version in the case of the Dirac field.} Given the Fock space, quantum states and observables of the theory can be defined and computations of outcomes of physical measurements can be performed. One has therefore achieved to give a quantum version of the classical Klein-Gordon theory on a null hypersurface.

A fundamental remark is that many choices of decomposition of $\mathscr{S}_{0}^{\mathbb{C}}$, i.e. choices of projector $K$ are possible. For instance on $\scri_R^+$ we can use the positive frequencies $\Om$, i.e the orthonormal modes $\frac{e^{- i \Om U}}{\sqrt{4 \pi \Om}}$, or the space spanned by the positive frequencies $\om$, i.e the orthonormal modes $\frac{e^{- i \om u}}{\sqrt{4 \pi \om}}$ using another notion of time $u = u(U) = \ln U$ in the mode decomposition. The two vacua and Fock spaces will generically be different and be related to different boundary conditions at late time. To lift the huge ambiguity one can use the UV and IR behavior of the field \cite{wald1994quantum}. In the UV one can impose the \emph{Hadamard condition}\footnote{This condition imposes the field's two-point function in the state $\ket{\Psi}$ to be expressed as
\begin{equation}
    \label{eq: Hadamard}
    \langle \phi(x) \phi(x') \rangle_\Psi = \frac{\alpha(x,x')}{\sigma} + \beta(x', x) \ln \sigma + H_\Psi(x, x')
\end{equation}
where $\sigma$ is the squared geodesic distance between $x$ and $x'$ with $x = (U,V, x^A$) collectively. In a neighborhood of a non expanding null hypersurface, one can always write
\begin{equation}
    \label{eq: geodesic distance}
    \frac{1}{\sigma^2} = \frac{1}{(U - U')(V - V') + (x^A - x'^A)^2} \, ,
\end{equation}
where $(U, V)$ are null coordinates. The coefficients $\alpha(x', x)$ and $\beta(x', x)$ only depend on the spacetime geometry, while $H_\Psi(x, x')$ is state dependent and can be chosen to vanish in the Minkowski vacuum.} which allows one to define a stress-energy tensor and therefore to use the semi-classical Einstein's equations, as we did in Section \ref{sec: max scri and einstein}. In the IR, one imposes boundary conditions (which on $\scri_R$ are late time conditions) on the field.\footnote{A standard result \cite{fulling1981singularity, wald1994quantum} states that in a closed Universe (namely a globally hyperbolic spacetime which possesses a compact Cauchy surface), there basically exist only one class of Hilbert spaces which satisfy the Hadamard condition.} This provides a prescription for the final state to which the quantized field will relax. 
For example a field can be asked to evolve towards a thermal state which, in a Schwarzschild black hole background, selects the class of Hilbert-spaces built from the Hartle-Hawking vacuum state.


\section{The algebraic approach to QFT in curved spacetimes}
\label{app: AQFT}

The spacetime approach to QFT led to non-unitary related notions of Hilbert-spaces, undermining our desire for universality.\footnote{The appearance of non-unitarily related Hilbert-spaces is not tight to QFT in curved spacetimes; the van Hove model \cite{VANHOVE1952145} (reviewed in \cite{Fewster:2020aeq}) is an example in flat space where non-unitarity arises because of a coupling between a scalar field and a distributional external source.} Even though the UV and IR behavior of the field can help choosing between them, we are forced to admit that the very notion of a Hilbert-space is not fundamental. The canonical commutation relations of the field with its momentum,\footnote{Or equivalently the canonical commutation relations of the ladder operators in second quantization.} like e.g. in \eqref{eq: null commut relations}, are however the same whatever the Hilbert-space representative and therefore fundamentally characterize the theory we try to quantize. The Hilbert-space formalism for QFT emphasizes the notion of states while one could like to focus more on the observables of the theory and the algebraic relations (like the commutation relations) between them. This route is the one followed by algebraic quantum field theory.

To any dynamical classical theory one can associate a symplectic space, namely the solution space $\mathscr{S}$, with symplectic structure $\Omega$. When suitably smeared against test functions $ f \in \mathscr{S}$, the field $\pi$ induces,\footnote{Here we treat the null case in which first the solution space and the space of test functions coincide and second the use of the momentum was necessary to render the symplectic structure \eqref{eq: null symplec product real} non-degenerate.} by exponentiation, a set of bounded operators
\begin{equation}
    \label{eq: weyl elements}
    \forall f \in \mathscr{S}, \; W(f) := e^{i\pi(f)} \, .
\end{equation}
As a consequence of the canonical commutation relations \eqref{eq: null commut relations} expressed in their algebraic form, the  $W(f)$ satisfy the following relations
\begin{equation}
    \label{eq: weyl relations}
    W(f)W(g) = e^{\frac{i}{8}\Omega(f,g)}W(f+g) \quad \text{and} \quad W(f)^{\ast} = W(-f) \, ,
\end{equation}
where the first one is proven thanks to the Baker-Campbell-Haussdorf formula while the second one is immediate.\footnote{ \label{footnoteweylresc} The usual Weyl algebra relations are often written 
\be
    W(f) W(g) = e^{\frac{i}{2}\Omega(f,g)}W(f+g) \quad \text{and} \quad W(f)^{\ast} = W(-f)
\ee
instead of \eqref{eq: weyl relations} but it is because the Weyl unitaries $W(f)$ are usually defined as 
\be
    W(f) := e^{i \Om(f, \phi)}, \qquad \frac{\Om(f, \phi)}{2} = \pi(f)
\ee
instead of \eqref{eq: weyl elements}. However, the two formulations are exactly equivalent.} The set of $W(f)$ for $f \in  \mathscr{S}$ defines the \emph{Weyl algebra}.\footnote{The theory of operators and $\mathbb{C}^\ast$-algebra proves that to any symplectic vector is associated a unique Weyl algebra, up to a $\ast$-isomorphism. Therefore such algebras really constitute the algebraic counterpart of the physical canonical commutation relations.} The latter is naturally endowed with the structure of a $\mathbb{C}^{*}$-algebra, namely a normed algebra over $\mathbb{C}$, complete for its norm topology and which satisfies the $\ast$-property.\footnote{Namely 
\begin{equation}
    \forall A \in \mathcal{A}, ||A^\ast A|| = ||A|| \, ||A^\ast|| = ||A||^2 \, .
\end{equation}} These algebras admits Hilbert-space representations via the GNS construction theorem (see below). Therefore the classical theory naturally induces an algebra from which one can build an Hilbert-space.

In the Hilbert-space formulation of quantum physics, pure states are seen as rays of a Hilbert-space while mixed states are density matrices i.e. positive and trace-class operators of the Hilbert-space normalized to $1$. Observables are self-adjoint operators on the Hilbert-space, which pre-exists: there is a hierarchy in which states precede observables. Algebraic Quantum Field Theory (or AQFT), on the other hand, reverses this hierarchy. It starts from observables, which are defined as the self-adjoint elements of a unital $\mathbb{C}^{\ast}$-algebra\footnote{The $\mathbb{C}^{\ast}$ structure is not strictly necessary though, having at hand a unital $\ast$-algebra is enough to define the notion of observables.} $\mathcal{A}$ (see the definition below) which characterizes the physical system at hand.\footnote{Hence the reason why we are interested in the Weyl algebra.} This subset of $\mathcal{A}$ is dubbed the \emph{algebra of observables}. \emph{Algebraic states} are maps $\omega : \mathcal{A} \to \mathbb{C}$ which are positive\footnote{The observables form the subspace of $\mathcal{A}$ which are self-adjoint. However a general element $A \in \mathcal{A}$ has no reason \emph{a priori} to be self-adjoint itself.}
\begin{equation}
    \label{eq: positivity property}
    \forall A \in \mathcal{A}, \; \omega(A^{\ast}A) \geq 0 \, ,
\end{equation}
and normalized
\begin{equation}
    \label{eq: normalization alge states}
    \omega(1_\mathcal{A}) = 1 \, .
\end{equation}
A state will be \emph{mixed} if it is the combination of other states i.e. $\omega = \lambda \omega_1 + \mu \omega_2$ with $\lambda + \mu = 1$, otherwise the state is \emph{pure}. 

The main examples of $\mathbb{C}^{\ast}$-algebras are the algebra of bounded operators of a Hilbert-space $\mathcal{B}(\mathscr{H)}$ and the Weyl algebra \eqref{eq: weyl relations}.\footnote{In the special case of $\mathcal{B}(\mathscr{H})$ one can explicitly show the relation between a state $\ket{\Psi} \in \mathscr{H}$ and an algebraic state $\omega_{\Psi} : \mathcal{B}(\mathscr{H}) \to \mathbb{C}$ via
\begin{equation}
    \label{eq: from ket to omega}
    \forall A \in \mathcal{B}(\mathscr{H}), \; \omega_{\Psi}(A) = \bra{\Psi}A\ket{\Psi} \, .
\end{equation}
The same can be done for mixed states $\rho$ via $\omega_{\rho}(A) = \Tr{\rho A}$. In \eqref{eq: from ket to omega}, $A$ is directly an operator on a Hilbert-space, but for a general $\mathbb{C}^{\ast}$-algebra like the Weyl algebra, there is no notion of Hilbert-space \emph{a priori}, hence the need for the subsequent representation theorem.} The Weyl algebra in particular, as it emanates from the symplectic structure, contains all the information about the theory we want to quantize. We define the process of quantizing a theory as the construction of a Hilbert-space in which lie the quantum states and the observables. From the algebraic viewpoint we seek a representation of the $\mathbb{C}^\ast$-algebra $\mathcal{A}$ in terms of a Hilbert-space. This can be achieved using the GNS construction (for Gelfand, Naimark and Segal) \cite{gelfand1943imbedding, segal1947irreducible}.\footnote{A proof of the GNS construction can be found in \cite{Fewster:2020aeq}.} From a state $\omega$ we get a triple $(\mathscr{H}, \mathscr{D}, \pi)_{\omega}$ (which depends on $\omega$) and a unit vector $\ket{\Omega_{\omega}} \in \mathscr{D}$ such that $\mathscr{D} = \{ \pi(A)\ket{\Omega_{\omega}} \lvert A \in \mathcal{A} \}$ and $\forall A \in \mathcal{A}, \; \omega(A) = \bra{\Omega_{\omega}}A\ket{\Omega_{\omega}}$. The representation $(\mathscr{H}, \mathscr{D}, \pi)_{\omega}$ is unique up to unitary equivalence. The state $\omega$ is pure iff the representation is irreducible. Via the map $\pi$, the abstract algebra $\mathcal{A}$ can now be seen as a subalgebra of the linear operators of $\mathscr{H}_{\omega}$ i.e. $\pi(\mathcal{A}) \subset \mathcal{B}(\mathscr{H}_\omega)$. However, in the following and through the main body of the paper, the notation for the algebra elements $A \in \mathcal{A}$ is also used for their representatives $\pi(A) \in \mathcal{B}(\mathscr{H}_\om)$  in order to lighten the writing. Finally, due to point $1.$ one says that $\ket{\Omega_{\omega}}$ is \emph{cyclic} i.e. $\mathscr{H} = \overline{\{\pi(A)\ket{\Omega_{\omega}}, A \in \mathcal{A}\}}$.

The state $\omega$ is usually referred as the \emph{vacuum state}. In many practical purposes, and for all the vacuum states considered in the main body of this paper, the algebraic state $\om$ is taken to be a quasifree (i.e. gaussian) state, defined so that the expectations values of the Weyl unitary elements are given by \footnote{Usually, the quasifree state is defined via the expectation value
\be
    \om(e^{i \Om(f, \phi)}) =  e^{-\frac{1}{2}\braket{f,f}_K}
\ee
but here the unitary elements of the Weyl algebra are defined by $W(f) = e^{i \pi(f)} = e^{ i \frac{\Om(f, \phi)}{2}}$, that is equivalent to rescale $f \rightarrow \frac{f}{2}$, hence the factor $\f18$ instead of $\f12$ in \eqref{eq: def gaussian state}, see also footnote \ref{footnoteweylresc}.}
\begin{equation}
    \label{eq: def gaussian state}
    \omega(e^{i\pi(f)}) = e^{-\frac{1}{8}\braket{f,f}_K} \, ,
\end{equation}
and where a choice of positive definite inner product $\braket{\cdot , \cdot}_K$ had to be made. It can be shown that if $\braket{\cdot , \cdot}_K$ is a positive definite inner product constructed from the Klein-Gordon inner product \eqref{eq: complex null symplectic product}, then the positivity condition \eqref{eq: positivity property} is satisfied for any element $A$ of the Weyl algebra. To make the connection between the definition of a quasifree state \eqref{eq: def gaussian state}, given by a choice of inner product, and the definition of the vacuum state $\ket{\Om^K} = \ket{\Omega_\omega}$ via a positive-frequency decomposition—namely, as the state annihilated by the operators $a(Kf)$ —more explicit, it suffices to compute
\begin{equation}
    \label{eq: proof Gaussian}
     W(\pi(f)) = \bra{\Om_\omega} e^{i \pi(f)} \ket{\Om_\omega} = \bra{\Om_\omega} e^{i \frac{a^\dag(Kf)}{2}} e^{i \frac{a(Kf)}{2}} \ket{\Om_\omega} e^{\frac18 [i a(Kf), i a^\dag(Kf)]} = e^{- \frac18 \langle f, f \rangle_K}
\end{equation}
given that 
\begin{equation}
    \label{eq: commutator}
    [a(Kf), a^{\dagger}(Kg)] = \Omega^{\mathbb{C}}(Kf,Kg) := \langle f , g \rangle_K \, .
\end{equation}

We note that for a given vacuum state $\omega$, we get a whole class of Hilbert-spaces all related by unitary transformations. However, from two different vacuum states $\om_1$ and $\om_2$, we generally obtain non-unitarily related representations. To illustrate this point we give the following example in Minkowski spacetime. Consider the Hilbert space $\mathscr{H}_M$ built from the Minkowski vacuum $\om_M$ on $\scri_R^+$, by using the modes $\frac{e^{- i \om u_+}}{\sqrt{4 \pi \om}}$, $\om > 0$, as positive frequency modes. One could have also considered the Hilbert space $\mathscr{H}_R$ coming from the state $\om_R$ above a cut $u_+ = u_i$, which is the vacuum state for the modes $\frac{e^{- i \om_R u_R}}{\sqrt{4 \pi \om_R}}$ (where $u_R = \kappa \ln (u_+ - u_i)$ and $\om_R > 0$). Due to the Unruh effect, $\ket{0}_M = \ket{\Omega^{u_+}}$ corresponds to an infinite number of particles, while $\ket{0}_R = \ket{\Omega^{u_R}}$ is a true vacuum (i.e. a pure state) with no particles. No unitary transformations can create an infinite amount of particles from zero, therefore the two GNS constructions are nonequivalent.

Note finally that the link between the algebraic and spacetime approaches is made via the support of the test functions we use to define the $W(f)$. When the latter lies into a spacetime region $\mathcal{O}$ we construct the algebra of observables of that region, $\mathcal{A}(\mathcal{O})$ which is a subalgebra of the algebra of the total space $\mathcal{M}$. When $\mathcal{M}$ is globally hyperbolic, the net of local algebras $\{\mathcal{A}(\mathcal{O}) | \mathcal{O} \subset_{\text{open}} \mathcal{M}\}$ satisfies the axioms of isotony ($\mathcal{O} \subset \mathcal{O}' \Longrightarrow \mathcal{A}(\mathcal{O}) \subset_{\text{subalgebra}} \mathcal{A}(\mathcal{O}')$) and causality (if $\mathcal{O}$ and $\mathcal{O}'$ are causally disjoints then $[\mathcal{A}(\mathcal{O}), \mathcal{A}(\mathcal{O}')] = 0$). Restriction of an algebra of observables to one of its subalgebra translates in spacetime as restricting our attention to a subregion and this amounts to consider test functions whose support lies in the latter instead of the full space. Many results can be derived from a restriction to a subalgebra of observables, like the Unruh effect derived in Appendix \ref{app: unruh}.


\section{Modular Theory and Araki's relative entropy}
\label{app: modular theory}

In this Appendix we present basics on the theory of von Neumann algebras and on Tomita-Takesaki's modular theory \cite{takesaki1972conditional, takesaki2003theory}. This mathematical framework allows to rigorously define the notions of modular Hamiltonian and relative entropy, which we widely use in the main content. We assume the reader to be familiar with $\ast$-algebras and the GNS representation theorem, both reviewed in Appendix \ref{app: AQFT}. We do not pretend to be comprehensive, just to introduce the tools underpinning the proof of the dual GSL. More details can be found e.g. in the reviews \cite{witten2018aps, hollands2018entanglement, Sorce:2023fdx, Sorce:2023gio}.

Along the Appendix we shall exemplified the notions using the algebra of observables $\mathcal{A}$ of a given region of spacetime which we assume to be decomposed into two complementary regions $I$ and $II$. $\mathcal{A}_I$ will be the algebra of region $I$ and $\mathcal{A}_{II}$ the one of region $II$. The total GNS Hilbert space which represents an arbitrary algebra $\mathcal{A}$ with vacuum state $\ket{\Omega}$ is denoted as usual $\mathscr{H}_{\Omega}$, and $\mathcal{A}$ can be seen as a subalgebra of\footnote{Recall that observables are built by \eqref{eq: weyl elements} so that we get bounded operators.} $\mathcal{B}(\mathscr{H}_{\Omega})$ via the representation map $\pi$ of the GNS theorem. As said in Appendix \ref{app: AQFT}, the state $\ket{\Omega}$ is cyclic for $\mathcal{A}$. If the vacuum state $\ket{\Om}$ is defined through a positive frequency decomposition using a global notion of time $U$ covering $\scri_R$ (or times $\{ U^{(l,m)} \}$ attached to each sector $(l,m)$), a simple extension of the Reeh-Schlieder theorem \cite{reeh1961bemerkungen} states that for a subalgebra $\mathcal{A}_i^\scri \subset \mathcal{A}^\scri$ of operators of an open region $\mathcal{D}_i^{\scri} = (U_i, + \infty) \times S^2 \subset \scri_R$, the vacuum state $\ket{\Omega}$ is also cyclic i.e. \footnote{In the two dimensional case, if $U$ is charting $\scri_R$, then the cyclic vacuum state $\ket{\Om}$ obtained from the positive frequency decomposition associated to the time $U$ is invariant under boosts and translations $U \rightarrow aU + b$ and the generator of the time translations has a positive spectrum. 
This is basically the set of hypothesis needed to prove the Reeh-Schlieder theorem. In the four dimensional case, the vacuum state $\ket{\Om}$ considered can be expressed as a tensor product of two dimensional vacua on each sector $(l,m)$, see \eqref{vallmten}, all satisfying the positivity conditions and invariant under $U^{(l,m)} \rightarrow a_{lm} U^{(l,m)} + b_{lm}$, so that the two-dimensional argument can immediately be extended to the case at hand.}
\be
     \overline{\pi\left(\mathcal{A}_i^\scri\right) \ket{\Om}} = \overline{\pi\left(\mathcal{A}^\scri\right) \ket{\Om}} = \mathscr{H}_\Om \, ,
\ee
so that we get the same Hilbert spaces from the two GNS constructions. Recall that we use the same notation for the abstract algebra $\mathcal{A}$ and for its representation and similarly we denote $A$ instead of $\pi(A)$ an arbitrary element of $\mathcal{A}$ acting on a vector of $\mathscr{H}_\Omega$. Assuming a GNS construction has been performed, we restrict our attention to subalgebras $\mathcal{A}$ of $\mathcal{B}(\mathscr{H})$ for a given Hilbert-space $\mathscr{H}$.

\subsubsection*{Von Neumann algebras}

We start by recalling some basics on von Neumann algebras. Such algebras are characterized by the fact of being equal to their double commutant. Recall that the \emph{commutant} $\mathcal{A}'$ of an algebra $\mathcal{A}$ is the set $\mathcal{A}' = \{B \in \mathcal{B}(\mathscr{H}) | \forall A \in \mathcal{A}, \; [A,B] = 0\}$. A von Neumann algebra therefore satisfies $\mathcal{A}'' = \mathcal{A}$. From a physical system, one gets a von Neumann algebra by taking the double commutant of the GNS representation of the Weyl algebra $\pi(\mathcal{A}_W)''$. Note that the commutant of an algebra of observables of a region $\mathcal{O}$ (not necessarily a von Neumann algebra though) can be related to the algebra of observables of the complementary region $\mathcal{O}'$:  we have $\mathcal{A}_{\mathcal{O}}^{'} \subset \mathcal{A}_{\mathcal{O}'}$. Instances for which we have the equality $\mathcal{A}_{\mathcal{O}}^{'} = \mathcal{A}_{\mathcal{O}'}$ are said to satisfy \emph{Haag's duality} \cite{Haag:1996hvx}. We assume for our examples that $\mathcal{A}_I^{'} = \mathcal{A}_{II}$. As also $\mathcal{A}_{II}^{'} = \mathcal{A}_{I}$ both of them are von Neumann algebras.\footnote{Note that even though the restricted algebras are von Neumann algebras, this is not always the case for the total algebra of observables $\mathcal{A}$.} When considering $\scri_R$ and cutting it at $U=0$ we also have that $\mathcal{A}_0^\scri$ and its commutant $\mathcal{A}_0^{\scri \, '}$ are von Neumann algebras. If the cut is at an arbitrary $U = U_i$ then, of course, the commutant of the algebra $\mathcal{A}_i^\scri$ is the algebra of observables of the region $U < U_i$, and both algebras are von Neumann. 

Von Neumann algebras can be classified in several types depending on their factors. We will not enter the details here (we refer to e.g.  \cite{Witten:2021unn, Sorce:2023fdx}) but for physical applications one can remember that \emph{type I} von Neumann algebras have pure states and density-matrices, \emph{type II} only density matrices and \emph{type III} neither of them, as notions such as traces no longer exist. For example when we restrict an algebra of observables $\mathcal{A} \subset \mathcal{B}(\mathscr{H}_\Omega)$ to one of its subalgebra $\mathcal{A}_I$, the only difference is that $\mathcal{A}_I''$ is a type III von Neumann algebra ($\ket{\Om}$ is not pure), while $\mathcal{A}''$ is a type I von Neumann algebra (as $\ket{\Om}$ is pure wrt the global algebra).

\subsubsection*{Cyclic and separating state}

The second ingredient of modular theory is the notion of cyclic and separating vector in a Hilbert-space. From now on we consider the Hilbert space at hand to be the GNS space representing $\mathcal{A}$ with vacuum state $\ket{\Omega}$ i.e. $\mathcal{A} \subset \mathcal{B}(\mathscr{H}_{\Omega})$. 
A \emph{cyclic} vector $\ket{\Psi} \in \mathscr{H}_{\Omega}$ is such that the space $\{A\ket{\Psi} \, , \, A \in \mathcal{A}\}$ is dense in $\mathscr{H}_{\Omega}$. It is in particular the case for $\ket{\Omega}$. A \emph{separating} vector $\ket{\Psi} \in \mathscr{H}_{\Omega}$ is such that\footnote{The restriction of the Hartle-Hawking state $\ket{\Omega_H}$ on the future horizon $\mathcal{H}_{L}^+$ is an example of cyclic and separating state.} 
$A\ket{\Psi} = 0 \Longrightarrow A = 0$. If we assume that $\mathcal{A}$ is a von Neumann algebra, then a vector $\ket{\Psi} \in \mathscr{H}_{\Omega}$ is cyclic for $\mathcal{A}$ iff it is separating for its commutant $\mathcal{A}'$.

Consider $\ket{\Omega} \in \mathscr{H}_{\Omega}$ a vacuum state and the algebra $\mathcal{A}$. Then $\ket{\Omega}$ is cyclic and separating for the subalgebra\footnote{Again it is a consequence of the Reeh-Schlider theorem and it implies the Unruh effect, see latter.} $\mathcal{A}_I$ and so it is for the commutant $\mathcal{A}_I^{'} = \mathcal{A}_{II}$. For algebras admitting trace operators and density matrices, we can give an enlightening decomposition of such a state in a basis of the total Hilbert space $\mathscr{H}_{\Omega} = \mathscr{H}_{I} \otimes \mathscr{H}_{II}$ as 
\begin{equation}
    \label{eq: cyclic separating}
    \ket{\Omega} = \sum_{i \in J \, ; \, i' \in J'} c_{ii'} \ket{i}_{I} \otimes \ket{i'}_{II}
\end{equation}
with $c_{ii'} \neq 0$ and $\{\ket i\}_I$ (resp. $\{\ket i'\}_{II}$) is an orthonormal basis of $\mathscr{H}_I$ (resp. $\mathscr{H}_{II}$) indexed by a set $J$ (resp. $J'$) which is at most countably infinite. It is a pure state from the point of view of the algebra $\mathcal{A}$. Its restriction to the subalgebra $\mathcal{A}_I$ is obtained via partial tracing
\begin{equation}
    \label{eq: partial trace}
    \hat{\rho}^{\mathcal{A}_I}_{\Omega} = \Tr_{\mathcal{A}_{II}} \, \ket{\Omega} \bra{\Omega} \, ,
\end{equation}
where the use of density matrices is mandatory as we deal now with a mixed state. One says that $\ket{\Omega}$ is the \emph{purification} of $\hat{\rho}_{\Omega}^{\mathcal{A}_I}$ in $\mathscr{H}_{\Omega}$. The paragon of a cyclic and separating state is the vacuum state \eqref{eq: 4d vac density matrix} which is pure from the point of view of the algebra on the complete null hypersurface $\mathcal{N}$ but becomes thermal upon restriction to region $I$ or region $II$. It is therefore cyclic wrt to the full algebra $\mathcal{A}^\mathcal{N}$ but becomes cyclic \emph{and} separating for $\mathcal{A}_I^{\mathcal{N}}$. 

\subsubsection*{Modular operator}

Consider $\mathcal{A}$ a von Neumann algebra on $\mathscr{H}_{\Omega}$, and a state $\ket{\Psi} \in \mathscr{H}_{\Omega}$ which is cyclic and separating for $\mathcal{A}$.\footnote{To fix ideas one can take $\ket{\Psi} = \ket{\Omega}$ and $\mathcal{A} = \mathcal{A}_I$ the restricted algebra. However there is no reason why the vacuum should be the only cyclic and separating state.} Modular theory starts by the introduction of the \emph{Tomita operator} $S_{\Psi}$ on $\mathscr{H}_{\Omega}$ which is anti-linear and satisfies
\begin{equation}
    \label{eq: Tomita operator}
    \forall A \in \mathcal{A}, \; S_{\Psi} \left( A \ket{\Psi} \right) = A^{\dagger} \ket{\Psi} \, .
\end{equation}
The requirement of cyclicity is essential for $A\ket{\Psi}$ to belong to $\mathscr{H}_{\Omega}$ while asking the vector to be separating ensures \eqref{eq: Tomita operator} to be consistent.\footnote{If $\ket{\Psi}$ was not separating it would exist a non-zero operator $A_0$ such that $A_0 \ket{\Psi} = 0$ and then $S_{\Psi} A_0 \ket{\Psi} = S_{\Psi}(0) = 0$ by anti-linearity on the one hand while $S_{\Psi} A_0\ket{\Psi} = A_{0}^{\dagger}\ket{\Psi} \neq 0$ on the other hand.} The Tomita operator depends on the choice of cyclic and separating vector $\ket{\Psi}$. In general, $S_{\Psi} \notin \mathcal{B}(\mathscr{H}_{\Omega})$ so in particular it does not belong to $\mathcal{A}$. Under the action of the Tomita operator, $\ket{\Psi}$ is invariant ($S_{\Psi}\ket{\Psi} = \ket{\Psi}$) and we also have that $S_{\Psi}^{2} = \mathbb{1}_{\mathscr{H}_{\Omega}}$, the identity operator on $\mathscr{H}_{\Omega}$. One can define the Hermitian conjugate $S_{\Psi}^{\dagger}$ and observe that it is just the Tomita operator of the commutant algebra $\mathcal{A}^{'}$.

Given a Tomita operator and its Hermitian conjugate, we introduce the \emph{modular operator} of $\ket{\Psi}$ wrt to $\mathcal{A}$ as 
\begin{equation}
    \label{eq: modular operator}
    \Delta_{\Psi} := S_{\Psi}^{\dagger} S_{\Psi} \, .
\end{equation}
This operator is positive-definite on $\mathscr{H}_{\Omega}$ because $\ket{\Psi}$ is cyclic and separating. Going back to the case where trace and density matrices are defined, we consider a state $\ket{\Psi}$ of the form \eqref{eq: cyclic separating}, cyclic an separating for $\mathcal{A}_I$ 
(it is so also for $\mathcal{A}_{II}$). 
The mixed state $\hat{\rho}_\Psi^{\mathcal{A}_I} \in \mathcal{B}(\mathscr{H}_{I})$
obtained via \eqref{eq: partial trace} can be written as
\begin{equation}
    \label{eq: state basis decomposition}
    \hat{\rho}_\Psi^{\mathcal{A}_I} = \sum_{i \in J} c_{ii}^{2}\ket{i}\bra{i}, \qquad c_{ii} \neq 0 \, ,
\end{equation}
and a similar expression holds for $\hat{\rho}_\Psi^{\mathcal{A}_{II}}$. One can easily show that  
\begin{equation}
    \label{eq: modular in terms of rho}
    \Delta_{\Psi} = \hat{\rho}_\Psi^{\mathcal{A}_I} \otimes \left(\hat{\rho}_\Psi^{\mathcal{A}_{II}}\right)^{-1}
\end{equation}
with $\left(\hat{\rho}_\Psi^{\mathcal{A}_{II}}\right)^{-1}$ the inverse of $\hat{\rho}_\Psi^{\mathcal{A}_{II}}$. 

Given any two cyclic and separating vectors $\ket{\Omega}$ and $\ket{\Psi}$ of $\mathscr{H}_\Omega$, one can also consider the \emph{relative Tomita operator}, denoted $S_{\Psi | \Omega}$, which satisfies
\begin{equation}
    \label{eq: relative Tomita}
    \forall A \in \mathcal{A}, \; S_{\Psi|\Omega} \left( A\ket{\Psi} \right) = A^{\dagger}\ket{\Om} \, .
\end{equation}
Of course $S_{\Omega|\Omega} = S_{\Omega}$.\footnote{Although $\ket{\Om}$ must be cyclic and separating for consistency, $\ket{\Psi}$ can actually be any state. However, if $\ket{\Psi}$ is not cyclic and separating, the relative entropy we shall define later in \eqref{eq: relative entropy} can be infinite.} Given  $S_{\Psi|\Omega}$ we can immediately introduce a \emph{relative modular operator}
\begin{equation}
    \label{eq: relative modular Ham}
    \Delta_{\Psi|\Omega} := S^{\dagger}_{\Psi|\Omega}S_{\Psi|\Omega} \, .
\end{equation}
When density matrices exists, consider again $\mathscr{H}_{\Omega} = \mathscr{H}_{I} \otimes \mathscr{H}_{II}$ and now two (possibly mixed) states $\hat{\rho}_{\Omega}^{\mathcal{A}_I} \in \mathscr{H}_{I}$ and $\hat{\rho}_{\Psi}^{\mathcal{A}_{II}} \in \mathscr{H}_{II}$ whose purifications in $\mathscr{H}_{\Omega}$ are given by the cyclic and separating states
\begin{equation}
    \label{eq: pure states for rho and sigma}
    \ket{\Omega} = \sum_{i \in I; i'\in I'}c_{ii'}\ket{i}\otimes \ket{i'} \quad \text{and} \quad \ket{\Psi} = \sum_{j \in J; j'\in J'} c_{jj'}\ket{j}\otimes \ket{j'}
\end{equation}
for some bases $(\ket{i})_{i\in I}$ and $(\ket{j})_{j\in J}$, which may be different. Therefore we have $\hat{\rho}_{\Omega}^{\mathcal{A}_I} = \Tr_{\mathcal{A}_{II}}(\ket{\Omega} \bra{\Omega})$ and $\hat{\rho}_{\Psi}^{\mathcal{A}_{II}} = \Tr_{\mathcal{A}_{I}}(\ket{\Psi}\bra{\Psi})$. Then we get
\begin{equation}
    \label{eq: relative modular in terms of density matrix}
    \Delta_{\Psi|\Omega} = \hat\rho_{\Omega}^{\mathcal{A}_I} \, \otimes \, \left(\hat{\rho}_{\Psi}^{\mathcal{A}_{II}}\right)^{-1} \, .
\end{equation}

\subsubsection*{Modular Hamiltonian}

We still consider a cyclic and separating vector $\ket{\Psi} \in \mathscr{H}_{\Omega}$. The modular operator \eqref{eq: modular operator} being positive definite one can write it as
\begin{equation}
    \label{eq: def mod hamiltonian}
    \Delta_\Psi = e^{-K_{\Psi}} \Longleftrightarrow K_{\Psi} = -\ln \Delta_\Psi \, ,
\end{equation}
where $K_\Omega$ is the \emph{modular Hamiltonian} of $\ket{\Psi}$. Generically $K_{\Psi} \notin \mathcal{B}(\mathscr{H}_{\Omega})$. When we restrict to the subalgebra $\mathcal{A}_I$, using \eqref{eq: modular in terms of rho} we can  generically write the total modular Hamiltonian as the combination of two terms, one coming from the algebra $\mathcal{A}_I$ (and acting on $\mathscr{H}_I$) and one coming from the complementary algebra $\mathcal{A}_{I}' =  \mathcal{A}_{II}$ (and acting on $\mathscr{H}_{II}$)
\be
\label{eq: def one sided mod ham}
    K_\Psi =  K_\Psi^{\mathcal{A}_I} - K_\Psi^{\mathcal{A}_{II}}
\ee
where operators appearing in this sum are called \emph{one-sided modular hamiltonians}. The right-sided modular Hamiltonian $K_\Psi^{\mathcal{A}_{II}}$ commutes with the algebra $\mathcal{A}_I$ while the left-sided modular Hamiltonian $K_\Psi^{\mathcal{A}_{I}}$ commutes with its commutant algebra $\mathcal{A}_{II}$. In the cases of interest, the modular Hamiltonian will be expressed as the integral of local operators on $\scri_R$, so that we take the one-sided modular Hamiltonian $K^{\mathcal{A}_I}_\Psi$ by restricting the integral to the region $U > 0$ (times cross sections), see Appendix \ref{app: modular boost energy} where we give the explicit expressions of the one-sided modular hamiltonians in terms of the stress-energy tensor of the theory at hand.

\subsubsection*{KMS conditions}

A fundamental use of the modular operator and Hamiltonian, \eqref{eq: modular operator} and \eqref{eq: def mod hamiltonian}, is in the \emph{Kubo-Martin-Schwinger (KMS) conditions} \cite{kubo1957statistical, martin1959theory}. A state $\ket{\Psi}$ satisfies the KMS conditions wrt a von Neumann algebra $\mathcal{A}$ if
\begin{equation}
    \label{eq: KMS condition}
    \forall (A,B) \in \mathcal{A}^{2}, \; \bra{\Psi}\Delta_{\Psi}^{-1}A\Delta_{\Psi}B\ket{\Psi} := \bra{\Psi}e^{K_{\Psi}}A e^{-K_{\Psi}}B\ket{\Psi} = \bra{\Psi}BA\ket{\Psi} \, .
\end{equation}
When \eqref{eq: KMS condition} holds one says that $\ket{\Psi}$ is at thermal equilibrium, with temperature $T = 1$, with respect to the modular Hamiltonian $K_{\Psi}$, see \eqref{eq: def mod hamiltonian}.\footnote{Be careful not to overstate this observation, see the next paragraph.} Actually, the cyclic and separating vector $\ket{\Psi} \in \mathscr{H}_\Om$ satisfies the KMS conditions wrt $\mathcal{A}$.

Two important remarks are in order here. First, the fact that $\ket{\Psi}$ satisfies the KMS conditions is a reminiscent of the cyclicity of the trace for von Neumann algebras in which this operation does not always exist. Second, observe that any cyclic and separating vector field of a von Neumann algebra is a KMS state. We must be careful not to overinterpret the satisfaction of the KMS condition, saying e.g. that any quantum state is a thermal state. The thermal behavior is with respect to a peculiar operator, namely the modular Hamiltonian $K_{\Psi}$, which has no reasons \emph{a priori} to be related to a well-defined notion of energy in the physical system at hand.\footnote{That is to say that the modular Hamiltonian needs not be the N\oe ther charge associated to ``time" translations.} However, if $\ket{\Psi}$ is for instance the Minkowski vacuum, the modular Hamiltonian turns out to be the boost Hamiltonian and so the previous statement gives us a straightforward proof of the Unruh effect. The exact same reasoning applies to the restriction of the Hartle-Hawking state on the future horizon $\mathcal{H}_L^+$ and future null infinity $\scri_R^+$.

To end this section on basics of modular theory, it is important to state Tomita's theorem (see e.g. \cite{Sorce:2023gio} for a nice discussion on the proof), namely that the modular flow preserves the von Neumann algebra $\mathcal{A}$ i.e. $\forall t \in \mathbb{R}, \; \Delta_{\Psi}^{it}\mathcal{A}\Delta_{\Psi}^{it} \subset \mathcal{A}$. Therefore, we understand why, geometrically, the modular Hamiltonian is a "boost". It is because it needs to preserve the algebra of observables $\mathcal A(\mathcal{O})$ of some spacetime region $\mathcal{O}$, so it must preserve $\mathcal{O}$. In the main content we considered subregions of $\scri_R$ above a certain cut $U = U_i$ and observed that indeed the modular Hamiltonian of our vacuum states were preserving that region, as they were associated to the boost vector field $\xi = \kappa(U-U_i)\partial_U$.

We have now introduced all the tools required to define the relative entropy.

\subsubsection*{The relative entropy}

Through the notion of relative entropy, the relative modular operator \eqref{eq: relative modular Ham} allows to build a quantity one can use to answer the question \emph{how close are two quantum states?} (wrt to some algebra of observables). Given two cyclic and separating states $\ket{\Psi}$ and $\ket{\Omega}$ for a von Neumann algebra $\mathcal{A}$, Araki's definition \cite{araki1975relative} of \emph{relative entropy} is
\begin{equation}
    \label{eq: relative entropy}
    S(\Psi|| \Omega) := \bra{\Psi}\left(-\ln \Delta_{\Psi|\Omega} \right) \ket{\Psi} \, .
\end{equation}
Often, $\ket{\Omega}$ is the vacuum state. Note that if one of the states is not separating then the relative modular operator may contain some zero eigenvalues, rendering the above definition divergent. Several immediate properties are that $S(\Omega||\Omega) = 0$, $S(\Psi \lvert \lvert \Om) \geq 0$ and when there exist an unitary element of the commutant $A' \in \mathcal{A}'$ (i.e. $A'^{\dagger}A' = 1_{\mathcal{A'}}$) such that $\ket{\Psi} = A' \ket{\Omega}$ then $S(\Omega || A'\Omega) = 0$. 

The fundamental property for this work though is the following. If $\mathcal{B} \subset \mathcal{A}$ is a subalgebra of $\mathcal{A}$ then we have for any two states $\ket{\Psi}$ and $\ket{\Omega}$ that
    \begin{equation}
        \label{eq: monotonicity}
        S_{\mathcal{A}}(\Psi || \Omega) \geq S_{\mathcal{B}}(\Psi||\Omega) \, .
    \end{equation}
This is called the \emph{monotonicity of the relative entropy} and it formalizes the fact that if we have at our disposal a smaller set of observables (the subalgebra $\mathcal{B}$) then it will be harder to distinguish between the two states $\ket{\Psi}$ and $\ket{\Omega}$. For systems in which one can deal with density matrices, the relative entropy is decreasing under the action of a Complete Positive Trace Preserving (CPTP) map $\Lambda$
\be \label{montonicityofrelentfini}
    S(\hat \rho_\Psi^\mathcal{A} \lvert \lvert \hat \rho_\Om^\mathcal{A}) \geq  S(\Lambda (\hat \rho_\Psi^\mathcal{A}) \lvert \lvert \Lambda (\hat \rho_\Om^\mathcal{A}))
\ee
and can be interpreted as the fact that the two states $\hat \rho_\Psi^\mathcal{A}$ and $\hat \rho_\Om^\mathcal{A}$ are less distinguishable after the application of the channel $\Lambda$. 

Finally we should be sure that the definition \eqref{eq: relative entropy} gives back the usual expression in the case where traces are well defined. We use again the states $\ket{\Omega}$ and $\ket{\Psi}$ which are cyclic and separating for both restricted algebras $\mathcal{A}_I$ and $\mathcal{A}_{II}$. Recall then that $\hat{\rho}_{\Omega}^{\mathcal{A}_I} = \Tr_{\mathcal{A}_{II}}(\ket{\Omega} \bra{\Omega})$ and $\hat{\rho}_{\Psi}^{\mathcal{A}_{II}} = \Tr_{\mathcal{A}_I}(\ket{\Psi}\bra{\Psi})$ so that for any observable $A$ we have $\langle A \rangle_{\Psi} = \Tr\left(\hat{\rho}_{\Psi}^{\mathcal{A}_{II}}A\right)$ when $A \in \mathcal{A}_{II}$ while $\langle A \rangle_{\Psi} = \Tr\left(\hat{\rho}_{\Psi}^{\mathcal{A}_{I}}A\right)$ when $A \in \mathcal{A}_I$. In that case the modular operator takes the form \eqref{eq: relative modular in terms of density matrix} and so 
\begin{align}
    \label{eq: relative enropy traces}
    S(\Psi \lvert \lvert \Om) &= \bra{\Psi}\left(-\ln \hat \rho_{\Omega}^{\mathcal{A}_I} + \ln \hat \rho_{\Psi}^{\mathcal{A}_{II}}\right)\ket{\Psi} \\
    \nonumber
    &= \Tr \left(\hat \rho_{\Psi}^{\mathcal{A}_{II}} \ln \hat \rho_{\Psi}^{\mathcal{A}_{II}}\right) - \Tr \left(\hat\rho_{\Psi}^{\mathcal{A}_{I}} \ln \hat \rho_{\Omega}^{\mathcal{A}_I}\right) \\
    \nonumber
    &= \Tr \left(\hat \rho_{\Psi}^{\mathcal{A}_{I}} \ln \hat \rho_{\Psi}^{\mathcal{A}_{I}}\right) - \Tr \left(\hat\rho_{\Psi}^{\mathcal{A}_{I}} \ln \hat \rho_{\Omega}^{\mathcal{A}_I}\right) \, ,
\end{align}
where we recognize the usual quantum relative entropy.\footnote{We have of course from our construction $\Tr \left(\hat \rho_{\Psi}^{\mathcal{A}_{II}} \ln \hat \rho_{\Psi}^{\mathcal{A}_{II}}\right) = \Tr \left(\hat \rho_{\Psi}^{\mathcal{A}_{I}} \ln \hat \rho_{\Psi}^{\mathcal{A}_{I}}\right)$.} Therefore \eqref{eq: relative entropy} is the natural extension of the relative entropy to cases where traces and density matrices do not exist.

\subsubsection*{Decomposition of the relative entropy}
\label{subsec: decompo relat entropy}

Let $\ket{\Omega}$ and $\ket{\Psi}$ in $\mathscr{H}_{\Omega}$ be two cyclic and separating states for the restricted algebra $\mathcal{A}_I$. In this paragraph we relate the relative entropy between them  to the one-sided modular Hamiltonian of $\ket{\Omega}$.

Consider first that both states $\ket{\Omega}$ and $\ket{\Psi}$ can be written as density matrices $\hat \rho_{\Omega}^{\mathcal{A}_I}$ and $\hat \rho^{\mathcal{A}_I}_{\Psi}$. We also assume that traces exists. The state $\ket{\Omega}$ satisfies the KMS conditions \eqref{eq: KMS condition} so it can be written as a generalized Gibbs state of the form
\begin{equation}
    \label{eq: Gibbs state}
    \hat{\rho}_{\Omega}^{\mathcal{A}_I} = \frac{e^{-K_{\Omega}^{\mathcal{A}_I}}}{\Tr \left( e^{-K_{\Omega}^{\mathcal{A}_I}} \right)}
\end{equation}
with $K_{\Omega}^{\mathcal{A}_I}$ the one-sided modular Hamiltonian. On the one hand, we can write the relative entropy as in \eqref{eq: relative enropy traces}
\begin{align}
    \label{eq: relative entropy finite dim}
    S\left(\hat \rho_{\Psi}^{\mathcal{A}_I} \Big| \Big| \hat \rho_{\Omega}^{\mathcal{A}_I} \right) &= \Tr\left( \hat \rho_{\Psi}^{\mathcal{A}_I} \ln \hat \rho_{\Psi}^{\mathcal{A}_I} \right) - \Tr \left( \hat \rho_{\Psi}^{\mathcal{A}_I} \ln \hat \rho_{\Omega}^{\mathcal{A}_I} \right) \\ \nonumber
    &= \Tr \left(\hat \rho_{\Psi}^{\mathcal{A}_I} \ln \hat \rho_{\Psi}^{\mathcal{A}_I} \right) + \ln \left[ \Tr \left(e^{-K^{\mathcal{A}_I}_{\Omega}} \right) \right] + \langle K^{\mathcal{A}_I}_{\Omega} \rangle_{\Psi} \, ,
\end{align}
where $\langle \cdot \rangle_{\Psi}$ denotes the vev in the state $\ket{\Psi}$ and we have used that $\Tr \hat \rho_{\Psi}^{\mathcal{A}_I} = 1$. On the other hand we have
\begin{equation}
    \label{eq: trace pho omega}
    \Tr \left(\hat \rho_{\Omega}^{\mathcal{A}_I} \ln \hat \rho_{\Omega}^{\mathcal{A}_I} \right) = -\langle K^{\mathcal{A}_I}_{\Omega} \rangle_{\Omega} - \ln \left[ \Tr \left(e^{-K^{\mathcal{A}_I}_{\Omega}} \right) \right] = - \ln \left[ \Tr \left(e^{-K^{\mathcal{A}_I}_{\Omega}} \right) \right]
\end{equation}
as $\langle K^{\mathcal{A}_I}_{\Omega} \rangle_{\Omega} = 0$ in virtue of $\Delta_{\Omega} \ket{\Omega} = \ket{\Omega}$. Combining \eqref{eq: relative entropy finite dim} with \eqref{eq: trace pho omega} we finally get
\begin{equation}
    \label{eq: relative in terms of von neumann}
    S\left(\hat \rho_{\Psi}^{\mathcal{A}_I} \Big| \Big| \hat \rho_{\Omega}^{\mathcal{A}_I} \right) = \Tr \left(\hat \rho_{\Psi}^{\mathcal{A}_I} \ln \hat \rho_{\Psi}^{\mathcal{A}_I} \right) - \Tr \left(\hat \rho_{\Omega}^{\mathcal{A}_I} \ln \hat \rho_{\Omega}^{\mathcal{A}_I} \right) + \langle K^{\mathcal{A}_I}_{\Omega} \rangle_{\Psi} = -S^{\text{v.N}, \mathcal{A}_I}_{\Psi|\Omega} + \langle K^{\mathcal{A}_I}_{\Omega} \rangle_{\Psi}
\end{equation}
where we see appearing the \emph{renormalized von Neumann entropy} of $\ket{\Psi}$ with respect to $\ket{\Omega}$
\begin{equation}
    \label{eq: def von neumann entropy}
    S^{\text{v.N}, \mathcal{A}_I}_{\Psi|\Omega} = \Tr \left(\hat \rho_{\Omega}^{\mathcal{A}_I} \ln \hat \rho_{\Omega}^{\mathcal{A}_I} \right) -  \Tr \left(\hat \rho_{\Psi}^{\mathcal{A}_I} \ln \hat \rho_{\Psi}^{\mathcal{A}_I} \right) \, .
\end{equation}
The renormalization of the von Neumann entropy is needed in quantum field theory because of universal ultraviolet and infrared divergences. However, in some Hilbert space $\mathscr{H}_\Om$, we expect that there exists a dense subspace of states sharing the same ultraviolet and infrared behaviors, so that the \textit{difference} of von Neumann entropies makes sense, even if for any state it is itself divergent. Therefore, \eqref{eq: def von neumann entropy} might be well defined even if both terms on the rhs are themselves divergent. In \eqref{eq: relative entropy finite dim}, \eqref{eq: trace pho omega}, \eqref{eq: relative in terms of von neumann} and \eqref{eq: def von neumann entropy} the traces are computed with respect to the complementary algebra to $\mathcal{A}_I$ namely $\mathcal{A}_{II}$.

Even if we cannot write the states $\ket{\Psi}$ and $\ket{\Omega}$ as density matrices with respect to the algebra $\mathcal{A}_I$, e.g. when dealing with $\scri_R$, the relative entropy \eqref{eq: relative entropy} is always well defined, so we can evaluate it. Same for the one-sided modular Hamiltonian \eqref{eq: def one sided mod ham}, as long as $\ket{\Psi}$ has a finite normal-ordered boost energy. Then, we take a state $\ket{\Psi} \in \mathscr{H}_\Om$ in the domain of $K^{\mathcal{A}_I^\scri}_\Om$ and \emph{define} a quantity $S_{\Psi \lvert \Om}^{\text{v.N.}, \mathcal{A}_I^\scri}$ by
\be \label{renovonentqft}
    S_{\Psi \lvert \Om}^{\text{v.N.}, \mathcal{A}_I^\scri} := S^{\mathcal{A}_I^\scri}(\Psi \lvert \lvert \Om) - \langle K_\Om^{\mathcal A_I^\scri} \rangle_\Psi
\ee
and interpret it as a (generalized) renormalized von Neumann entropy. Since $\langle K_\Om^{\mathcal{A}_I^\scri} \rangle_\Om = 0$, \eqref{renovonentqft} vanishes for $\ket{\Psi} = \ket \Om$. By doing this, we have decomposed Araki's relative entropy into two finite components as long as $\ket \Psi$ belongs to the domain of the one-sided modular Hamiltonian $K_\Om^{\mathcal{A}_I^\scri}$. Therefore \eqref{eq: relative in terms of von neumann} is the general decomposition of the relative entropy in terms of one-sided modular Hamiltonian and renormalized von Neumann entropy, valid whatever the type of von Neumann algebras we deal with. It remains to relate the one-sided modular Hamiltonian to a physical quantity of interest, namely the (normal-ordered) stress-energy tensor of the theory, this is to be done in Appendix \ref{app: modular boost energy}.


\section{Modular Hamiltonian and boost energy}
\label{app: modular boost energy}

In this Appendix we prove that the modular Hamiltonian for a vacuum state $\ket{\Omega}$ wrt a von Neumann algebra of observables $\mathcal{A}$ (introduced in \eqref{eq: def mod hamiltonian}) can be written as an integral, over the spacetime region whose observables lies in $\mathcal{A}$, of the normal-ordered boost energy.

Consider that $\mathcal{A}$ is the algebra of observables at $\scri_R$ and that we decompose the latter in two regions: region $I$ for $U > U_0$ and region $II$ for $U < U_0$. Like in Appendix \ref{app: modular theory} we consider $\mathscr{H}_{\Omega}$ the GNS Hilbert space representing $\mathcal{A}^\scri$ and we have that $\mathcal{A}^\scri$ decomposes into $\mathcal{A}_I^\scri$ for region $I$ and $\mathcal{A}_{II}^\scri = \left(\mathcal{A}_{I}^\scri\right)'$  for region $II$. The state $\ket{\Omega}$ is cyclic and separating for $\mathcal{A}_I^\scri$ and $\mathcal{A}_{II}^\scri$. The Tomita operator $S_{\Omega}^I$ acts on operators of $\mathcal{A}_I^\scri$ so in particular it acts on the annihilation operator $\hat a_{\omega, I}$ \footnote{Note that $\hat{a}_{\omega, I}$ is not in the Weyl algebra, since it is unbounded. However, we can obtain it in a particular limit of bounded operator.} (defined in Appendix \ref{app: unruh}
eq. \eqref{eq: omega annihilation}) via
\begin{equation}
    \label{eq: Tomita region I}
    S^{I}_{\Omega}\left(\hat a_{\omega, I} \ket{\Omega} \right) = \hat a^{\dagger}_{\omega, I}\ket{\Omega} \, ,
\end{equation}
and similarly for its Hermitian conjugate $\left(S^{I}_{\Omega}\right)^{\dagger}$ which 
is the Tomita operator for the algebra $\mathcal{A}_{II}^\scri$ in region $II$. Therefore
\begin{equation}
    \label{eq: Tomita region II}
    \left(S^{I}_{\Omega}\right)^{\dagger}\left(\hat a_{\omega, II} \ket{\Omega} \right) = \hat a^{\dagger}_{\omega, II}\ket{\Omega} \, .
\end{equation}
Recall also the relations \eqref{eq: first anihil two regions} and \eqref{eq: second anihil two regions} which we display again here for convenience
\begin{align}
    \label{eq: first anihil two regions E}
    \left(\hat{a}_{\omega, II} - e^{-\pi \omega}\hat{a}^{\dagger}_{\omega, I} \right) \ket{\Omega} &= 0 \\
    \label{eq: second anihil two regions E}
    \left(\hat{a}_{\omega, I} - e^{-\pi \omega}\hat{a}^{\dagger}_{\omega, II} \right) \ket{\Omega} &= 0 \, .
\end{align}
They give a relation between the ladder operators of region $I$ and those of region $II$ once they act on the total vacuum state $\ket{\Omega}$. Using \eqref{eq: Tomita region I}, \eqref{eq: Tomita region II}, \eqref{eq: first anihil two regions E} and \eqref{eq: second anihil two regions E} we can show the following relation for the modular operator \eqref{eq: modular operator}
\begin{align}
    \label{eq: one-part state eigenstate}
    \Delta_\Om (a_{\om,I}^\dag \ket{\Om}) &= \left(S_\Om^{I}\right)^{\dag} S_\Om^I (a_{\om,I}^\dag \ket{\Om}) = \left(S_\Om^{I}\right)^{\dag}(a_{\om, I} \ket{\Om}) \nn \\
    &= \left(S_\Om^{I}\right)^{\dag}(e^{-\pi \om} a_{\om, II}^\dag \ket{\Om}) = e^{- \pi \om} \left(S_\Om^{I}\right)^{\dag} (a_{\om, II}^\dag \ket{\Om}) = e^{- \pi \om} S_\Om^{II}(a_{\om, II}^\dag \ket{\Om}) \nn \\ &=  e^{- \pi \om} a_{\om, II} \ket{\Om} = e^{- 2 \pi \om} a_{\om,I}^\dag \ket{\Om}
\end{align}
i.e. the state $\hat a^{\dagger}_{\omega, I}\ket \Omega$ is an eigenstate of the modular operator with eigenvalue $e^{-2\pi \omega}$. This works exactly the same in region $II$ for the state $\hat a^{\dagger}_{\omega, II}\ket \Omega$. By induction we have that
\begin{align}
    \label{eq: N-part state eigenstate I}
    \Delta_{\Omega} \left( \hat a^{\dagger}_{\omega, I}\right)^{N}\ket \Omega &= e^{-2\pi N \omega} \left( \hat a^{\dagger}_{\omega, I}\right)^{N}\ket \Omega \\
    \label{eq: N-part state eigenstate II}
    \Delta_{\Omega} \left( \hat a^{\dagger}_{\omega, II}\right)^{N}\ket \Omega &= e^{-2\pi N \omega} \left( \hat a^{\dagger}_{\omega, II}\right)^{N}\ket \Omega
\end{align}
and we have a similar relation for the states $\hat a_{\om, I}^N \ket{\Om}$ and $\hat a_{\om, II}^N \ket{\Om}$, which are eigenvectors of $\Delta_\Om$ with eigenvalues $e^{2 \pi N \om}$, so that the modular Hamiltonian $K_{\Omega}$ can be expressed as
\begin{equation}
\label{eq: expr total mod ham}
    K_{\Omega} = 2\pi \sum_{\omega} \omega \left(\hat N_{\omega,I} - \hat N_{\omega, II} \right) \, .
\end{equation}
From that we define the one-sided modular Hamiltonian of region $I$ as
\begin{equation}
    \label{eq: expr one sided}
    K^{\mathcal{A}_I^\scri}_{\Omega} = 2\pi \sum_{\omega} \om \left( \hat N_{\omega, I} - \langle \hat N_{\omega, I} \rangle_{\Omega} \right)
\end{equation}
where we have subtracted the vev in the global vacuum $\ket \Omega$ to get a finite quantity. A similar expression to \eqref{eq: expr one sided} can be written for $K^{\mathcal{A}_{II}^\scri}_{\Omega}$ using $\hat N_{\omega, II}$.

Now consider the affine stress-energy tensor of region $I$ and take its doubly null time component $T_{uu}$ where $u = \ln(U - U_0)$. In terms of ladder operators it reads
\begin{align}
    \label{eq: stress tensor in terms of ladder}
    T_{uu} = \partial_{u}\phi \partial_{u}\phi =& -\int_{0}^{+\infty} \int_{0}^{+\infty} \frac{\sqrt{\omega \omega'}}{4\pi} \left(\hat{a}_{\omega, I}\hat{a}_{\omega', I}e^{-i(\omega + \omega') u} + \hat{a}_{\omega, I}^{\dagger}\hat{a}_{\omega', I}^{\dagger} e^{i(\omega + \omega')u}\right) \text{d}\omega \text{d}\omega' \nn \\
    &+ \int_{0}^{+\infty} \int_{0}^{+\infty} \frac{\sqrt{\omega \omega'}}{4\pi} \left(\hat{a}_{\omega, I}^{\dagger}\hat{a}_{\omega', I}e^{i(\omega - \omega') u} + \hat{a}_{\omega, I}\hat{a}_{\omega', I}^{\dagger} e^{i(\omega - \omega')u}\right) \text{d}\omega \text{d}\omega' \, .
\end{align}
The one-sided integral of the normal-ordered stress-tensor is defined by 
\begin{equation}
    \label{eq: modumar Hamiltonian}
    \hat{K}^{I} = \int_{-\infty}^{+\infty} :T_{uu}:_{\Om} \text{d}u = \int_{U_0}^{+\infty} (U - U_0) :T_{UU}:_{\Om} \text{d}U \, ,
\end{equation}
which, in terms of ladder operators gives, from \eqref{eq: stress tensor in terms of ladder}
\begin{equation}
    \label{eq: normal ordered stress tensor}
    \hat{K}^{I} = \int_{0}^{+\infty} \om \left( \hat{a}^{\dagger}_{\omega, I} \hat{a}_{\omega, I} - \langle \hat{a}^{\dagger}_{\omega, I} \hat{a}_{\omega, I} \rangle_\Om  \right)\text{d}\omega \, = \frac{K_\Om^{\mathcal{A}_I^\scri}}{2 \pi} .
\end{equation}
In \eqref{eq: normal ordered stress tensor} we see appearing the number operator of region $I$, namely $\hat N_{\omega, I} = \hat{a}^{\dagger}_{\omega, I} \hat{a}_{\omega, I}$. Similarly, one could have led an exactly similar analysis in region $II$ i.e. $U < U_0$ using the coordinate $u_- = -\ln(U_0 - U)$ and obtain
\begin{equation}
    \label{eq: modular region II}
    \hat{K}^{II} = -\int_{-\infty}^{U_0} (U_0 - U) :T_{UU}:_\Om \text{d}U \, ,
\end{equation}
so that in terms of ladder operators of region $II$ we get an expression of the form
\begin{equation}
    \label{eq: normal ordered stress tensor II}
    \hat{K}^{II} = \int_{0}^{+\infty} \om \left( \hat{a}^{\dagger}_{\omega, II} \hat{a}_{\omega, II} -  \langle \hat{a}^{\dagger}_{\omega, II} \hat{a}_{\omega, II} \rangle_\Om \right)\text{d}\omega = -\frac{K_\Om^{\mathcal{A}_{II}^\scri}}{2 \pi} \, .
\end{equation}
Finally we get the subsequent fundamental result
\begin{equation}
    \label{eq: funda relation normal ordered mod ham}
    2\pi \int_{-\infty}^{+\infty} (U-U_0) :T_{UU}:_{\Omega} \text{d}U = 2\pi \int_{0}^{+\infty} \omega \left( \hat N_{\omega, I} - \hat N_{\omega, II} \right) \text{d}\omega \, .
\end{equation}
Combining the latter equation with \eqref{eq: expr total mod ham} we find the advertised relation between the modular Hamiltonian and the normal-ordered stress-energy tensor, namely
\begin{equation}
    \label{eq: normal ordered and stress tensor}
    K^{\mathcal{A}^\scri}_{\Omega} = 2\pi \int_{-\infty}^{+\infty} (U-U_0) :T_{UU}:_{\Omega} \text{d}U \, .
\end{equation}
The one-sided modular Hamiltonian \eqref{eq: expr one sided} is obtained by restricting the time integral of \eqref{eq: normal ordered and stress tensor} to the spacetime domain corresponding to $\mathcal{A}_I^\scri$ i.e. in the case at hand the region $U > U_0$, see \eqref{eq: normal ordered stress tensor}.

This proof can be immediately extended to the four-dimensional case, and to any notion of time. In particular, if we start with the $\kappa_l$-vacuum states defined from the choice of times \eqref{ulmsoftdef}, one can apply the previous analysis to each sector $l$ and immediately find the expression \eqref{eq: kappalvacmodham}. 


\section{Dual GSL from the Minkowski vacuum}
\label{App: MinkowskivacGSL}

If there is no black hole, $\scri^+_R$ is a complete characteristic data surface for free massless fields (for half of the modes of courses as explained in Section \ref{sec: max scri and einstein}). As emphasized in Section \ref{sec: quantization}, the natural state to consider for massless fields is the Minkowski vacuum $\om_M$ attached to the inertial (affine) time $u := u_+$. Further details on its properties -most notably its symmetry structure- and on the Hilbert space $\mathscr{H}_{0_M}$ obtained as the GNS representation induced by $\om_M$, are provided in subsection \ref{subsect: minkowskivac}. This is the framework adopted, for example, in \cite{Kapec:2016aqd, Bousso:2016vlt, Hollands:2019ajl}. 
To obtain the generalized second law in this set-up, one first has to  restrict the vacuum state $\ket{0_M}$ to a subalgebra $\mathcal{A}_i^\scri \subset \mathcal{A}_0^\scri$ attached to the region $\mathcal{D}_i^\scri = (u_i, + \infty) \times S^2$. The one-sided modular Hamiltonian of the Minkowski vacuum associated to the algebra $\mathcal{A}_i^\scri$ is given by 
\be \label{onesmodhminkvac}
    K_{0_M}^{\mathcal{A}_i^\scri} = 2 \pi \int_{\mathcal{D}_i^\scri} (u - u_i) :T_{uu}:_{0_M} \text{d}u \w \eps_S = 2 \pi \int_{\mathcal{D}_i^\scri} (u - u_i) T_{uu} \text{d}u \w \eps_S 
\ee
that is the one-sided boost energy. It can be shown that the integral boost energy is related to the variation of the renormalized area $\bar{A}$ \cite{Kapec:2016aqd, Ciambelli:2024swv}. To check that it is true, we consider a null Rindler horizon at $v = v_0$. Its spacelike cross sections are orthogonal to the normal $\boldsymbol{n} = \p_v$ which points towards a fixed angular direction $x^A$ on the celestial sphere. Taking the limit $v_0 \rightarrow \infty $, the Rindler plane approaches, in the conformal compactification, a single null generator of $\scri^+_R$ associated with the chosen angle $x^A$. \footnote{Therefore, one studies the radiative data on the celestial sphere geodesic by geodesic, as in \cite{Rignon-Bret:2024mef}.} In this limit, all outgoing radiation reaching $\scri^+_R$ at the geometric angle $x^A$ eventually crosses the corresponding null Rindler horizon. The evolution of its cross-sectional area is therefore governed by the local null Raychaudhuri equation, which can be used to relate the accumulated energy flux to the variation of the renormalized area
\be \label{eq: rayrenarea}
    \p_u^2 \bar{a}(x^A) = - 8 \pi T_{uu}(x^A) \eps_S, \qquad \bar{a}(x^A) = a(x^A) - a_0(x^A)
\ee
at first order, where $a$ is the area of cross section of the plane and  $a_0$ corresponds to the (infinite) area of the Rindler plane when there is no radiation, so that $\bar{a}$ can be interpreted as a renormalized area. The application of the second law to null Rindler planes necessarily requires such a renormalization, since the physical area of these planes is, strictly speaking, infinite. By considering the family of Rindler planes normal to all solid angles $x^A$ on the celestial sphere and summing the corresponding contributions, we then obtain that 
\be \label{totalrenarea}
    \p_u^2 \bar{A} = - 8 \pi \int_{S} T_{uu} \eps_S \qquad \bar{A} = -\int_S \bar{a}(x^A)
\ee
where $S$ is a cross section $\scri^+_R$. Multiplying both sides of the equation \eqref{totalrenarea} by the factor $-(u - u_i)$ we get
\be
    \frac{d}{du} \left[\frac{1}{4}\left(\bar{A} - (u - u_i) \frac{d \bar{A}}{du} \right) \right] = - 2 \pi \int_S (u - u_i) T_{uu} \eps_S
\ee
that is exactly similar to \eqref{thirdrayperphor}, but with $u$ the affine coordinate. We have related the boost Hamiltonian \eqref{onesmodhminkvac} to the variation of renormalized area. Then, by considering a cyclic and separating vector $\ket{\Psi} \in \mathscr{H}_{0_M}$, we can run the derivation of the second law similarly to what we did in Section \ref{sec: GSL proof} and find that 
\be \label{genrind}
    \Delta \left(\frac{\bar{A}}{4G} + S\right) \geq 0
\ee
since the one-sided modular Hamiltonian of the Minkowski vacuum is given by \eqref{onesmodhminkvac}. We recover therefore the result of \cite{Kapec:2016aqd}. Note that \eqref{genrind} is not surprising since it can be obtained directly from the standard proof of the generalized second law on Rindler horizons, as in \cite{Wall:2010cj}.


\bibliographystyle{style}
\bibliography{bibliographe.bib}

\end{document}